\documentclass[peerreview,draftcls,onecolumn,12pt,letterpaper]{IEEEtran}
\ifCLASSINFOpdf
\else
\fi

\usepackage{graphicx}

\usepackage{url}
\usepackage{subfigure}        
\usepackage[usenames]{color}            
\usepackage{amsfonts}%
\usepackage{amssymb}%
\usepackage{mathrsfs}
\usepackage{amsmath}%

\usepackage[latin1]{inputenc}

\newtheorem{definition}{Definition}
\newtheorem{theorem}{Theorem}
\newtheorem{proposition}{Proposition}
\newtheorem{corollary}{Corollary}
\newtheorem{lemma}{Lemma}
\newtheorem{pf}{Proof}
\newtheorem{remark}{Remark}
\newtheorem{ex}{Example}

\begin{document}
%
\title{Shannon-Kotel'nikov Mappings for Analog Point-to-Point Communications}

\author{Pål~Anders~Floor,
        Tor~A.~Ramstad,
\thanks{P. A. Floor is with the Colour Laboratory, Department of Computer Science, Norwegian University of Science and Technology
(NTNU), Gj{\o}vik, Norway (e-mail: paal.anders.floor@ntnu.no). T. A. Ramstad  is Prof. Emeritus at the Department of Electronic Systems, Norwegian University of Science and Technology
(NTNU), Trondheim,
Norway (e-mail: tor.ramstad@ntnu.no).}
\thanks{This work was supported by NTNU via the project CUBAN and the Research Council of Norway (NFR) via the project MELODY nr. 187857/S10.}
\thanks{Parts of this paper have previously been presented at SPAWC 2006~\cite{floor_spawc06}, NORSIG 2006~\cite{floor_norsig06} and ITW 2007~\cite{floor_itw07}.}
}

\maketitle

\begin{abstract}
In this paper an approach to joint source-channel coding (JSCC)
named Shannon-Kotel'nikov mappings (S-K mappings) is discussed.
S-K mappings are continuous, or piecewise continuous direct source-to-channel
mappings operating directly on amplitude continuous and discrete time signals. Such mappings include several existing JSCC schemes as special cases.
Many existing approaches to analog- or hybrid discrete analog JSCC provide both excellent performance as well as robustness to variable noise level at low delay and relatively low complexity. However, a general theory explaining their performance and behaviour, as well as guidelines on how to construct close to optimal mappings, do not currently exist. Therefore, such mappings are often based on educated guesses inspired by configurations that are known in advance to produce good solutions through numerical optimization methods. The objective of this paper is to develop a theoretical framework for analysis of analog- or hybrid discrete analog S-K mappings which enables calculation of distortion when applying them on point-to-point links, reveal more about their fundamental nature, and provide guidelines for their construction at low (and arbitrary) complexity and delay. Such guidelines will likely help constrain solutions to numerical approaches and help explain why machine learning approaches obtain the solutions they do. The overall task is difficult and we do not provide a complete framework at this stage: We focus on high SNR and memoryless sources with an arbitrary continuous unimodal density function and memoryless Gaussian channels.  We also provide example mappings based on surfaces which are chosen based on the provided theory.
\end{abstract}

\begin{IEEEkeywords}
 Joint source channel coding, analog mappings, distortion analysis, differential geometry, OPTA.
\end{IEEEkeywords}

%
\IEEEpeerreviewmaketitle

\section{Introduction}\label{sec:Introduction}
\IEEEPARstart{O}{ver} the last decades more and more attention has been directed towards miniature devices, for example in-body sensors and miniature electronic modules replacing ceratin neural network function in the brain. For this reason, and several others, it has become important to study  communication systems with low complexity and delay with the highest possible performance. Further, it is crucial to determine performance limits of such schemes.

In this paper we investigate a general set of analog or hybrid discrete-analog joint source-channel coding (JSCC) schemes named  \emph{Shannon-Kotel'nikov mappings} (S-K mappings). S-K mappings operate directly on analog information sources and are known to perform well at low complexity and delay~\cite{Skoglund2006,hekland_floor_ramstad_T_comm,Akyol_rose_ramstad_itw10,Hu_garcia_lamarca_tcom,Akyol2014_TIT,Saidutta_JSAC2021}.

Shannons' \emph{separation theorem} or \emph{information transmission theorem} (see e.g.~\cite[pp. 224-227]{blahut91}) for communication
of a single source over a point-to-point link states that source coding and channel coding can be performed
separately, without any loss compared to a joint technique. To prove that separation is optimal, arbitrary complexity and delay is assumed. With a constraint on complexity and delay, separate source and channel coding (SSCC) does not necessarily result in the best possible performance, as some examples illustrate: It was shown in~\cite{Goblick1965,BergerTufts67} that for an independent and identically distributed (i.i.d.) source and an additive white Gaussian noise (AWGN) channel, both of the same bandwidth, the information theoretical bound\footnote{By \emph{information theoretical bound} we refer to a bound derived with no restriction on complexity and delay.} is achieved by a simple linear source-channel mapping operating on a symbol-by-symbol basis. This result was generalized in~\cite{BergerTufts67,ramst06}  to special combinations of correlated sources and channels with memory. Furthermore, it was shown in~\cite{SchalkBlue}, that with an ideal feedback channel, the information theoretical bound is achieved when the channel-source bandwidth ratio is an integer. This was extended to simple sensor networks in~\cite{Kim_Ramstd_Feedback2010}. However, with limited (or no) feedback, the asymptotic bounds cannot be obtained at finite complexity and delay when source and channel are of different bandwidth or dimension, or in general, when the source and channel are not \emph{probabilistically matched}~\cite{Gastp03}. An open question is what the best possible performance is for this case under complexity and delay constraints. Efforts dealing with this issue are Kostina and Verdú~\cite{Kostina_TIT2013,Kostina_Verdu_ITW2012} and Merhav~\cite{Merhav_N,Merhav_Vector_2020}.

Several analog and semi-analog JSCC schemes for the bandwidth mismatch case, operating at low and arbitrary complexity and delay, have been suggested in the literature: The \emph{analog matching} scheme in~\cite{Kochman_Zamir} is a structured semi-analog approach built on \emph{lattices} that achieves the information theoretical bounds in the limit of infinite complexity and delay for any colored Gaussian source transmitted on any colored Gaussian channel. However, the performance of analog matching scheme in the finite complexity and delay regime is, to our knowledge, unknown at present. Schemes that are known to perform well at low complexity and delay are the \emph{hybrid digital-analog} (HDA) schemes in~\cite{McRae1971,Coward00b,mittal02,Skoglund2006,Kleiner_rimoldi,Prabhakaran_puri_ramachndran_hda}, certain analog mappings like the \emph{Archimedes Spiral}~\cite{SYChung00,ramstad-telektronikk,Thomas1975} and mappings found by machine learning~\cite{Saidutta_JSAC2021}.

The approach to JSCC studied in this paper, namely S-K mappings, is inspired by many earlier works: First of all, Shannon suggested the use of continuous mappings through space curves as a way of getting close to the information theoretical bounds~\cite{shann49}. Simultaneously, Kotel'nikov developed a theory for analyzing distortion of certain amplitude continuous and time discrete systems realized as parametric curves in $N$ dimensions\footnote{These are basically bandwidth (or dimension) expanding systems with \emph{pulse position modulation} as a special case.} in~\cite{kotelnikov59}. The efforts of Goblick~\cite{Goblick1965}, Berger et al.~\cite{BergerTufts67} and Vaishampayan~\cite{vaish89,Vaisha03a} are pioneering works on this subject. The effort by Gastpar et. al.~\cite{Gastp03} is another important contribution and Merhav's efforts~\cite{Merhav_N,Merhav_Vector_2020} provides insight into the underlying workings of such scheme through analysis based on statistical mechanics. Other important works includes the development of \emph{power constrained channel optimized vector quantizers} (PCCOVQ)~\cite{vaish89,lervi94e,fulds97a}, the HDA schemes in~\cite{McRae1971,Coward00b,mittal02,Skoglund2006}, the linear \emph{block pulse amplitude modulation} (BPAM) scheme in~\cite{LeePetersen76,vaish89}  and the use of parametric curves for both bandwidth expansion~\cite{Vaisha03a} and compression~\cite{SYChung00,ramstad-telektronikk}. Other recent efforts dedicated to analog or semi-analog mappings are found in~\cite{hekland_floor_ramstad_T_comm,HeklandThesis,Cai_mod06,FloorThesis,Kochman_Zamir,Ramstad_NonGauss2008,Hu_garciafrias_lamarca_tcom,Akyol2014_TIT,Chen_wornell}. Lately machine learning was applied to find the optimal structure of such mappings~\cite{Saidutta_JSAC2021}. These efforts illustrate that such schemes perform well at low complexity and delay, some providing excellent performance not matched by any other known scheme. 

Besides Goblikc's~\cite{Goblick65},  Gastpar's~\cite{Gastp03} and Merhav's approaches~\cite{Merhav_N,Merhav_Vector_2020}, there are, as far as we know, no theory providing means to analyze such mappings nor guideline for construction on a general basis. The objective of this paper is therefore to introduce a theoretical framework based on \emph{differential geometry}, encompassing many analog and hybrid discrete-analog schemes. This approach seeks to complement that of Merhav and Gastpar. The proposed theoretical framework facilitate calculation and analysis of the overall distortion in order to reveal the fundamental nature of S-K mappings, as well as guidelines on their construction. The main reason for developing a theory is to gain knowledge on how to optimally construct such mappings in general, not having to rely on educated guesses, numerical optimization sensitive to initial conditions, or machine learning approaches in which little is known about why a certain result is produced.

Treating nonlinear mappings on a general basis is a difficult problem, and we do not present a complete theory at this point, rather introduce a set of tools providing insights  on the construction of S-K mappings. We limit the study to memoryless and independent analog sources drawn from an arbitrary unimodal density function. The sources are transmitted on memoryless, independent Gaussian point-to-point channels, possibly with limited feedback providing channel state information. Generally, the mappings apply when the channel-source dimension (or bandwidth) ratio is a positive rational number. Most of the results provided are proven under the assumption of \emph{high} signal-to-noise ratio (SNR). We focus on low complexity and delay but also consider how these mappings potentially perform by letting their \emph{dimensionality} increase. That is, what gains may be obtained if we increase the mappings dimensions in order to code blocks of samples. Finally, we provide particular examples on mappings chosen based on the provided theory. 

The paper is organized as follows: In Section~\ref{sec:prob_def} the problem is formulated, the information theoretical  limit \emph{OPTA} is introduced, S-K mappings are defined and key concepts from differential geometry are presented. In Section~\ref{sec:Dist_SK} a distortion framework for S-K mappings based on concepts from differential geometry is developed and guidelines for their construction are given. In Section~\ref{sec:AsymptAnalysis_SK} asymptotic analysis is considered and it is shown under which conditions S-K mappings may achieve optimality for Gaussian sources. Section~\ref{sec:Mapping_constr} provides examples on construction of S-K mappings using surfaces to illustrate the theory developed in preceding sections. In Section~\ref{sec:dis_con} a discussion is given.

\section{Problem formulation and preliminaries}\label{sec:prob_def}
Assume a source $\mathbf{x}\in\mathbb{R}^M$, drawn from a continuous unimodal multivariate probability density function (pdf) $f_{\mathbf{x}}(\mathbf{x})$, with i.i.d. components $x_i$. $\mathbf{x}$ is mapped through an  S-K mapping (defined in Section~\ref{ssec:intr_sk_mapp}) to a vector $\mathbf{z}\in \mathbb{R}^N$ which is transmitted over a memoryless channel with average power $P$, so that $1/N \sum_{i=1}^N E\{z_i^2\}\leq P$, and additive Gaussian noise $\mathbf{n}\in \mathbb{R}^N$ with joint pdf $f_{\mathbf{n}}(\mathbf{n})$ with i.i.d. components $n_i\sim \mathcal{N}(0,\sigma_n^2)$.  The channel output $\hat{\mathbf{z}}=\mathbf{z}+\mathbf{n}$ is mapped through an S-K mapping at the receiver to reconstruct the source.

As a measure of performance, the end-to-end mean squared error per source sample between the input- and reconstructed vector, $D_t=(1/M)  E\{\|\mathbf{x}-\hat{\mathbf{x}}\|^2\}$, is considered and compared to the \emph{optimal performance theoretically attainable} (OPTA)~\cite{BergerTufts67}.

\subsection{OPTA}\label{sec:opta}

OPTA in the i.i.d. case is obtained by equating the rate-distortion function for the relevant source with the relevant channel capacity. The equation is solved with respect to the signal-to-distortion ratio (SDR), which becomes a function of the channel signal-to-noise ratio (SNR)~\cite{BergerTufts67}. For the case of Gaussian sources and channels, OPTA is explicitly given by
\begin{equation}\label{eq:OPTA_BWrelation}
\frac{\sigma_x^2}{D_t} =
\left(1+\frac{P}{\sigma_{n}^2}\right)^{f_c/f_s}=\left(1+\frac{P}{\sigma_{n}^2}\right)^{N/M},
\end{equation}
where $\sigma_x^2$ is the source variance, $\sigma_{x}^2/D_t$ is the SDR and $P /\sigma_n^2$ is the channel SNR. Assuming Nyquist sampling and an ideal Nyquist channel, the ratio between channel signalling rate $f_c$, and source sampling rate $f_s$, can be obtained by combining $M$ source samples with $N$ channel samples. That is, $f_c/f_s\approx N/M = r$, where $r$ is a positive rational number ($r\in \mathbb{Q}_+$), named \emph{dimension change factor}. If $r>1$, the channels dimension is higher than that of the source and this can be utilized for noise reduction. If $r \in [0,1)$, the source dimension, and hence the information, has to be reduced in a lossy way before transmission.
We denote the operation where a source of dimension $M$ is mapped onto a channel of dimension $N$ an $M$:$N$ mapping.

\subsection{Shannon-Kotel'nikov mappings}\label{ssec:intr_sk_mapp}
S-K mappings operate
directly on amplitude continuous, discrete time signals. Let $\mathcal{S}$ denote a general S-K mapping and $\mathbf{S}$ a specific realization. We have the following definition:

\begin{definition}\label{def:sk_mapp}\emph{Shannon-Kotel'nikov mapping}\\
An S-K mapping $\mathcal{S}$ is a continuous or piecewise continuous nonlinear or linear mapping between $\mathbb{R}^M$ (source space) and $\mathbb{R}^N$ (channel space). There are three cases to consider:

1. Equal dimension $M = N$: $\mathcal{S}$ is a bijective\footnote{MMSE decoding is needed at low SNR in order to obtain optimality, effectively weakening this condition.}  mapping.

2. Dimension expansion $M<N$: $\mathcal{S}\subseteq \mathbb{R}^N$ is a mapping that can be realized by a hyper surface described by the parametrization\footnote{This is not a restriction, i.e. the mapping does not need to be described by a parametrization.}
\begin{equation}\label{e:par_surf_eq}
\mathbf{S}(\mathbf{x})=[S_1(\mathbf{x}),S_2(\mathbf{x}),\cdots,S_N(\mathbf{x})],
\end{equation}
where each source vector $\mathbf{x}$ should have a unique representation $\mathbf{S}(\mathbf{x})\in \mathcal{S}$. $\mathcal{S}$ is then an M dimensional (locally Euclidean) manifold embedded in $\mathbb{R}^N$.

3. Dimension reduction $M>N$:  $\mathcal{S}\subseteq \mathbb{R}^M$ is a mapping that can be realized by a hyper surface described by the parametrization
\begin{equation}\label{e:par_surf_eq2}
\mathbf{S}(\mathbf{z})=[S_1(\mathbf{z}),S_2(\mathbf{z}),\cdots,S_M(\mathbf{z})],
\end{equation}
where each channel vector $\mathbf{z}$ should have a unique representation $\mathbf{S}(\mathbf{z})\in \mathcal{S}$. $\mathcal{S}$ is then an $N$ dimensional (locally
Euclidean) manifold embedded in $\mathbb{R}^M$. \hspace{5.5cm}$\square$
\end{definition}

Case 1 is trivial for Gaussian i.i.d. sources, i.e. OPTA is obtained by a linear mapping with MMSE decoding at the receiver (often referred to as uncoded transmission)~\cite{Goblick1965}. This paper is concerned with the case $M\neq N$ (case 2 and 3). However, the $M=N$ case fall out as special cases for some of the results given. Piecewise continuity is considered in order to include hybrid discrete-analog (HDA) schemes.
%

\subsection{Relevant concepts from differential geometry}\label{ssec:DiffGeom}
The theory of S-K mappings is based on concepts from differential geometry which may be unknown to some readers. A brief introduction to necessary concepts are given here with more details provided in Appendix~\ref{sec:app_DiffGeom_Details} and~\cite{floor2021tools} which is available online. All concepts presented are taken from Kreyszig's book~\cite{Kreyszig_DiffGeom91}. We use variables $u \in\mathbb{R}$ and $u_i\in\mathbb{R}$ here to keep the discussion general, not specifically referring  to source- or channel variables.

Consider a parametric curve ($1$:$N$ or $M$:$1$ mappings), $\mathcal{C}:\mathbf{S}(u)=[S_1(u), S_2(u), \cdots,S_n(u)]\in\mathbb{R}^n $.
In the following we denote the derivatives with respect to (w.r.t) to a general parameter, $u$, as ${\mathbf{S}}'$, ${\mathbf{S}}''$ etc. In the special case of the parameter being the \emph{arch length}, we denote the derivatives $\dot{\mathbf{S}}$, $\ddot{\mathbf{S}}$ etc. That is, when we parameterize the curve via
\begin{equation}\label{e:CurveLength}
\ell(u)= \int_{u_0}^u \sqrt{\mathbf{S}'\cdot\mathbf{S}'} \mbox{d}u =\int_{u_0}^u \|\mathbf{S}'\| \mbox{d}u.
\end{equation}
Then $\|\dot{\mathbf{S}}\|=\|\mathbf{t}\|=1$, $\forall u$, with $\mathbf{t}$ the curves tangent vector(s) (See Appendix~\ref{sec:app_usp}).

For a curve, $\mathbf{S}(u)$, one can define \emph{curvature} w.r.t. arch length at $u_0$ as $\kappa_0=\|\ddot{\mathbf{S}}(u_0)\|$~\cite[p. 34]{Kreyszig_DiffGeom91}.  Consider arc length parametrization with an amplification $\alpha$, which we name \emph{scaled arc length parametrization}. Then $\|{\mathbf{S}}'(u_0)\|=\alpha\|\dot{\mathbf{S}}(u_0)\|=\alpha$, $\forall u_0$, and $\kappa(u_0)=\|\mathbf{S}_0''(u_0)\|/\|\mathbf{S}_0'\|^2$ according to Appendix~\ref{sec:app_usp}. The \emph{torsion}~\cite[p. 37-40]{Kreyszig_DiffGeom91} is defined as $\tau(x) = {\big|\dot{\mathbf{S}} \ \ddot{\mathbf{S}} \ \dddot{\mathbf{S}}\big|}/{\|\ddot{\mathbf{S}}\|^2}.$
When $\tau = 0$, $\forall x$, we have a plane curve. Whenever $\tau\neq 0$, the curve will \emph{twist} up into space ($\mathbb{R}^n$).

For surfaces, $\mathcal{S}$, with parametric representation as~(\ref{e:par_surf_eq}),~(\ref{e:par_surf_eq2}), we denote  partial derivatives as
\begin{equation}\label{e:PartDer_Not}
\mathbf{S}_\alpha = \frac{\partial \mathbf{S}}{\partial u^\alpha}, \ \ \mathbf{S}_{\alpha\beta} = \frac{\partial^2 \mathbf{S}}{\partial u^\alpha \partial u^\beta}.
\end{equation}
The use of subscripts and superscripts here relates to \emph{Einstein's summation convention} which is described in Appendix~\ref{sec:app_FundForms_Einstein}.

The curvature of a surface $\mathcal{S}$ depends on the choice of \emph{coordinate curves} on $\mathcal{S}$: A curve, $\mathcal{C}$, on surface $\mathcal{S}: \mathbf{S}(u^1,u^2)$ is represented by the parametrization $\mathcal{C}:  \ u^1=u^1(t),  \ u^2=u^2(t)$,
which is $\in C^1$ (the set of differentiable functions), where $t\in \mathbb{R}$. The coordinate curves, $u^1 = $constant and  $u^2 = $constant, corresponds  to parallel curves in the $u^1,u^2$-plane. One must always choose \emph{allowable coordinates} which conditions are provided in~\cite[p.2]{floor2021tools}.

The \emph{normal curvature}, $\kappa_n$, of $\mathcal{S}$ at point $P$ is given by $\kappa_n ={b_{\alpha\beta} \mbox{d}{u^\alpha} \mbox{d}{u^\beta}}/{g_{\alpha\beta} \mbox{d}{u^\alpha} \mbox{d}{u^\beta}}$ (see Appendix~\ref{ssec:app_curvature}, Eqn.~(\ref{e:normals_CurveSurf_angle3})),
where $g_{ij}=\mathbf{S}_i\cdot\mathbf{S}_j$ are the components of the \emph{metric tensor}, or \emph{first fundamental form} (FFF) of $\mathcal{S}$,  and $b_{\alpha\beta}=\mathbf{S}_{\alpha\beta} \cdot  \mathbf{n}$ are components of the \emph{second fundamental form} (SFF) of $\mathcal{S}$,  with $\mathbf{n}$, the \emph{unit normal} to $\mathcal{S}$  at $P$ (see Appendix~\ref{ssec:app_FFF_SFF} for details).

A special case of particular interest is the extremal values of $\kappa_n$, named \emph{lines of curvature} (LoC). If one chooses LoC as coordinate curves then the curvature of $\mathcal{S}$ in those directions, the \emph{principal curvature}, are given by
$\kappa_i = b_{ii}/g_{ii}$, $\forall i$ (see Appendix~\ref{ssec:app_curvature} for details). For general coordinate curves, $\kappa_i $ are the roots of~(\ref{e:PrincipalCurvature_Gen}) in Appendix~\ref{sec:app_FundForms_Einstein}. The normal curvature $\kappa_n$ for any (tangent) direction can be represented in terms of $\kappa_1$ and $\kappa_2$ according to \emph{theorem of Euler}~\cite[p. 132]{Kreyszig_DiffGeom91} (see also~\cite{floor2021tools}) as $\kappa_n=\kappa_1\cos^2\alpha+\kappa_2\sin^2\alpha$, with $\alpha$ the angle between an arbitrary direction at $P$ and the direction corresponding to $\kappa_1$.
%

\section{Distortion analysis for S-K mappings}\label{sec:Dist_SK}
In this section we quantify distortion for S-K mappings.

\subsection{Dimension expanding S-K mappings.}\label{sec:mn_dimexp}
In this section Kotel'nikovs theory from~\cite[pp.62-99]{kotelnikov59} on $1$:$N$ mappings is generalized to include vector sources, enabling analysis of more
general mappings. The results presented in this section are extensions of~\cite{floor_spawc06}.

\begin{figure}[h!]
  \includegraphics[width=0.7\columnwidth]{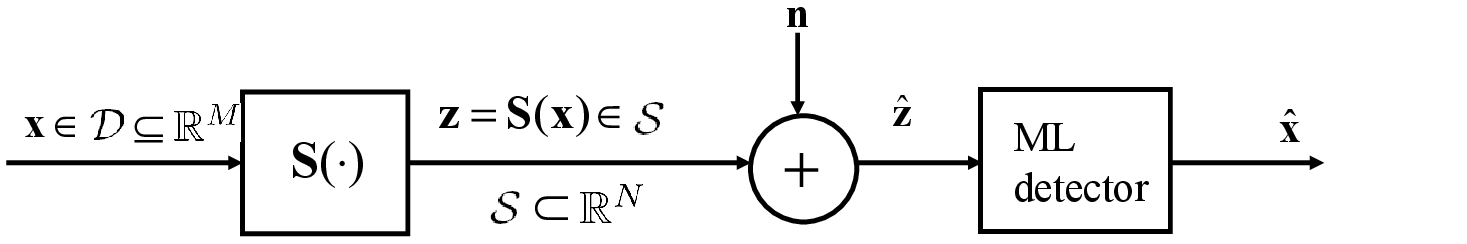}
  \caption{Dimension expanding ($M<N$) communication system for S-K mappings.}\label{fig:exp_sk_system}
\end{figure}

Fig.~\ref{fig:exp_sk_system} depicts the block diagram for a dimension expanding S-K communication system.
Consider source vector $\mathbf{x}\in \mathcal{D} \subseteq \mathbb{R}^M$, with domain $\mathcal{D}$. The source is represented by a  a \emph{signal hyper surface} in the channel space $\mathbf{x}\mapsto \mathbf{S}(\mathbf{x})\in\mathcal{S}\subset \mathbb{R}^N$, (see Definition~\ref{def:sk_mapp}). Applying a specific realization of $\mathcal{S}$, $\mathbf{S}$, the likelihood function of the received signal $\hat{\mathbf{S}}=\mathbf{S}(\mathbf{x})+\mathbf{n}$ is
\begin{equation}\label{e:ml}
f_{\hat{\mathbf{S}}|\mathbf{x}}(\hat{\mathbf{S}}|\mathbf{x})=\bigg(\frac{1}{2\pi\sigma_n^2}\bigg)^{N/2}e^{-\frac{\|\hat{\mathbf{S}}-\mathbf{S}(\mathbf{x})\|^2}{2\sigma_n^2}},
\end{equation}
The maximum likelihood (ML) estimate is then defined as~\cite{therrien}\footnote{Ideally MMSE estimation should be considered but is difficult to deal with analytically for such mappings. This will result in a loss at low SNR. See for example~\cite{Hu_garciafrias_lamarca_tcom}.}
\begin{equation}\label{e:ml_est}
\hat{\mathbf{x}}=\max_{\mathbf{x}\in\mathbb{R}^M} f_{\hat{\mathbf{S}}|\hat{\mathbf{x}}}(\hat{\mathbf{S}}|\mathbf{x}),
\end{equation}
and is maximized by the vector $\mathbf{x}$ that minimizes $\|\hat{\mathbf{S}}-\mathbf{S}(\mathbf{x})\|$. I.e., the ML estimate of $\mathbf{x}$ corresponds to the point on $\mathbf{S}$ closest to the received vector in Euclidean distance.

Ideally one could formulate the exact distortion for any such scheme once a specific representation $\mathbf{S}$ is chosen. However, this is inconvenient when it comes to analysis of the behavior of such mappings as it is usually very hard, if at all possible, to find closed form solutions. For this  reason we use an approach suggested by Kotel'nikov in~\cite[pp.62-99]{kotelnikov59}. Kotel'nikov reasoned that there are two main contributions to the total distortion using such mappings: \emph{low intensity noise} and \emph{strong noise}. Low intensity noise is when the error in the reconstruction at the decoder varies gradually with the magnitude of the noise samples. Distortion due to low intensity noise can be analyzed without reference to a specific $\mathbf{S}$ when the noise can be considered \emph{weak}. The resulting distortion is named \emph{weak noise distortion}, denoted by $\bar{\varepsilon}_{wn}^2$, as defined in section~\ref{sec:mn_veak_noise}. Strong noise is known as \emph{anomalous errors} in the literature, and results from a \emph{threshold effect}\footnote{A thorough treatment of threshold effects, going beyond what we present here, is given in~\cite{Merhav_Vector_2020,Merhav_N_it_2011}}~\cite{shann49}. The resulting distortion is named \emph{anomalous distortion} and denoted by $\bar{\varepsilon}_{an}^2$.

\subsubsection{Weak noise distortion}\label{sec:mn_veak_noise}
\begin{figure}[h]
    \begin{center}
        \subfigure[]{
            \includegraphics[width=.40\columnwidth]{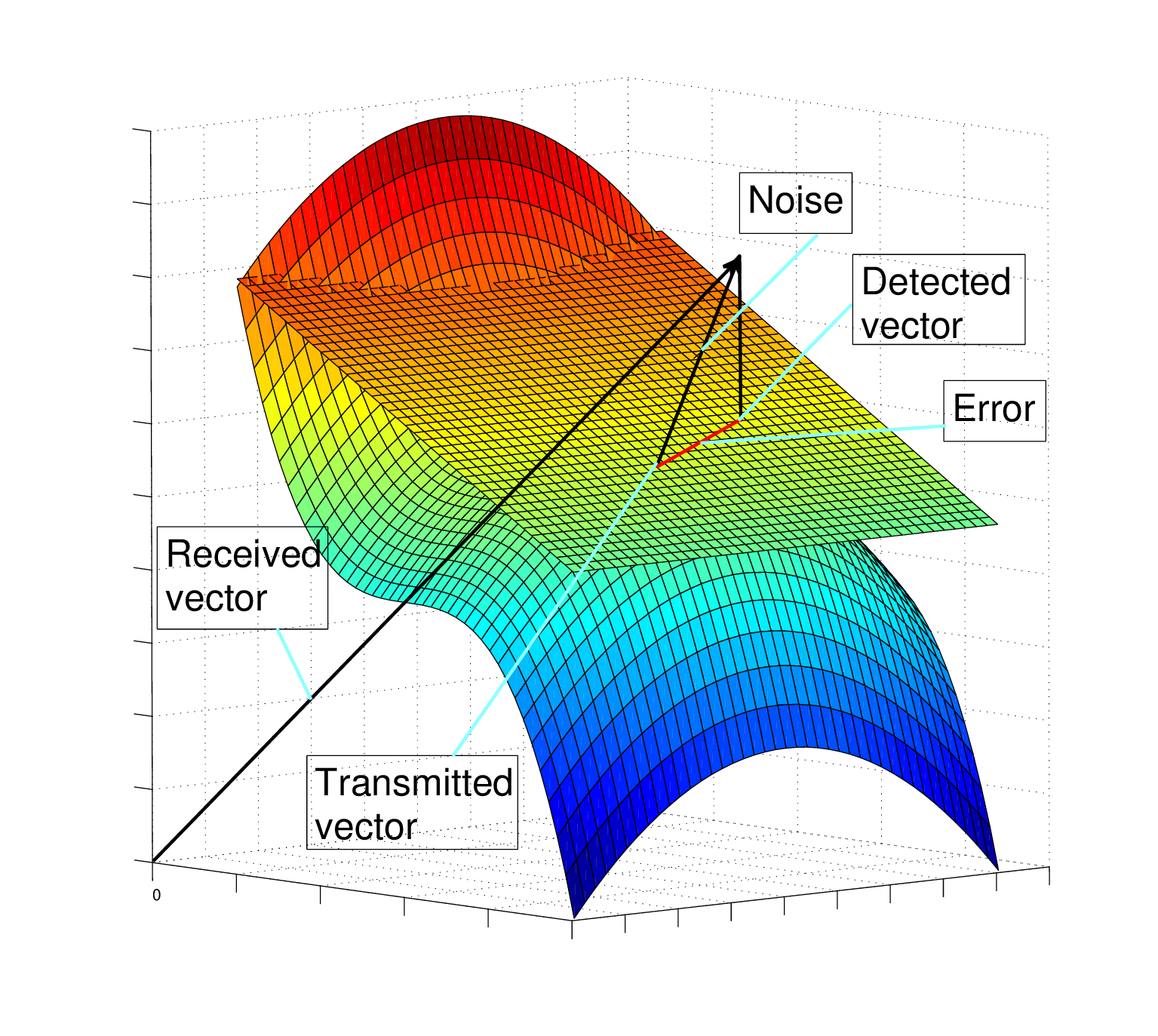}
        \label{fig:smaln_gen}}
        \hfil
        \subfigure[]{
            \includegraphics[width=.40\columnwidth]{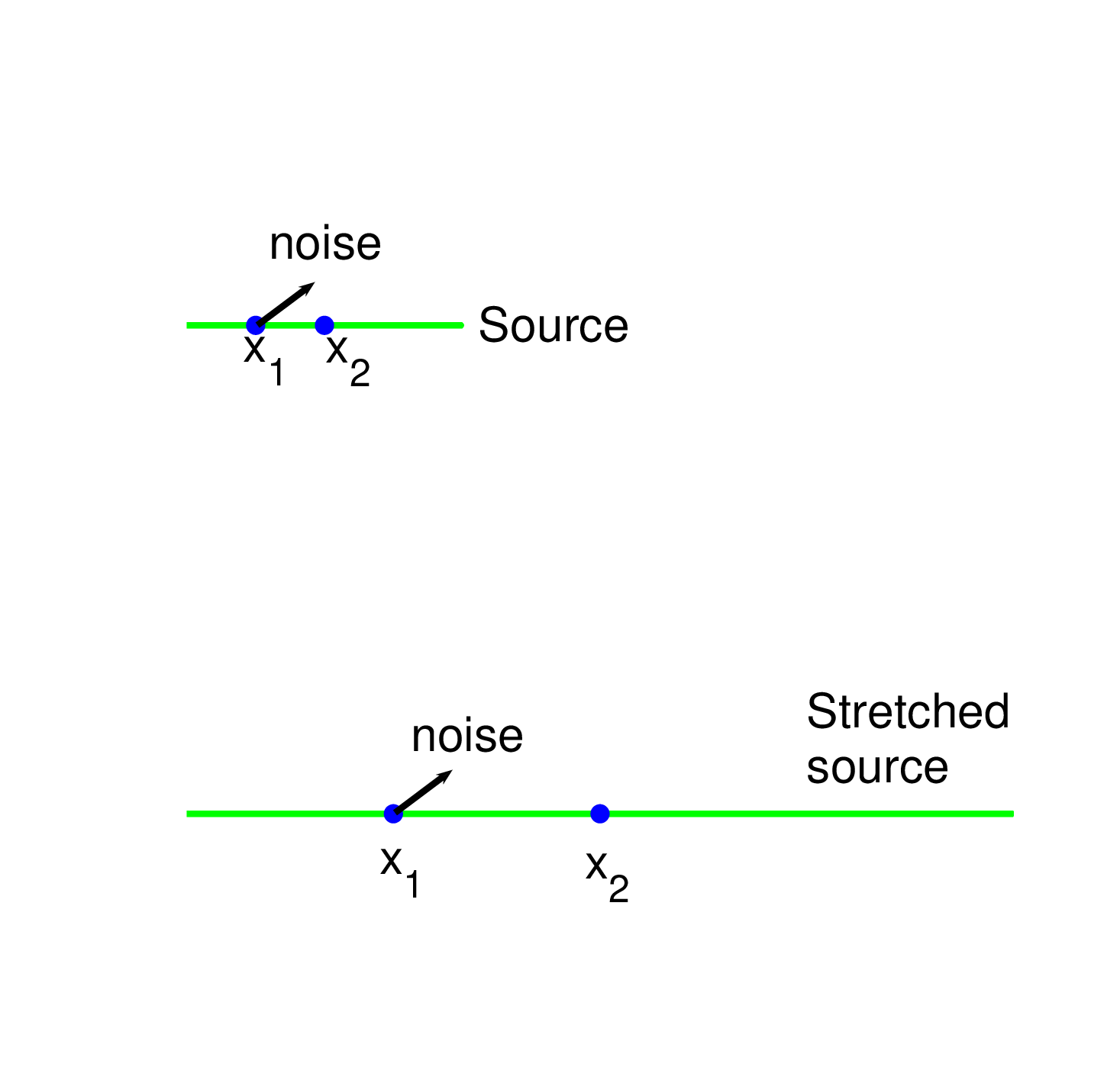}
        \label{fig:kot_concept}}
    \end{center}
    \caption{Dimension expanding S-K mappings.~\ref{fig:smaln_gen} ML estimate approximation for a $2$:$3$ mapping under the weak noise regime.~\ref{fig:kot_concept} Kotel'nikovs concept of analog error reduction for $1$:$N$ mappings.}\label{fig:Dim_exp_wn_concept}
\end{figure}

To analyze non-linear mappings without reference to a specific structure the concepts introduced in Section~\ref{ssec:DiffGeom} and the Taylor expansion apply. We begin by quantifying \emph{weak noise distortion}:  Let $\mathbf{S}_{lin}(\mathbf{x})$ denote 1st order Taylor approximation of $\mathbf{S}(\mathbf{x})$ at $\mathbf{x}_0$
\begin{equation}\label{e:lin_apr_exp}
\mathbf{S}_{lin}(\mathbf{x})= \mathbf{S}(\mathbf{x}_0)+J(\mathbf{x}_0)(\mathbf{x}-\mathbf{x}_0),
\end{equation}
where $J(\mathbf{x}_0)$ denotes the \emph{Jacobian} (see Appendix~\ref{sec:app_FundForms_Einstein}) of $\mathbf{S}$ evaluated at $\mathbf{x}_0$. Fig.~\ref{fig:smaln_gen} shows how the ML estimate is computed by the approximation in~(\ref{e:lin_apr_exp}) for the $2$:$3$ case. We have the following proposition providing the exact distortion under linear approximation:

\begin{proposition}\label{th:weak_noise}\emph{Minimum weak noise distortion}\\
For any continuous i.i.d. source $\mathbf{x}\in \mathbb{R}^M$ with unimodal pdf $f_{\mathbf{x}}(\mathbf{x})$ communicated on an i.i.d. Gaussian channel of dimension $N$ using a continuous dimension expanding S-K mapping $\mathbf{S}$ where $S_i\in
C^r(\mathbb{R}^M), \ r\geq 1, \  i=1,..,N$,  the minimum distortion under the linear approximation in~(\ref{e:lin_apr_exp}) is given by
\begin{equation}\label{e:mseort_mean_exp}
\bar{\varepsilon}_{wn}^2=\frac{\sigma_n^2}{M}\iint
\cdots
\int_{\mathcal{D}}\sum_{i=1}^M\frac{1}{g_{ii}(\mathbf{x})}f_{\mathbf{x}}(\mathbf{x})\mbox{d}\mathbf{x},
\end{equation}
obtained when the metric tensor (or FFF) $G$ of $\mathbf{S}$ (Appendix~\ref{sec:app_FundForms_Einstein}) is diagonal with entries
$g_{ii}=\|\partial\mathbf{S}(\mathbf{x})/\partial x_i\|^2$. I.e., the squared norm of the tangent vector along the i'th coordinate curve.
\end{proposition}

\emph{Proof:} See Appendix~\ref{sec:app_pf_th1}.\hspace{11.0cm}$\square$

The name \emph{weak noise distortion} is due to Definition~\ref{def:weak_noise_exp} given later in this section. Eqn.~(\ref{e:mseort_mean_exp}) states that weak noise distortion becomes smaller by increasing the $g_{ii}$'s. This is equivalent to making tangent vectors at any given point of $\mathbf{S}$ longer, and is obtained by stretching $\mathbf{S}$ like a rubber-sheet. Bending, or cutting, of the signal hyper surface does not reduce weak noise distortion.

The concept is illustrated in Fig.~\ref{fig:kot_concept} for the $1$:$N$ case when $\mathcal{S}$ is a curve.
Stretching of the curve makes source vectors appear longer compared to a given noise vector, or equivalently, the more the source is stretched at the transmitter trough $\mathcal{S}$ the more the noise will be attenuated at the receiver, resulting in smaller distortion. This result by itself implies that the source space should be stretched indefinitely. However, as will be seen in Section~\ref{ssec:sph}, under a channel power constraint, this cannot be done without introducing large anomalous errors.

\begin{remark} Proposition~\ref{th:weak_noise} is extendable to piecewise continuous mappings since one can integrate over each
surface element, then sum all the contributions afterwards (see example in~\cite{McRae1971}).
\end{remark}

The following corollary is a special case of Proposition~\ref{th:weak_noise}:

\begin{corollary}\label{cor:shape_preserv_exp}\emph{Shape preserving mapping}\\
When $\mathbf{S}$ has a diagonal metric $G$ with $g_{ii}(\mathbf{x})=g_{jj}(\mathbf{x})=\alpha, \forall \mathbf{x},i,j$, and with $\alpha$ a constant, then
\begin{equation}\label{e:wn_uniform}
\bar{\varepsilon}_{wn}^2 =\frac{\sigma_n^2}{\alpha^2}.
\end{equation}
That is, all source vectors are equally scaled when mapped through $\mathbf{S}$ and the noise will affect all values of $\mathbf{x}$ equally.
\end{corollary}

\emph{Proof:} Insert  $g_{ii}(\mathbf{x})=g_{jj}(\mathbf{x})=\alpha$ in~(\ref{e:mseort_mean_exp}).\hspace{8.8cm}$\square$

A shape preserving mapping can be seen as an amplification factor $\alpha$ from source to channel. Although a shape preserving mapping leads to simple analysis, its not necessarily optimal in general. A result obtained in~\cite[294-297]{sakrison68} using variational calculus can be used for $1$:$N$ mappings to find the optimal $g_{11}(x)$ for a given source pdf. 

In order to determine the error made in the distortion estimate under linear approximation, we need to consider 2nd order Taylor expansion. We have the following proposition:

\begin{proposition}\label{th:DimExp_2ndOrder}\emph{Weak noise error under 2nd order Taylor approximation}\\
Under 2nd order Taylor approximation, the special case of $1$:$N$ mappings (curves) has an error in the absence of anomalies given by
\begin{equation}\label{e:WeakNoiseDist_2ndOrder_Final}
{\varepsilon}_{wn}^2 \approx \frac{\sigma_n^2}{\|\mathbf{S}_0'\|^2}\bigg(1+\frac{1}{4}\sigma_n^2 \frac{\|\mathbf{S}_0''\|^2}{\|\mathbf{S}_0'\|^4}\bigg) = \frac{\sigma_n^2}{\|\mathbf{S}_0'\|^2}\bigg(1+\frac{1}{4}\sigma_n^2 \kappa^2(x_0)\bigg),
\end{equation}
valid for any S-K mapping $\mathbf{S}(x)\in C^n$, $n\geq 2$. The last equality is true under scaled arc length parametrization. Further, for any dimension expanding  S-K mapping as defined in Definition~\ref{def:sk_mapp}, with LoC coordinate curves, the error is given by
\begin{equation}\label{e:WeakNoiseDist_2ndOrder_M_N}
{\varepsilon}_{wn}^2(\mathbf{x}_0) \approx \frac{\sigma_n^2}{M}\sum_{i=1}^M \bigg\{\frac{1}{g_{ii}(\mathbf{x}_0)}\bigg(1+\frac{\sigma_n^2}{4} {\frac{b_{ii}^2(\mathbf{x}_0)}{g_{ii}^2(\mathbf{x}_0)}}\bigg)\bigg\} = \frac{\sigma_n^2}{M} \sum_{i=1}^M \bigg\{\frac{1}{g_{ii}(\mathbf{x}_0)}\bigg(1+\frac{\sigma_n^2}{4} \kappa_i^2(\mathbf{x}_0)\bigg)\bigg\}.
\end{equation}
Here, $\kappa_i=b_{ii}/g_{ii}$ is the curvature along coordinate curve $i$ with $b_{ii}$ the diagonal components of the \emph{second fundamental form} (SFF) as described in Appendix~\ref{sec:app_FundForms_Einstein}.
\end{proposition}

\emph{Proof:} See Appendix~\ref{sec:app_pf_prop_WN_2nd_order}.\hspace{11.0cm}$\square$


\begin{remark} We only treat 2nd order Taylor approximation here to simplify analysis. It will be seen in Section~\ref{sec:mn_dimred} that higher order terms are even less influential as $\sigma_n$ is raised to a power twice that of the order, which at high SNR ($\sigma_n << 1$) leads to a negligible contribution.
\end{remark}


Note that in the absence of anomalies, one can characterize distortion for S-K mappings in general without choosing a specific $\mathcal{S}$ in advance, as it is expressed solely w.r.t. FFF and SFF components. This makes it easier to evaluate the distortion analytically for such mappings.

From~(\ref{e:WeakNoiseDist_2ndOrder_M_N}) alone a linear mapping seems convenient as $\kappa_i=0, \forall i $. However, at high SNR linear mappings perform poorly, and with~(\ref{e:WeakNoiseDist_2ndOrder_M_N}) in mind, one would seek nonlinear mappings with the smallest possible $\kappa_i(\mathbf{x})$. Therefore the weak noise regime, as defined next, is a good approximation for any \emph{reasonably} chosen mapping at high SNR.
%

\begin{definition}\label{def:weak_noise_exp}\emph{Weak noise regime (dimension expansion)}\\
Let $\mathbf{x}_0$ denote the transmitted vector and $\mathbf{S}(\mathbf{x}_0)$ its representation in the channel space. We say that we are in the \emph{weak noise regime} whenever the 2nd order term in~(\ref{e:WeakNoiseDist_2ndOrder_M_N}) (the term containing $\kappa_i$), is negligible compared to the 1st order term. That is,~(\ref{e:lin_apr_exp}) is a close approximation to  $\mathbf{S}$ and the weak noise distortion in~(\ref{e:mseort_mean_exp}) provides an accurate approximation to the actual distortion in the absence of anomalies.\hspace{14.5cm}$\square$
\end{definition}

\begin{ex} When is Definition~\ref{def:weak_noise_exp} satisfied? There are at leats three cases:

i) SNR$\rightarrow \infty$ ($\sigma_n\rightarrow 0$): The linear approximation in~(\ref{e:lin_apr_exp}) is exact as $\mathcal{S}$ is locally Euclidean.

ii) $\mathbf{S}$ is linear or HDA: Then $\kappa_i=0, \forall i$. A linear mapping is optimal when SNR$\rightarrow 0$~\cite{LeePetersen76,Akyol12}. HDA systems  are composed of piecewise line- or (hyper) plane patches.

iii) Small maximal principal curvature $\kappa_{\text{max}}$: The smaller $\kappa_{\text{max}}$ is, the larger $\sigma_n$ can be before~(\ref{e:mseort_mean_exp}) becomes inaccurate. This is also inline with solutions resulting from numerical optimization algorithms, which tend to bend less and less the lower the SNR is~\cite{fulds97a,floor_iswcs07}.
\end{ex}

\subsubsection{Anomalous distortion}\label{ssec:sph}
\begin{figure}[h]
    \begin{center}
        \subfigure[]{
            \includegraphics[width=.41\columnwidth]{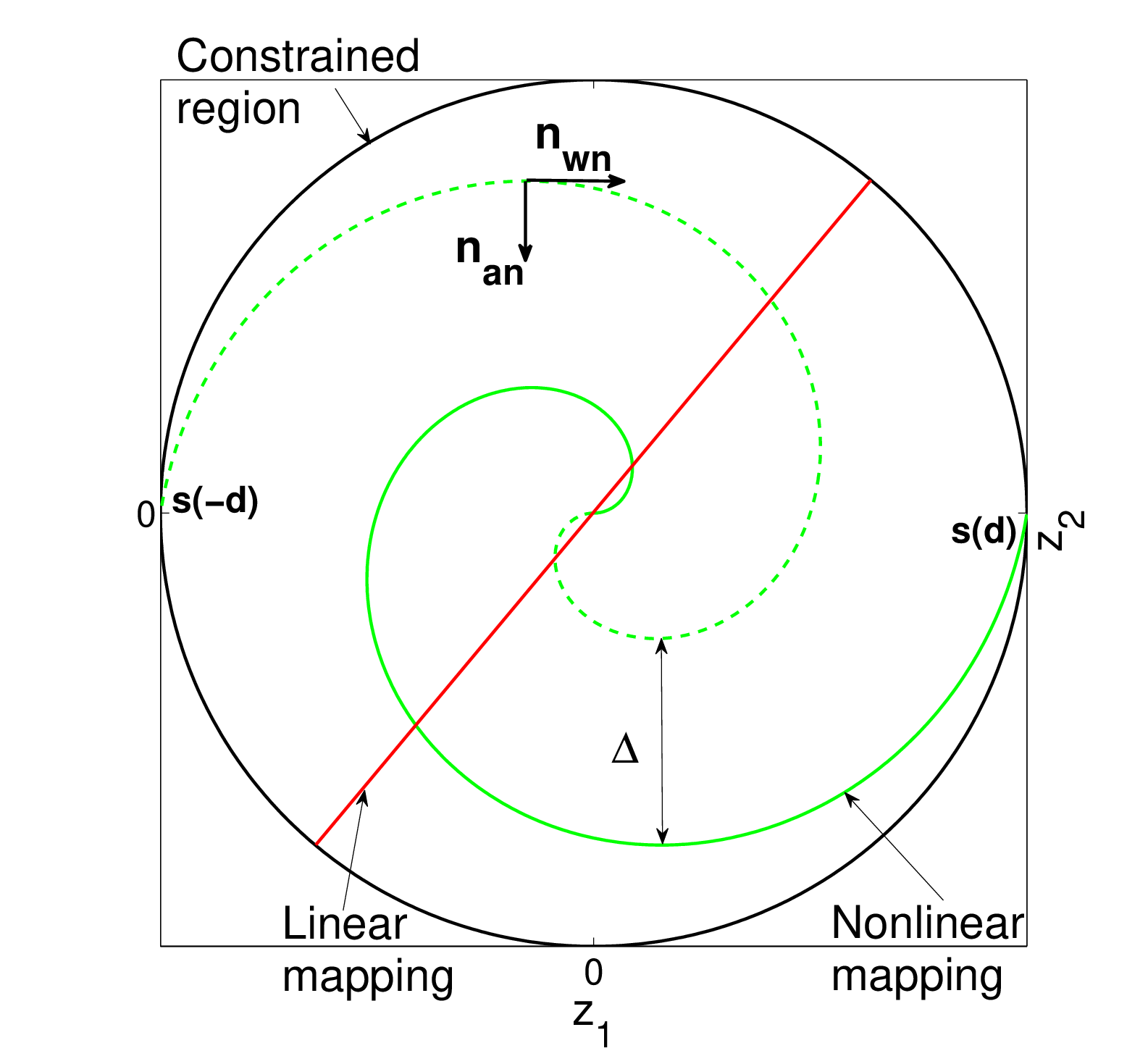}
        \label{fig:dim_exp_norm}}
        \hfil
        \subfigure[]{
            \includegraphics[width=.43\columnwidth]{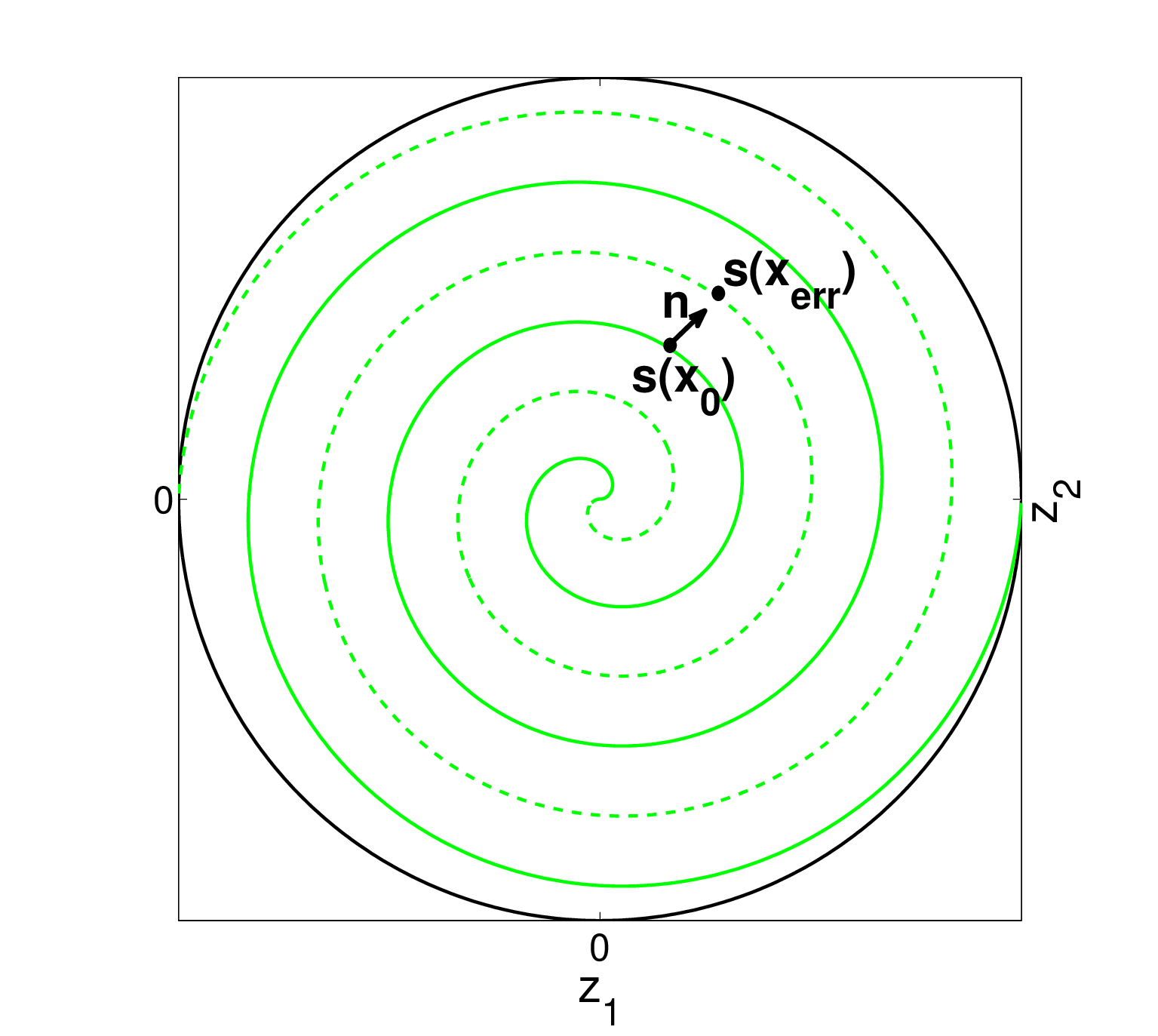}
        \label{fig:dim_exp_anom}}
    \end{center}
    \caption{Example of $1$:$2$ S-K mappings ($\pm d$ denote the boundary of $\mathcal{D}$).~\ref{fig:dim_exp_norm} Linear and nonlinear mappings (negative source values represented by dashed curve).~\ref{fig:dim_exp_anom} As the spiral arms gets close, noise may take the transmitted vector $\mathbf{S}(x_0)$
closer to another fold of the curve leading to large decoding errors.}\label{fig:thres}
\end{figure}

With a channel power constraint, $\mathcal{S}$ must be constrained to lie within some $N-1$ sphere\footnote{The definition of an $N$-sphere is $\mathbb{S}^{N} = \{\mathbf{y} \in
\mathbb{R}^{N+1}|d(\mathbf{y},0)=\text{constant}\}$~\cite[p.7]{Spivak99}, where $d$ is the distance from any point $\mathbf{y}$ on $\mathbb{S}^N$ to the origin of $\mathbb{R}^{N+1}$. E.g. the \emph{sphere} embedded in $\mathbb{R}^3$ is denoted $\mathbb{S}^2$, the ``2-sphere''.}, $\mathbb{S}^{N-1}$. In order to make weak noise distortion small, the relevant hyper surface must first be stretched, then bent and twisted to "fit" within this sphere. Fig.~\ref{fig:dim_exp_anom} illustrates how this can be done in the $1$:$2$ case.
Take a decomposition of the noise $\mathbf{n}$ into a tangential component to the signal curve $\mathbf{n}_{wn}=\mathbf{n}_{||}$, and a normal component  $\mathbf{n}_{an}=\mathbf{n}_{\perp}$ as depicted in Fig.~\ref{fig:dim_exp_norm}. $\mathbf{n}_{wn}$ contributes to weak noise, whilst $\mathbf{n}_{an}$ contributes to \emph{anomalous errors}, which are large errors occurring whenever $\|\mathbf{n}_{an}\|$ crosses a certain threshold. Then the transmitted vector $\mathbf{S}(x_0)$ representing $x_0$, will be detected as the vector $\mathbf{S}(x_{err})$ on another fold of the curve. This happens if the distance, $\Delta$, between the spiral arms is chosen too small w.r.t. $\sigma_n$. Although $\mathbf{S}(x_{err})$ is not far away from $\mathbf{S}(x_0)$ in the channel space, the value it represents, $x_{err}$, is far away from $x_0$ in source space, leading to large reconstruction errors. The occurrence of anomalous errors depends on $\sigma_n$, and the minimum distance $\Delta_{min}$ between folds of $\mathbf{S}$ as well as its curvature. For anomalous errors to occur with small probability, $\Delta_{min}$ should be chosen as large as possible. There is thus a tradeoff between reducing weak noise distortion (where $\Delta_{min}$ should be as small as possible) and anomalous distortion. The exception is at low SNR where linear mappings may do just as well~\cite{LeePetersen76,Akyol12}. In this case anomalous errors  do not occur, and we will always be in the weak noise regime of Definition~\ref{def:weak_noise_exp}  as $\kappa_i = 0, \forall i$  in~(\ref{e:WeakNoiseDist_2ndOrder_Final}) (see Fig.~\ref{fig:dim_exp_norm}).

To quantify anomalous distortion it is convenient to consider \emph{canal surfaces}~\cite[pp. 266-268]{Kreyszig_DiffGeom91}. We begin with curves ($1$:$N$ mappings).

\begin{definition}\label{def:CanalSurf_curve}\emph{Canal surface}\\
A canal surface is the \emph{envelope}, $E$, to the family, $F$, of congruent spheres (or $N-1$ hyper-spheres $\mathbb{S}^{N-1}$), and is the set of all \emph{characteristics} to $F$, defined by~\cite[p. 263]{Kreyszig_DiffGeom91}
\begin{equation}\label{e:CaSu_charachteristic}
S_c(z_i,x)=0, \ \ \frac{\partial S_c(z_i,x)}{\partial x}=0 \ \  i=1,\cdots,N,
\end{equation}
where $S_c=0$  defines a surface in $\mathbb{R}^3$ (or hypersurface in $\mathbb{R}^N$). The characteristic is a curve in $\mathbb{R}^3$ (or a hypersurface of dimension $N-2$ in $\mathbb{R}^N$). The \emph{characteristic points} of  the canal surface are the intersection of the characteristics given by~\cite[p. 266]{Kreyszig_DiffGeom91}
\begin{equation}\label{e:CaSu_CharPoints}
S_c(z_i,x)=0, \ \ \frac{\partial S_c(z_i,x)}{\partial x}=0, \ \ \frac{\partial^2 S_c(z_i,x)}{\partial x^2}=0, \ \  i=1,\cdots,N.
\end{equation}\hspace{16.2cm}$\square$
\end{definition}

An important special case is the family, $F$, of spheres with constant radius $r$ and center on a curve $C: \mathbf{y}(x)$, which can be represented as $S_c(\mathbf{z},x)=(\mathbf{z}-\mathbf{y}(x))\cdot(\mathbf{z}-\mathbf{y}(x))-r^2=0$.
In this case the characteristics of $F$ are circles and the characteristic points are points of intersection of these circles. This concept can be directly applied to $1$:$N$ S-K mappings in Gaussian noise  by setting $\mathbf{y}=\mathbf{S}(x)$, with $\mathbf{z}$ the \emph{channel coordinates} and $x$ the source values.

The extension to $M$:$N$ mappings is straight forward: The \emph{canal hypersurface} of an M-dimensional $\mathcal{S}$ embedded in $\mathbb{R}^N$ is the envelope of the congruent hyper-spheres $\mathbb{S}^{N-M-1}$. We refer to a canal hypersurface simply as ``canal surface'' in the following.

Canal surfaces are important for S-K mappings as they under certain conditions can guarantee low probability for anomalous errors.

\begin{lemma}\label{lem:canal_surf_cond_surfaces}
Consider $M$:$N$ dimension expanding $\mathcal{S}$. Let $\rho_{\text{min}}=1/\kappa_{\text{max}}$, with $\kappa_{\text{max}}$ the maximal principal curvature of $\mathcal{S}$, and $r$ the radius of the hyper-sphere $\mathbb{S}^{N-M-1}$. Further, let $\Delta_{min}$ be the minimum distance between any fold of $\mathcal{S}$ for any $\mathbf{x}$: Then the corresponding canal surface, the envelope of $\mathbb{S}^{N-M-1}$, will not intersect itself at any point. That is, the canal surface will have no characteristic points  $\Longleftrightarrow$  i) $\Delta_{min}> 2 r$ and ii) $\rho_{\text{min}} > r$ for all points of $\mathcal{S}$. 
\end{lemma}

\emph{Proof:} See Appendix~\ref{sec:app_pf_th_Casu}.\hspace{11.0cm}$\square$

\begin{remark} Note that condition ii) is incorporated into condition i). The reason we explicitly state ii) is to constrain the curvature of $\mathcal{S}$ so that it can be removed from the analysis later.
%
\end{remark}

\begin{ex}\label{ex:CanalSurface1_3}
We give an example on a $1$:$3$ mapping. Fig.~\ref{fig:tube} depicts a canal surface surrounding a curve in channel space $\mathbb{R}^3$.
\begin{figure}[h]
    \begin{center}
    \includegraphics[width=0.5\columnwidth]{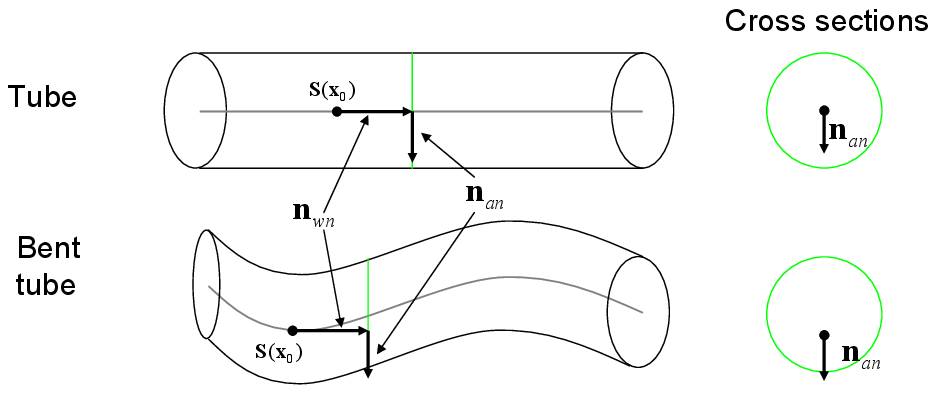}
    \end{center}
\caption{Canal Surfaces: Top figure: Linear signal curve $\mathbb{L}\times \mathbb{S}^{1}$. Bottom figure: nonlinear signal curve $\mathbf{S}\times \mathbb{S}^{1}$.} \label{fig:tube}
\end{figure}
The radius of the canal surface is linked to the noise vector $\mathbf{n}_{an}=\mathbf{n}_{\perp}$. Bending of the tube can increase the probability for anomalous errors, implying that straight lines have the lowest probability for such errors. From this perspective, non-linear  mappings seem to be sub-optimal. However, according to Lemma~\ref{lem:canal_surf_cond_surfaces}, one can circumvent this if the radius of curvature of $\mathbf{S}$ is small enough. The \emph{significant probability mass}\footnote{\emph{significant probability mass} refers to all events except those with very low probability. E.g., like the ``$4\sigma$ loading'' used in~\cite[pp. 124-125]{Jayant_Noll_1984} when constructing scalar quantizers.} of the normalized noise vector $\mathbf{n}_{an}$ is located within a circle of radius $\rho_n=\sqrt{2 b_n^2\sigma_n^2/3}$, with $b_n$ related to the variance of $\mathbf{n}_{an}$ (typically $b_n>4$ incorporates about $99.99\%$ of the probability mass when the dimension of $\mathbf{n}_{an}$ is small).  Therefore, if i) is satisfied, and if
\begin{equation}\label{e:canal_surface_noise_cond_1_N}
 \rho_s > r = \rho_n \geq \sqrt{\frac{N-1}{N} b_n\sigma_n^2},
 \end{equation}
then no characteristic points exists, and the canal surface will not intersect itself.
\end{ex}

We provide a definition of anomalous distortion valid in the vicinity of the optimal operational SNR. That is, we only consider jumps to the nearest point on another fold, $\mathbf{S}(x_{err})$, from a given transmitted point, $\mathbf{S}(\mathbf{x}_0)$ (jumps across several folds may happen as $\sigma_n$ grows, but this is far from optimal). Fig.~\ref{fig:thres} shows the terminology used in the following definition.

\begin{definition}\label{def:an_dist}\emph{Anomalous distortion}\\
Let $\mathbf{x}_0$ denote the transmitted vector and $\mathbf{S}(\mathbf{x}_0)$ its representation in the channel space. Let $\mathbf{n}_{an}$ denote the $K (\leq N)$ dimensional component of a decomposition of the noise vector $\mathbf{n}$ that points in the direction of the closest point $\mathbf{S}(\mathbf{x}_{err})$  on any other fold of $\mathbf{S}$ from $\mathbf{S}(\mathbf{x}_0)$ (as seen in Fig.~\ref{fig:dim_exp_norm}). $\mathbf{x}_{err}(\mathbf{x}_0)$ denotes the reconstructed vector in the case of this anomaly. Let $\Delta_{min}(\mathbf{x}_0)$ denote the Euclidean distance  between $\mathbf{S}(\mathbf{x}_0)$ and $\mathbf{S}(\mathbf{x}_{err})$. Further, let $\rho_{an}=\|\mathbf{n}_{an}\|$ with $f_{\rho_{an}}(\rho_{an})$ its pdf. The probability that $\mathbf{x}_{0}$ is detected as $\mathbf{x}_{err}$ is then\\
\begin{equation}\label{e:an_err_prob}
P_{an}(\mathbf{\mathbf{x}_0})=\int_{\Delta_{min}(\mathbf{x}_0)/2}^\infty f_{\rho_{an}}(\rho_{an})\mbox{d}\rho_{an}.
\end{equation}
The anomalous distortion close to the optimal operational SNR is then defined as
\begin{equation}\label{e:an_dist_gen}
\bar{\varepsilon}_{an}^2= E_{\mathbf{x}}\big\{P_{an}(\mathbf{\mathbf{x}})\|\mathbf{x}-\mathbf{x}_{err}(\mathbf{x})\|^2\big\}.
\end{equation}
\hspace{16cm} $\square$
\end{definition}

%

\subsection{$M$:$N$ Dimension Reducing S-K mappings.}\label{sec:mn_dimred}
\begin{figure}[h!]
  \includegraphics[width=0.7\columnwidth]{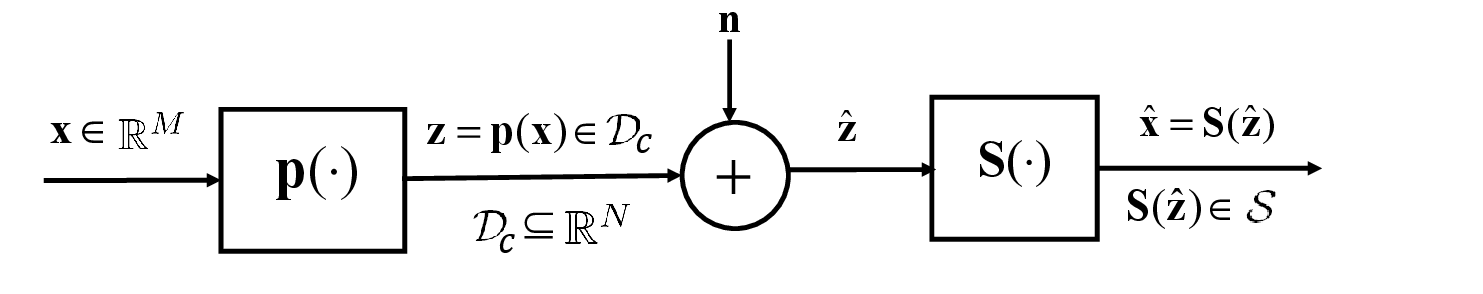}
  \caption{Dimension reducing ($M>N$) communication system for S-K mappings.}\label{fig:red_sk_system}
\end{figure}

Results presented in this section are extensions of~\cite{floor_norsig06}. Fig.~\ref{fig:red_sk_system} shows a block diagram for the dimension
reducing communication system under consideration.
As defined in Section~\ref{ssec:intr_sk_mapp}, a dimension reducing S-K mappings $\mathcal{S}$ is an
$N$ dimensional subset of the source space $\mathbb{R}^M$ that can be realized by a hyper surface $\mathbf{S}$ as in~(\ref{e:par_surf_eq2}), parameterized by
the channel signal $\mathbf{z}$. In this sense, the S-K mapping is a representation of the channel in the source space.

\begin{figure}[h]
    \begin{center}
        \subfigure[]{
            \includegraphics[width=.43\columnwidth]{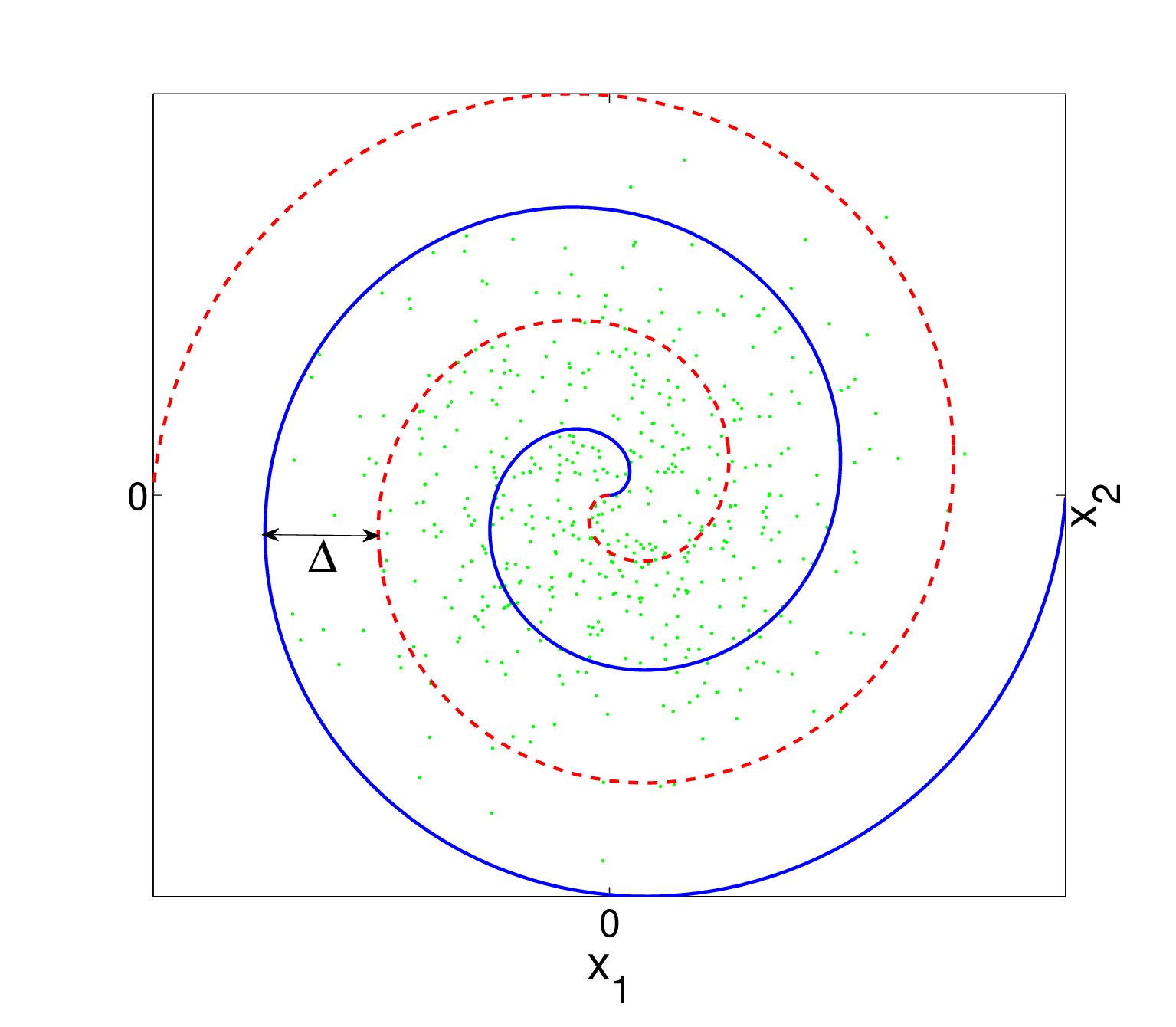}
        \label{fig:dim_red_concept}}
        \hfil
        \subfigure[]{
            \includegraphics[width=.43\columnwidth]{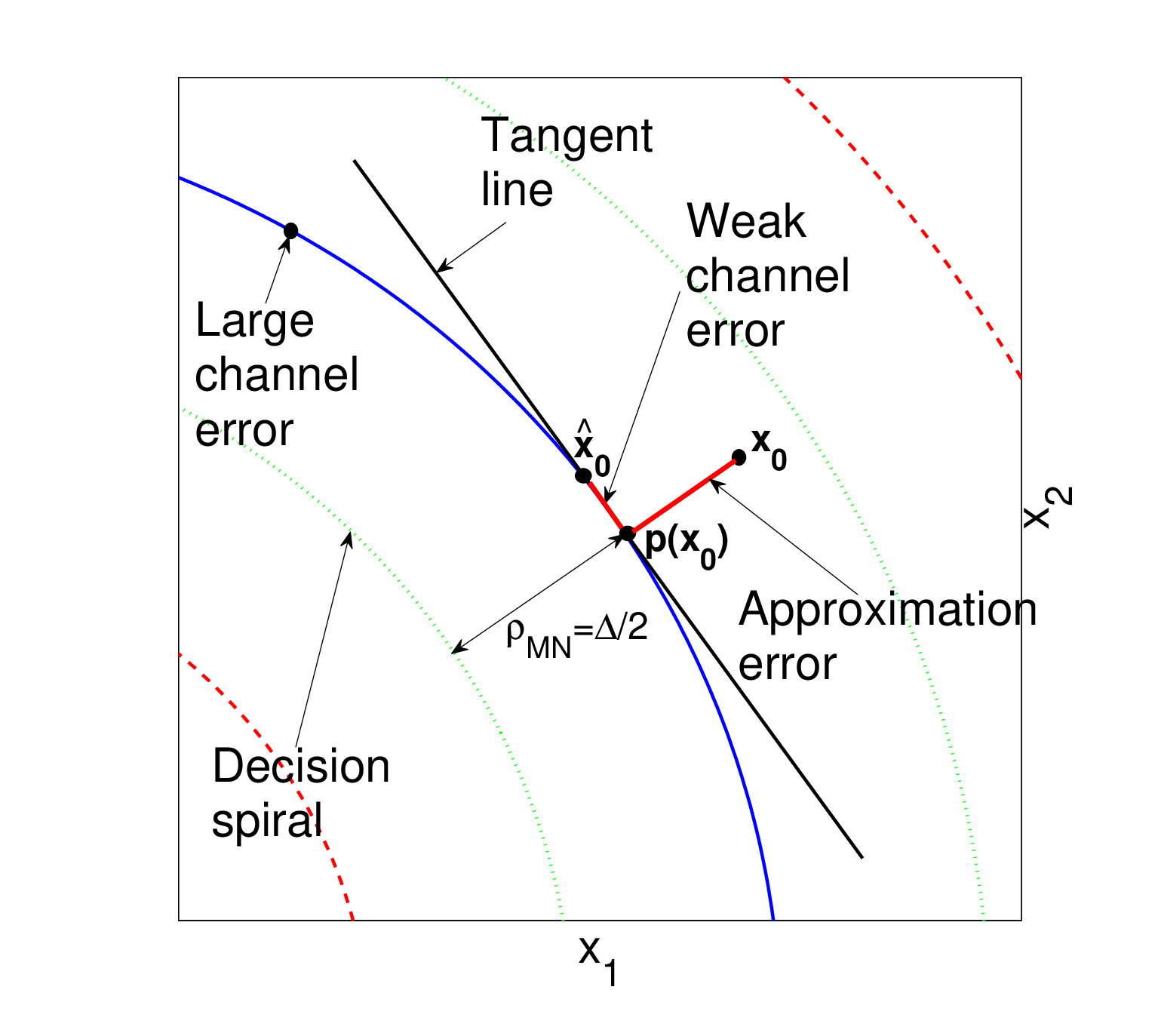}
        \label{fig:dim_red_linapp}}
    \end{center}
    \caption{Dimension reducing S-K mapping in the $2$:$1$ case.~\ref{fig:dim_red_concept} Covering of source space with parametric curve. The dashed lines represent negative channel values. Green dots are source vectors drawn from a 2D Gaussian distribution. ~\ref{fig:dim_red_linapp} Local behavior. The spiral segments are close to osculating circles. }\label{fig:dim_red_explenation}
\end{figure}

To reduce the dimension of a source under a channel power constraint, some of its information content will be lost. The source vectors $\mathbf{x}$ are approximated by their projection onto $\mathcal{S}$, an operation denoted
$\mathbf{q}(\mathbf{x})\in \mathbf{S} \subset \mathbb{R}^M$. The dimension is subsequently changed from $M$ to $N$ by a lossless
operator $\mathbf{d}_r:\mathbf{S}\rightarrow \mathcal{D}_c\subseteq\mathbb{R}^N$, where $\mathcal{D}_c$ is the domain of the channel signal determined by the channel power constraint. The total operation is named \emph{projection operation}, and denoted $\mathbf{p}=\mathbf{d}_r
\circ \mathbf{q}: \mathbf{x}\in \mathbb{R}^M \mapsto \mathbf{p}(\mathbf{x})\in \mathcal{D}_c \subseteq\mathbb{R}^N$. The vector $\mathbf{z}=\mathbf{p}(\mathbf{x})$ is transmitted over an AWGN channel with noise $\mathbf{n}\in\mathbb{R}^N$. Channel noise will lead to displacements of the projected source vector along $\mathcal{S}$. With a continuous $\mathcal{S}$, the distortion due to channel noise will be gradually increasing with $\sigma_n^2$, i.e. no anomalous errors will occur. However, anomalous errors may occur if $\mathcal{S}$ is piecewise continuous (like HDA schemes). Considering ML detection, the reconstructed vector is $\hat{\mathbf{x}}=\mathbf{S}(\hat{\mathbf{z}})$. The concept is illustrated for a $2$:$1$ mapping in Fig.~\ref{fig:dim_red_linapp}.

There are two main contributions to the total distortion for continuous $\mathcal{S}$: \emph{approximation distortion} from the lossy projection operation, and \emph{channel distortion} resulting from channel noise mapped through $\mathcal{S}$ at the receiver.

\subsubsection{Channel distortion}\label{ssec:ch_noise}
The received vector $\hat{\mathbf{z}}=\mathbf{z}+\mathbf{n}$ is mapped through $\mathcal{S}$ to reconstruct $\mathbf{x}$. When noise is sufficiently small, distortion can be modelled by considering the tangent space of $\mathcal{S}$. That is, one can consider the linear approximation $\mathbf{S}_{lin}(\mathbf{z}_0)$ of $\mathbf{S}(\mathbf{z})$ at $\mathbf{z}_0$,
\begin{equation}\label{e:lin_appr_red}
\mathbf{S}_{lin}(\mathbf{z}_0+\mathbf{n})=\mathbf{S}(\mathbf{z_0})+J(\mathbf{z_0})\mathbf{n}.
\end{equation}
The following proposition gives the exact distortion under linear approximation:

\begin{proposition}\label{th:channel_dist}\emph{Minimum Weak Channel Distortion}\\
For any continuous i.i.d. Gaussian channel of dimension $N$ and any dimension reducing S-K mapping $\mathbf{S}$ where $S_i\in
C^r(\mathbb{R}^M), \ r\geq 1, \ i=1,...,M$,  the distortion due to channel noise under the linear approximation in~(\ref{e:lin_appr_red}) is given by
\begin{equation}\label{e:mseort_mean_DimRed}
\bar{\varepsilon}_{chw}^2=\frac{\sigma_n^2}{M}\iint
\cdots \int_{\mathcal{D}_c}\sum_{i=1}^N
g_{ii}({\mathbf{z}})f_{{\mathbf{z}}}({\mathbf{z}})\mbox{d}{\mathbf{z}},
\end{equation}
where $f_{{\mathbf{z}}}({\mathbf{z}})$ is the channel pdf, and $g_{ii}$ the diagonal components of the metric tensor of $\mathbf{S}$.
\end{proposition}

\emph{Proof:} See Appendix~\ref{sec:app_pf_ChDist_lin}.\hspace{11.0cm}$\square$

The name \emph{weak channel distortion} is due to Definition~\ref{def:weak_noise_red} given below. Proposition~\ref{th:channel_dist} states that weak channel distortion increases in magnitude when $\mathcal{S}$ is stretched as the $g_{ii}$'s increases. To keep the channel distortion small, $\mathcal{S}$ should be stretched minimally\footnote{The opposite is sought in the dimension expansion case as an increase of $g_{ii}$ leads to larger attenuation of noise at the receiver side, whereas in the dimension reduction case, increase of $g_{ii}$ will amplify the noise at the receiver.}.

The following corollary is a special case of Proposition~\ref{th:channel_dist}:

\begin{corollary}\label{cor:shape_preserv}\emph{Shape preserving mapping}\\
When $\mathbf{S}$ has a diagonal metric with $g_{ii}(\mathbf{z})=g_{jj}(\mathbf{z})=\alpha, \forall z,i,j$, and $\alpha$ constant, then
\begin{equation}\label{e:e_ch_uniform}
\bar{\varepsilon}_{ch}^2=\frac{N \sigma_n^2}{M} \alpha^2.
\end{equation}
I.e. all channel vectors are equally scaled when mapped through $\mathbf{S}$, and thus noise will affect all source vectors $\mathbf{x}$ equally.
\end{corollary}

\emph{Proof:} Insert  $g_{ii}(\mathbf{z})=g_{jj}(\mathbf{z})=\alpha$ in~(\ref{e:mseort_mean_DimRed}).\hspace{8.5cm}$\square$

Under Corollary~\ref{cor:shape_preserv}, $\mathcal{S}$ can be seen as an amplification $\alpha$ from channel to source at the receiver.

As the channel noise becomes larger,~(\ref{e:mseort_mean_DimRed}) becomes inaccurate as illustrated in Fig.~\ref{fig:dim_red_linapp}. To determine the error under linear approximation, we consider 2nd order Taylor expansion:

\begin{proposition}\label{prop:DimRed_2ndOrder}\emph{Error under 2nd order Taylor approximation (dimension reduction)}\\
Under 2nd order Taylor approximation, in the special case of $M$:$1$ mappings, the error due to channel noise is given by
\begin{equation}\label{e:ChannelDist_2ndOrder_Final}
{\varepsilon}_{ch}^2(x_0) = \sigma_n^2\|\mathbf{S}'_0\|^2 + \frac{3\sigma_n^4}{4}\frac{\|\mathbf{S}''_0\|^2}{\|\mathbf{S}'_0\|^4}= \|\mathbf{S}'_0\|^2\sigma_n^2 + \frac{3\sigma_n^4 \kappa^2(x_0) }{4},
\end{equation}
valid for any S-K mapping $\mathbf{S}(x)\in C^n$, $n\geq 2$. The last equality is true under scaled arc length parametrization. Further, for any dimension reducing  S-K mapping as defined in Definition~\ref{def:sk_mapp}, with LoC coordinate curves, the error is given by
\begin{equation}\label{e:ChannelDist_2ndOrder_M_N}
{\varepsilon}_{ch}^2(\mathbf{x}_0) \approx \frac{\sigma_n^2}{M} \sum_{i=1}^N \bigg({g_{ii}(\mathbf{z}_0)}+\frac{3\sigma_n^2}{4} \kappa_i^2(\mathbf{z}_0)\bigg) = \frac{\sigma_n^2}{M} \sum_{i=1}^N \bigg({g_{ii}(\mathbf{z}_0)}+\frac{3\sigma_n^2}{4} \frac{b_{ii}^2(\mathbf{z}_0)}{g_{ii}^2(\mathbf{z}_0)})\bigg).
\end{equation}
\end{proposition}

\emph{Proof:} See Appendix~\ref{sec:app_pf_prop_WN_2nd_order_red}.\hspace{11.0cm}$\square$

Comparing with dimension expansion in Proposition~\ref{th:DimExp_2ndOrder} we see that distortion is scaled by the components of the SFF (or curvature) in a similar manner. The scaling w.r.t. $g_{ii}$ is different however, corresponding to the results in~(\ref{e:mseort_mean_exp}) and~(\ref{e:mseort_mean_DimRed}). 

\begin{remark} From the proof of Proposition~\ref{prop:DimRed_2ndOrder}, Appendix~\ref{sec:app_pf_prop_WN_2nd_order_red}, Eq.~(\ref{e:Derive_ChDist_M1_3rdOrder}), we have
\begin{equation}\label{e:ChDist_M1_3rdOrder}
{\varepsilon}_{ch}^2(x_0) = \sigma_n^2\|\dot{\mathbf{S}}(x_0)\|^2 + \frac{3\sigma_n^4}{4} \|\ddot{\mathbf{S}}(x_0)\|^2 + \frac{5\sigma_n^6}{12} \|\dddot{\mathbf{S}}(x_0)\|^2= \sigma_n^2 + \frac{3}{4}\kappa_0^2 \sigma_n^4 + \frac{5}{12}\kappa_0^2 \tau_0^2\sigma_n^6,
\end{equation}
for the channel error under 3rd order Taylor expansion. The last equality is a \emph{canonical representation}~\cite[p.48]{Kreyszig_DiffGeom91}, valid for any curve $\mathbf{S}\in C^3$, $r\geq3$. This shows that higher order terms become smaller as $\sigma_n$ decreases, at least for curves with small $\kappa_0$ and $\tau_0$. Referring to Section~\ref{sec:mn_dimexp}, this is the reason why we did not consider Taylor expansion beyond 2nd order there.
\end{remark}


\begin{definition}\label{def:weak_noise_red}\emph{Weak noise regime (dimension reduction)}\\
Let $\mathbf{z}_0$ denote the transmitted vector and $\mathbf{S}(\mathbf{z}_0)$ its representation in the source space. We are in the \emph{weak noise regime} whenever the second (or higher) order term in~(\ref{e:ChannelDist_2ndOrder_M_N}), i.e., the terms containing $\kappa_i$, is negligible compared to the 1st order term. That is,~(\ref{e:lin_appr_red}) is a close approximation to  $\mathbf{S}$ and the weak channel distortion in~(\ref{e:mseort_mean_DimRed}) provides an accurate approximation to the actual distortion due to channel noise.\hspace{13.5cm}$\square$
\end{definition}

\begin{remark} 
Generally the error in the ML-estimate increases with $\kappa_{max}$ (and $\tau$). However, for continuous mappings, $\kappa_{max}$ (and $\tau$) need to be non-zero in order to cover the sources space and thereby keep the approximation distortion low. One should therefore choose a mapping that fills the source space with the smallest possible $\kappa$ (and $\tau$). Alternatively, one may choose HDA systems consisting of parallel lines or planes where $\kappa_i=0$, at the expense of introducing anomalous errors.  Therefore the weak noise regime is a good approximation for any reasonably chosen mappings at high SNR.
\end{remark}

\subsubsection{Approximation distortion}\label{ssec:approx}
Approximation distortion results from the lossy operation $\mathbf{p}$. Its magnitude depends on the average distance source vectors have to $\mathcal{S}$. In order to make
the approximation distortion as small as possible, $\mathcal{S}$ should cover the source space so that every $\mathbf{x}$ is as close to it as possible. Covering of the source space is obtained by stretching, bending and twisting the transformed channel space $\mathcal{S}$ inside the subset of the source space with significant probability mass (an example for the $2$:$1$ case is provided in Fig.~\ref{fig:dim_red_concept}). This is in conflict
with the requirement of reducing channel distortion in which the stretching of $\mathcal{S}$ should be minimized. There is thus a tradeoff between the two distortion contributions.

Since approximation distortion is structure dependent, one cannot find a closed form expression for it in general. However, one can find a
general expression valid for certain \emph{simple} mapping structures that becomes exact as the dimension of the mapping becomes large.

\begin{definition}\label{def:uniform_sk}\emph{Uniform S-K mapping}\\
For an S-K mapping where at each point, $\mathbf{S}(\mathbf{z}_0)$, $\forall \mathbf{z}_0 \in \mathcal{D}_c$, there is a fixed distance $\Delta$ to the nearest point on another fold of $\mathbf{S}$, is named \emph{uniform S-K mapping}.
The maximal approximation error from $\mathbf{x}$ to $\mathbf{S}$ will then be $\Delta/2$ for any $\mathbf{x}$ to any point of $\mathbf{S}$.\hspace{5cm}$\square$
\end{definition}

\begin{remark} Note that for a uniform mapping, any vector being approximated to any point of $\mathcal{S}$ will be confined within a Canal Surface as defined in Section~\ref{ssec:sph}, Definition~\ref{def:CanalSurf_curve}.
\end{remark}

The $2$:$1$ S-K mapping shown in Fig.~\ref{fig:dim_red_concept} is a uniform mapping (except close to the origin). For uniform S-K mappings, a similar distortion lower bound  as that derived for vector quantizers in~\cite{gersho79} can be found for small $\Delta$, i.e., a \emph{sphere bound}~\cite{Conway_Sloane_99}. We have the following proposition:

\begin{proposition}\label{th:apx_dist}\emph{Sphere bound for approximation distortion}\\
For a uniform S-K mapping with distance $\Delta$ between closest points on neighboring folds, the approximation distortion is bounded by
\begin{equation}\label{e:a_noise}
\bar{\varepsilon}_q^2 \geq \frac{M-N}{4M(M-N+2)}\Delta^2.
\end{equation}
As this is a sphere bound, equality is achieved in the limit when $M,N\rightarrow \infty$~\cite{Conway_Sloane_99}, with $N/M=r$ a constant, when $\Delta$ is sufficiently small.
\end{proposition}

\emph{Proof:} See Appendix~\ref{sec:app_pf_Approx_UniSphere}.\hspace{11.0cm}$\square$

\begin{remark} Note that the bound in~(\ref{e:a_noise}) is exact in some low-dimensional cases. For example when $M=2$, $N=1$ using the Archimedes spiral, as this case is equivalent to a scalar quantizer.
\end{remark}


\section{Asymptotic analysis for S-K mappings}\label{sec:AsymptAnalysis_SK}
We investigate how S-K mappings perform as the \emph{dimensionality}\footnote{I.e., letting $M,N$
increase while $r=N/M\in\mathbb{Q}^+$ is kept constant.} (or \emph{block-length}) of the mappings increases. That is, can S-K mappings achieve OPTA as $M,N\rightarrow \infty$ in general?

\subsection{Asymptotic analysis for dimension expanding S-K mappings.}\label{sec:assym_exp}
We determine under which conditions dimension expanding S-K mappings may achieve OPTA for $\forall r\in\mathbb{Q}[1,\infty)$ in the limit $M,N\rightarrow \infty$. We only treat the case of Gaussian sources and channels. The results presented are extensions of~\cite{floor_itw07}.

As proving the existence of hyper surfaces satisfying a distortion criterion is hard, if at all possible, we use a geometrical argument and consider how large volume the transformed source will occupy in the channel space, a generalization of results presented in~\cite[pp.666-674]{wozandj65}.

We start with a proposition concerning anomalous errors in the asymptotic case $M,N\rightarrow\infty$:

\begin{proposition}\label{prop:asympt_an_dist}\emph{Asymptotic anomalous distortion}\\
Let the noise be normalized with the channel dimension $N$, ${\tilde{\mathbf{n}}}=\mathbf{n}/\sqrt{N}$, and let $\Delta_{min}$ denote the smallest distance to the closest point, $\mathbf{S}(\mathbf{x}_{err})$, on any other fold of $\mathbf{S}$ for any transmitted vector $\mathbf{S}({\mathbf{x}_0})$. Furthermore, with $\mathbf{n}_{an}$ the $K (\leq N)$ dimensional component of $\mathbf{n}$ pointing in the direction of $\mathbf{S}(\mathbf{x}_{err})$ from $\mathbf{S}(\mathbf{x}_0)$. Then $\bar{\varepsilon}_{an}^2\rightarrow 0$ as $K,N\rightarrow \infty$ if $\Delta_{min}> 2\sqrt{K/N}\sigma_n$.
\end{proposition}

\emph{Proof:} First consider normalized Gaussian noise vectors ${\tilde{\mathbf{n}}}$. By definition, these vectors have mean length $\sigma_n$. It is shown in~\cite[pp.324-325]{wozandj65} that the variance of $\|\tilde{\mathbf{n}}\|$ decreases as $N$ increases and that $\lim_{N\rightarrow \infty}
\|\tilde{\mathbf{n}}\| =\sigma_n$ with probability one. For $\mathbf{n}_{an}$, a $K (< N)$ dimensional subset of ${\tilde{\mathbf{n}}}$, we get $\|\mathbf{n}_{an}\|=\sqrt{K/N}\sigma_n$ with probability one.\hspace{8.3cm}$\square$

\begin{remark} Proposition~\ref{prop:asympt_an_dist} is equivalent to having no characteristic points for the canal surface as stated in Lemma~\ref{lem:canal_surf_cond_surfaces}. I.e., $1/\kappa_{max} \geq \rho_n \geq  \sqrt{b_n^2 \sigma_n^2(N-M)/N}$, with $K=N-M$, where $b_n\rightarrow 1$ as $M,N \rightarrow \infty$.
\end{remark}

\begin{figure}[h]
    \begin{center}
        \subfigure[]{
            \includegraphics[width=.43\columnwidth]{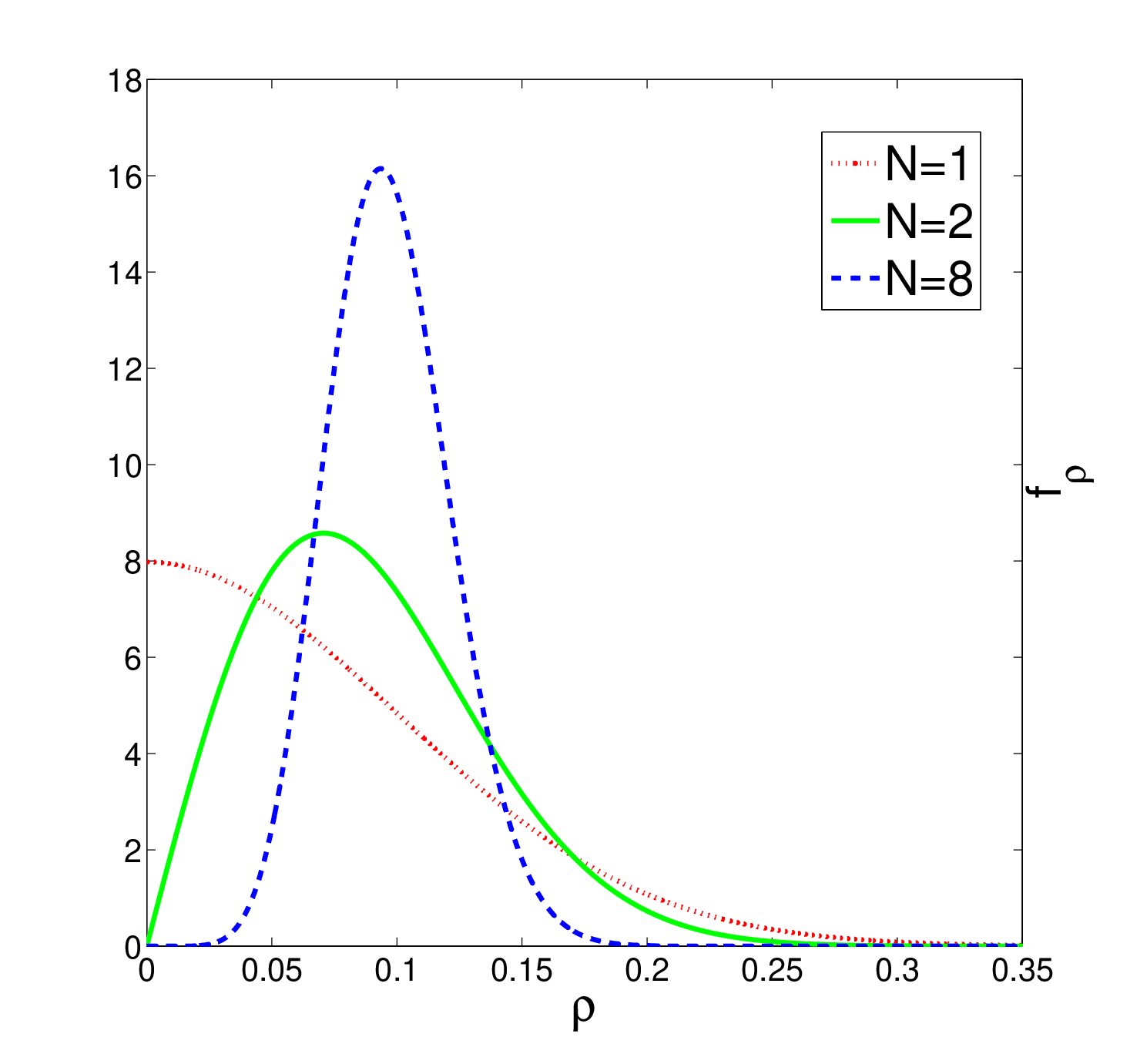}
        \label{fig:noise_pdf}}
        \hfil
        \subfigure[]{
            \includegraphics[width=.43\columnwidth]{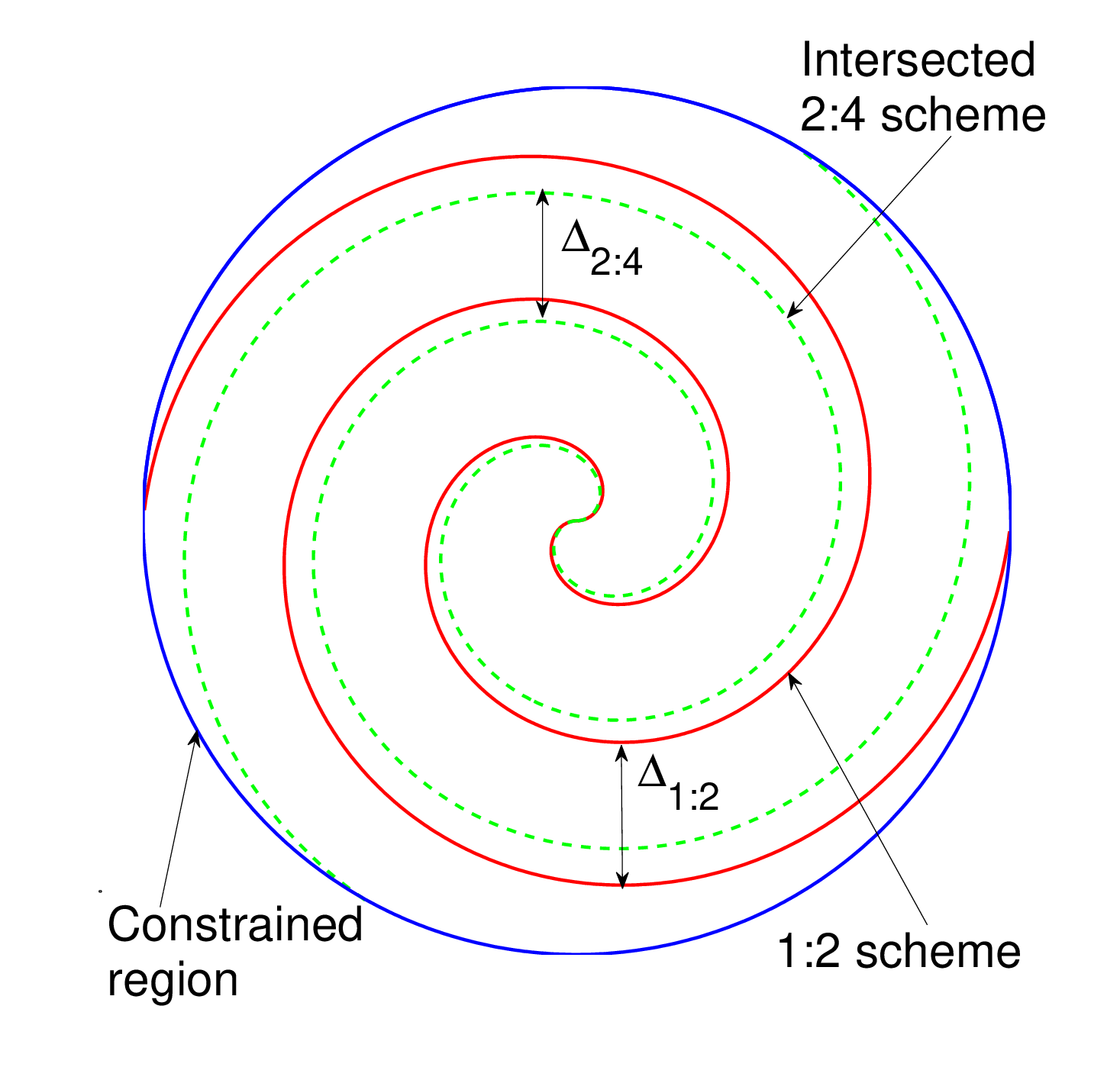}
        \label{fig:dim_gain_exp}}
    \end{center}
    \caption{~\ref{fig:noise_pdf} The pdf of $\rho=\|\tilde{\mathbf{n}}\|$ when $\sigma_n =
0.1$. ~\ref{fig:dim_gain_exp} Performance improvement by increasing mapping dimensionality: The green dashed curve illustrates an
intersected surface. As $\Delta_{2:4}<\Delta_{1:2}$ for the same anomalous error probability,
the $2$:$4$ mapping may be stretched a bit further. This increases the $g_{ii}$'s  and so $\bar{\varepsilon}_{wn}^2$ is reduced.}\label{fig:an_error_char}
\end{figure}

Proposition~\ref{prop:asympt_an_dist} is the key to improve performance by increasing mapping dimensionality: Consider Definition~\ref{def:an_dist}. The distribution of $\rho=\|\tilde{\mathbf{n}}\|$, $\tilde{\mathbf{n}}\in \mathbb{R}^N$, is given by ~\cite[p. 237]{Cramer51}
\begin{equation}\label{e:mspdf}
f_{\rho}(\rho)=\frac{2(\frac{N}{2})^{\frac{N}{2}}\rho^{N-1}}{\Gamma(\frac{N}{2})\sigma_n^N}
e^{-\frac{\frac{N}{2}\rho^2}{\sigma_n^2}},\hspace{1cm}N\geq1,
\end{equation}
where $\Gamma(\cdot)$ is the \emph{Gamma function}~\cite{Bateman53}. Fig.~\ref{fig:noise_pdf} shows~(\ref{e:mspdf}) for selected values of
$N$.
Note that the probability mass of $\rho$ becomes more located around $\sigma_n$ as $N$ increases. Considering this effect w.r.t. S-K mappings, a gain can be obtained when increasing dimensionality as $\Delta_{min}$ can be reduced: Consider $r=2$, which can be accomplished by both $1$:$2$ and $2$:$4$ mappings. Take a $2$:$4$ mapping with diagonal $G$ with  $g_{11}=g_{22}$, both chosen optimally. The $2$:$4$ mapping can then be ``packed''
more densely in the channel space as $f_\rho(\rho)$ narrows. That is, $\Delta_{2:4} < \Delta_{1:2}$. Fig.~\ref{fig:dim_gain_exp} illustrates. The $g_{ii}$'s can therefore be made larger with the $2$:$4$ mapping, effectively reducing $\bar{\varepsilon}_{wn}^2$ and thereby the gap to OPTA. Note that the intersected $2$:$4$ mapping in the figure is just an illustration, not an actual $2$:$4$ mapping (the whole 4-dimensional space has to be considered, as will become apparent from Proposition~\ref{prop:Map_Split} in Section~\ref{sec:Mapping_constr}).

\begin{remark} Linear mappings do not introduce anomalous errors, so they cannot benefit from increased dimensionality. Therefore they are sub-optimal whenever $M\neq N$ except when SNR$\rightarrow -\infty$.
\end{remark}

According to Proposition~\ref{prop:asympt_an_dist}, anomalous errors can be avoided as $M,N\rightarrow\infty$ by making $\Delta_{min} \geq 2\sqrt{K/N} \sigma_n$. We need to determine the smallest obtainable weak noise distortion under this condition without violating a channel power constraint. As will be seen in the following, for a fixed noise variance $\sigma_n^2$, this is the same as satisfying Lemma~\ref{lem:canal_surf_cond_surfaces}.

In order to determine the volume $\mathcal{S}$ occupies in the channel space it must be enclosed within an entity of dimension $N$. Arguments in~\cite[pp. 670-672]{wozandj65} reveal that for $1$:$N$ mappings this entity should be a tube with constant radius $\rho_{MN}\geq\|\mathbf{n}_{an}\|>\sqrt{b_{NM}^2\sigma_n^2(N-1)/N }$ ($b_{NM} \rightarrow 1$ as $N\rightarrow \infty$), with the signal curve at its center. That is, a $N-1$ dimensional \emph{tube} $\mathbf{S}\times \mathbb{S}^{N-2}$, with $\mathbb{S}^{N-2}$ an $N-2$ sphere with radius $\rho_{MN}$. This entity is a canal surface after Definition~\ref{def:CanalSurf_curve} in Section~\ref{ssec:sph}. Locally this canal surface can be approximated by $\mathbb{L}\times \mathbb{S}^{N-2}$, with $\mathbb{L}$ a line-segment. Referring back to Example~\ref{ex:CanalSurface1_3} we locally have $\mathbb{L}\times \mathbb{S}^{N-2}$ as long as the principal curvature $\kappa$ is small enough (for the same reason as in Definition~\ref{def:weak_noise_exp}).

To analyze $M:N$ mappings, $\mathbf{S}\times \mathbb{S}^{N-2}$ must be generalized to enclose M-dimensional hyper surfaces. This is obtained by considering canal hyper surfaces in Section~\ref{ssec:sph} and Definition~\ref{def:CanalSurf_curve}: Thus, we obtain the entity $\mathbf{S}\times\mathbb{S}^{N-M-1}$ which is locally described by $\mathbb{B}^M\times\mathbb{S}^{N-M-1}$. $\mathbb{S}^{N-M-1}$ is an $N-M-1$ sphere with radius $\rho_{MN}$ $\geq \sqrt{b_{NM}^2\sigma_n^2(N-M)/N}$ and $\mathbb{B}^M$  is an M-dimensional \emph{ball} with radius $\rho_M$.  I.e., a spherical region in $\mathbb{R}^M$ with a certain \emph{radius} $\rho_M$. $\rho_M$ will be unbounded in finite dimensional cases, and  as $M\rightarrow \infty$, $\rho_M\rightarrow \sigma_x$.
%

We have the following definition:

\begin{definition}\label{def:local_hycyl_exp}\emph{Local $\mathbb{B}^M\times \mathbb{S}^{N-M-1}$ regime}\\
The S-K mapping $\mathbf{S}$ is locally at the \emph{center} of $\mathbb{B}^M\times \mathbb{S}^{N-M-1}$ if: i) Definition~\ref{def:weak_noise_exp} is satisfied. ii) The distance to the closest point on a different fold of $\mathbf{S}$ is $\Delta_{min}=2\rho_{MN} \geq 2\sqrt{b_{NM}^2\sigma_n^2(N-M)/N}$  at every point $\mathbf{S}(\mathbf{x}_0), \forall \mathbf{x}_0\in\mathcal{D}$. iii) Lemma~\ref{lem:canal_surf_cond_surfaces} is satisfied, i.e., the canal surface $\mathbf{S}\times \mathbb{S}^{N-M-1}$  has no characteristic points.
\hspace{12.5cm} $\square$
\end{definition}

\begin{remark} Condition i) says that $\mathbf{S}$ must be approximately flat inside a sphere of radius $\sigma_n$ as $M,N\rightarrow \infty$ at every point of $\mathbf{S}$. That is, $\kappa_{max}$ must be small so that the 1st order term in~(\ref{e:WeakNoiseDist_2ndOrder_M_N}) dominates. Conditions ii) and iii) are to minimize the effect of anomalous errors. For example, Definition~\ref{def:local_hycyl_exp} is satisfied for a $1$:$3$ mapping if the cylinder in Fig.~\ref{fig:tube} is a valid model locally along the whole curve. To avoid sub-optimal utilization of the channel space, $\rho_{MN}$ should be chosen constant and as small as possible for a given SNR while satisfying Definition~\ref{def:local_hycyl_exp}.
\end{remark}

\begin{remark}For fixed SNR there is an optimal $\rho_{MN}$: If $\sigma_n$ increases the performance will deteriorate due to anomalous
errors, while if $\sigma_n$ decreases there will be un-utilized space available to stretch $\mathcal{S}$ further implying sub-optimal $\bar{\varepsilon}_{wn}^2$. In the latter case the slope of SDR vs SNR will follow that of a linear system according to~(\ref{e:WeakNoiseDist_2ndOrder_M_N}) as the first term dominates.
\end{remark}

We have the following proposition:

\begin{proposition}\label{th:asympt_dim_exp}\emph{Minimum asymptotic distortion for dimension expanding S-K mappings}\\
Assume that $f_\mathbf{x}(\mathbf{x})$ is Gaussian. Any shape preserving dimension expanding S-K mapping satisfying Definition~\ref{def:local_hycyl_exp}, will in the limit $M,N\rightarrow \infty$, for any $r=N/M \in \mathbb{Q}([1,\infty))$, have anomalous distortion $\bar{\varepsilon}_{an}^2 \rightarrow 0$ and potentially obtain weak noise distortion given by
\begin{equation}\label{e:min_weak_noise_dist}
\bar{\varepsilon}_{wn_{min}}^2=\sigma_x^2\bigg( 1+ \frac{P_N}{\sigma_n^2}\bigg)^{-r}
\end{equation}
\end{proposition}

\emph{Proof:} See Appendix~\ref{sec:app_pf_th2}.\hspace{11.3cm}$\square$

We summarize the conditions that dimension expanding S-K mappings must satisfy in the limit $M,N\rightarrow \infty$ to obtain the distortion in~(\ref{e:min_weak_noise_dist}):

1. Definitions~\ref{def:weak_noise_exp} and~\ref{def:local_hycyl_exp} should be satisfied: $\mathcal{S}$ should be nearly flat within a hyper-sphere of radius $\sigma_n$. The larger $\sigma_n$ is, the smaller $\kappa_{max}$ should be, so that the 1st term in~(\ref{e:WeakNoiseDist_2ndOrder_Final}) dominates.

2. Corollary~\ref{cor:shape_preserv_exp} should be satisfied: $\mathcal{S}$ should be shape preserving. This is a sufficient but not necessary condition.

3. At any point $\mathbf{S}(\mathbf{x}_0)\in \mathbf{S}$, $\Delta_{min}>2\sqrt{(1-1/r)\sigma_n^2}$ to avoid anomalous errors. That is, the canal surface $\mathbf{S}\times\mathbb{S}^{N-M-1}$ should satisfy Lemma~\ref{lem:canal_surf_cond_surfaces}.

4. $\mathcal{S}$ should fill the channel space as densely as possible while satisfying 1) and 3) for a given power constraint in order to stretch (amplify) the source as much as possible and thereby minimize $\bar{\varepsilon}_{wn}^2$. A mapping $\mathbf{S}( \mathbf{x})$ with $\rho_{MN}=\Delta_{min}$, $\forall \mathbf{x}$ is then sufficient.


\begin{ex} What S-K mapping would satisfy these conditions? Low dimensional equivalents to such mappings are shown for the $1$:$2$ case in Fig~\ref{fig:opt_exp_structures}.
\begin{figure}[h]
    \begin{center}
        \subfigure[]{
            \includegraphics[width=.43\columnwidth]{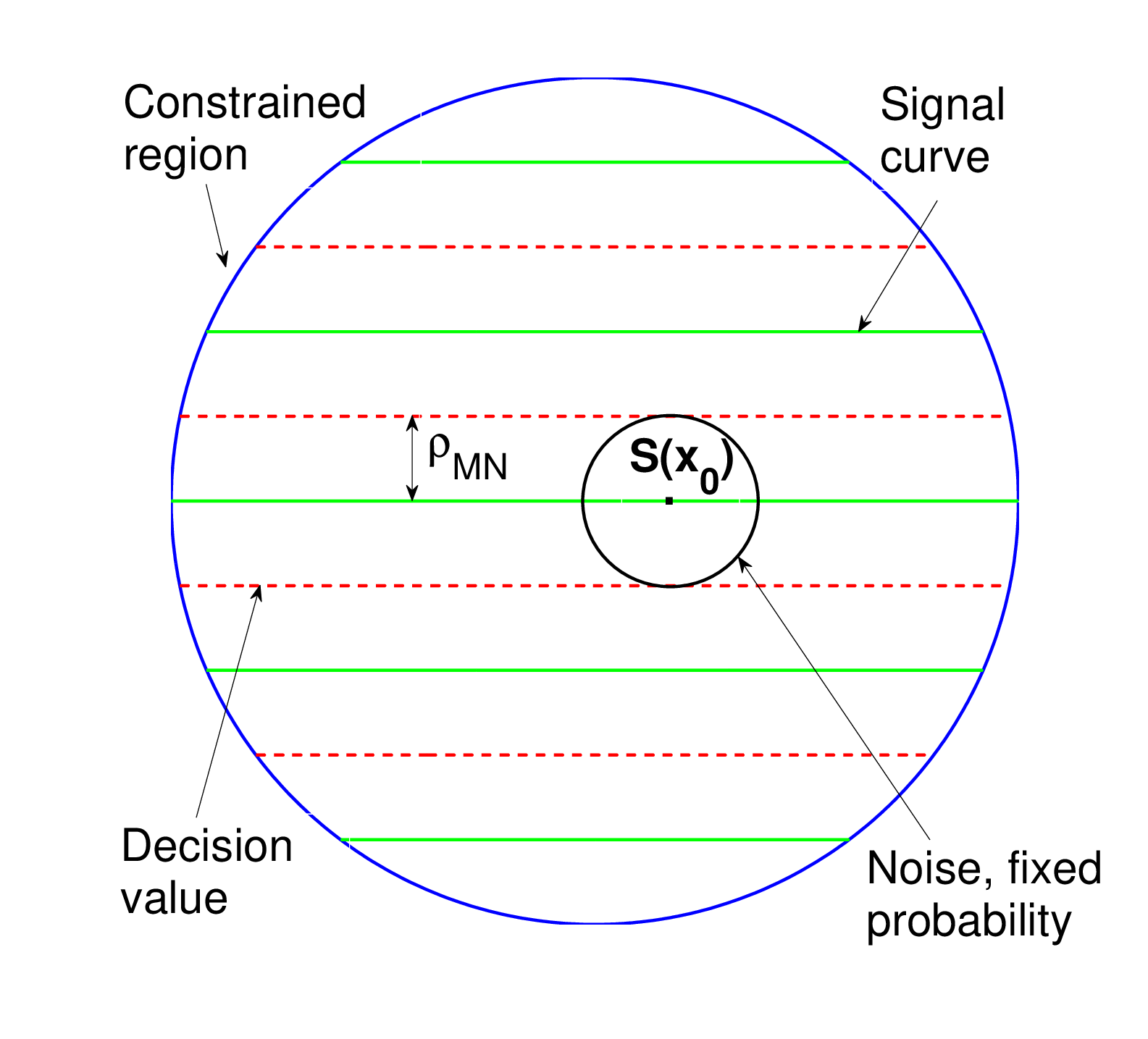}
        \label{fig:HDA_map}}
        \hfil
        \subfigure[]{
            \includegraphics[width=.43\columnwidth]{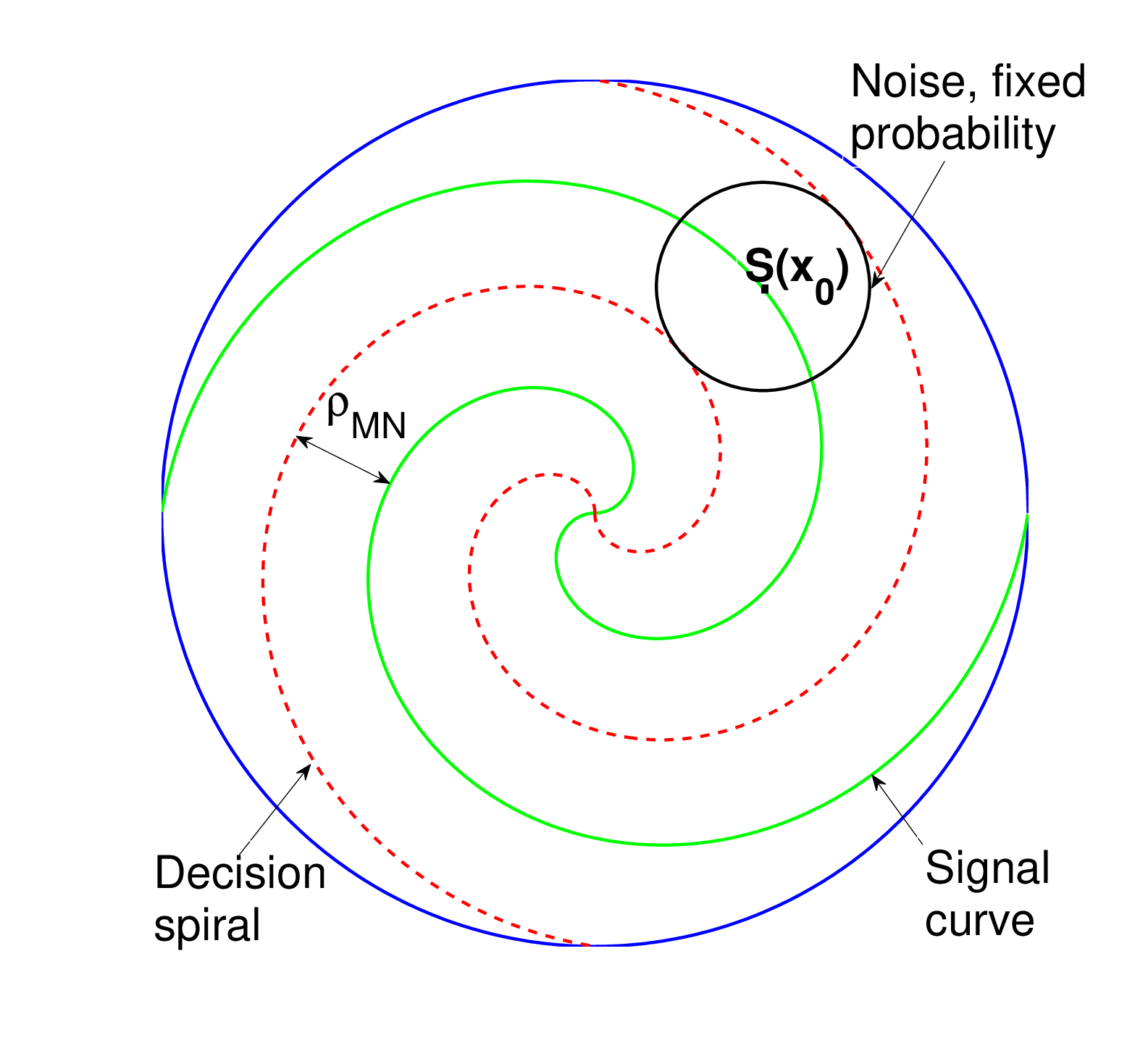}
        \label{fig:Spiral_map}}
    \end{center}
    \caption{Structures that potentially satisfy the necessary and sufficient conditions of Proposition~\ref{th:asympt_dim_exp}. $\rho_{MN}$ should decrease with increasing SNR. (a) Parallel line segments (HDA system) (b) Archimedes spiral.}\label{fig:opt_exp_structures}
\end{figure}
The mapping in Fig.~\ref{fig:HDA_map} potentially fulfill all condition as $\kappa=0$, its uniform,  and it fills the channel space properly. The spiral in Fig.~\ref{fig:Spiral_map} potentially satisfies 2-4, but also 1 as long as $\kappa << \sigma_n^2$. That is, the spiral must have smaller curvature as the SNR drops (obtained by choosing $\Delta_{min}$ lager). This is inline with earlier efforts~\cite{hekland_floor_ramstad_T_comm}. The question is if higher dimensional generalizations satisfying the above conditions can be constructed. The parallel lines mapping in Fig.~\ref{fig:HDA_map} is clearly the simplest to generalize. As will be shown in Section~\ref{sec:Mapping_constr} (Proposition~\ref{prop:Map_Split}) any such mapping cannot be decomposable into lower dimensional sub-mappings.
\end{ex}

\begin{remark} The case $M=N$ is a special case of Proposition~\ref{th:asympt_dim_exp}, where $\bar{\varepsilon}_{an}^2 = 0$, $\bar{\varepsilon}_{wn}^2$ follows~(\ref{e:mseort_mean_exp}) exactly $\forall M,N$, and one can set $g_{ii}=\alpha_i$ (following Corollary~\ref{cor:shape_preserv_exp}). Then~(\ref{e:min_weak_noise_dist}) is obtained even when $M$=$N$=$1$ under MMSE decoding~\cite{Akyol2014_TIT}.
\end{remark}

\subsection{Dimension reducing S-K mappings.}\label{sec:appr_n}
In this section we determine under which conditions dimension reducing S-K mappings may achieve OPTA for $\forall r\in\mathbb{Q}[0,1)$ in the limit $M,N\rightarrow\infty$. We only treat Gaussian sources.

We consider continuous mappings here to avoid anomalous errors. We then need to determine the optimal balance between approximation distortion and channel distortion (as in~\cite{hekland05}). The approximation distortion is determined by the way the $\mathcal{S}$ covers the source space, whereas the channel distortion is determined from the \emph{stretching} of $\mathcal{S}$ necessary to obtain this cover.

For the same reason as in Section~\ref{sec:assym_exp}, we use a volume approach.  Again we need to enclose $\mathcal{S}$ inside a canal surface, now of dimension $M-1$. By similar reasoning as in Section~\ref{sec:assym_exp}, we obtain the canal surface $\mathbf{S} \times \mathbb{S}^{M-N-1}$, now residing in the source space. This canal surface can  locally be approximated as $\mathbb{B}^N \times\mathbb{S}^{M-N-1}$, with $\mathbb{B}^N$ is a ball with radius $\rho_N$, a local representation of the transformed channel space in source space, and $\mathbb{S}^{M-N-1}$, a hyper-sphere with radius $\rho_{MN}$,  corresponding to the decision borders for approximation to a uniform $\mathcal{S}$ (Definition~\ref{def:uniform_sk}).
We have:

\begin{definition}\label{def:local_hycyl_red}\emph{Local $\mathbb{B}^N \times\mathbb{S}^{M-N-1}$ regime}\\
A S-K mapping, $\mathbf{S}$, resides locally at the \emph{center} of $\mathbb{B}^N \times \mathbb{S}^{M-N-1}$ if: i) Definition~\ref{def:weak_noise_red} is satisfied ii) Definition~\ref{def:uniform_sk}
is satisfied with $\Delta=2\rho_{MN}$, where $\rho_{MN}$ is the radius of $\mathbb{S}^{M-N-1}$.
\hspace{1.5cm} $\square$
\end{definition}

Condition i) states that $\mathbf{S}$ must be approximately flat inside a sphere of radius $\alpha \sqrt{b_N}\sigma_n$ at any point $\mathbf{S}(\mathbf{z}_0)$, where $\alpha$ is the amplification factor in~(\ref{e:e_ch_uniform}). Condition ii) ensures uniformity (Definition~\ref{def:uniform_sk}). Note that both i) and ii) will be satisfied iff the canal surface $\mathbf{S}\times \mathbb{S}^{M-N-1}$  has no characteristic points, which limits the maximal principal curvature $\kappa_{max} $. Take the $3$:$1$ case: we then have the canal surface in Fig.~\ref{fig:tube}, but where $\mathbf{n}_{an}$ now corresponds to the approximation error $\mathbf{x}_0-\mathbf{p}(\mathbf{x}_0)$ and $\mathbf{n}_{wn}$ corresponds to the channel error $\mathbf{S}(\hat{\mathbf{x}}_0)-\mathbf{p}(\mathbf{x}_0)$.

We have the Proposition.

\begin{proposition}\label{th:asympt_dim_red}\emph{Minimum asymptotic distortion for dimension reducing S-K mappings}\\
Assume that $f_\mathbf{x}(\mathbf{x})$ is Gaussian. Any shape preserving and continuous dimension reducing S-K mapping satisfying Definition~\ref{def:local_hycyl_red} will in the limit $M,N\rightarrow \infty$, for any $r=N/M \in \mathbb{Q}([0,1])$, potentially obtain the distortion
\begin{equation}\label{e:dist_red_assympt}
D_{min}=\bar{\varepsilon}_{q}^2+\bar{\varepsilon}_{ch}^2=\sigma_x^2\bigg( 1+ \frac{P_N}{\sigma_n^2}\bigg)^{-r}.
\end{equation}
\end{proposition}

\emph{Proof:} See Appendix~\ref{sec:app_pf_th_red}.\hspace{11.5cm}$\square$

We summarize the conditions that a dimension reducing S-K mapping should fulfill in order to satisfy Proposition~\ref{th:asympt_dim_red}:

1. Definitions~\ref{def:weak_noise_red} and~\ref{def:local_hycyl_red} must be satisfied: $\mathbf{S}$ should be approximately flat within a sphere of radius $\alpha\sigma_n$, implying that larger $\sigma_n$  necessitates smaller maximal principal curvature $\kappa_{max}$.

2. $\mathbf{S}$ should be uniform (Definition~\ref{def:uniform_sk}) and shape preserving (Corollary~\ref{cor:shape_preserv}).

3. $\mathbf{S}$ should be continuous to avoid anomalous errors.

4. For fixed approximation distortion, the canal surface of $\mathbf{S}$ should cover the source space with the least possible stretching and curvature to minimize channel distortion.

As for expanding mappings, $\mathbf{S}$ cannot be decomposable into lower dimensional sub-mappings according to Proposition~\ref{prop:Map_Split} in Section~\ref{sec:Mapping_constr}.

What $\mathcal{S}$ satisfies these conditions?  The mapping in Fig~\ref{fig:dim_red_concept} satisfies 2-4 in the finite dimensional case. However, as in the expansion case $\kappa << \sigma_n^2$ if point 1) should be satisfied. A similar mapping to the one shown in Fig.~\ref{fig:HDA_map}, now residing in the source space, clearly satisfies 1), 2) and 4) (as $\kappa=0$) but now 3) is violated. It has been shown that the generalization of such a $2$:$1$ mapping to arbitrary dimensionality can achieve the bound as SNR$\rightarrow \infty$~\cite{Floor_kim_ramstad_ITW2012,Floor_et_al_TCOM_MGMAC_2015}. Condition 3 is therefore not necessary, only sufficient. This also goes for condition 2).


\section{Mapping Construction}\label{sec:Mapping_constr}
Construction of $1$:$N$ or $M$:$1$ mappings follow more or less directly from results and conditions derived for curves throughout this paper as exemplified in~\cite{hekland_floor_ramstad_T_comm}. However, when it comes to surfaces, or hyper surfaces in general, more constraints have to be imposed to guarantee that the mapping is well-performing and follow the same slope as OPTA at high SNR. We consider surfaces in $\mathbb{R}^3$ (if not otherwise stated) in order to obtain simple and explicit results, which can be extended to higher dimensional surfaces and spaces more or less directly. 

Earlier investigations~\cite[pp. 88-89]{FloorThesis} indicated that a diagonal $G$ with $g_{ii}(x_i)=$constant $\forall i$, is convenient as it avoids nonlinear distortion, providing a shape preserving mapping (Corollary~\ref{cor:shape_preserv_exp} and~\ref{cor:shape_preserv})\footnote{A diagonal G arises naturally from~(\ref{e:mseort_mean_exp}) and~(\ref{e:mseort_mean_DimRed}) as only the $g_{ii}$'s contribute.}. Further, for general (source) distributions it can be convenient to choose
\begin{equation}\label{e:diagonal_indep_MT_pdfopt}
G(x_1,x_2)= \text{diag}[g_{11}(x_1), g_{22}(x_2)]
\end{equation}
where $g_{ii}(x_i)$ can be optimized for the relevant source pdf for each coordinate curve on $\mathcal{S}$ (like the method in~\cite[pp.296-297]{sakrison68} for $1$:$N$ mappings). However, as we show later, the metric in~(\ref{e:diagonal_indep_MT_pdfopt}) is not sufficient for a mapping  to follow the same slope as OPTA curve as SNR$\rightarrow \infty$.

%
%
Coordinate curves on $\mathcal{S}$ where $g_{ii}$ only depends on $x_i$ are possible only for certain sub-families of surfaces: An \emph{isometric mapping} between two surfaces $\mathcal{S}$ and $\mathcal{S}^\ast$ are length preserving under the same choice of coordinates. I.e., $g_{\alpha\beta}=g_{\alpha\beta}^\ast$~\cite[pp.176-177]{Kreyszig_DiffGeom91}. Any $\mathcal{S}$ that has a metric like~(\ref{e:diagonal_indep_MT_pdfopt}) can be mapped isometrically to the Euclidean plane, and Theorem 59.3 in~\cite[p.189]{Kreyszig_DiffGeom91} states that this has to be a \emph{developable surface}~\cite[p.189]{Kreyszig_DiffGeom91}:

\begin{definition} \emph{Developable surface:}\\
A \emph{ruled surface} (RS) is obtained by a set of straight lines, $\mathbf{z}(\ell)$, named \emph{generators} interrelated through a space curve $\mathbf{y}(\ell)$, named \emph{indicatrix}~\cite[p.181]{Kreyszig_DiffGeom91}, 
\begin{equation}\label{e:RuSu}
\mathbf{S}(\ell,t)=\mathbf{y}(\ell)+ t \mathbf{z}(\ell).
\end{equation}
$\mathbf{z}$ is a unit vector linearly independent of the tangent $\dot{\mathbf{y}}$, i.e., $\dot{\mathbf{y}}\times \mathbf{z}\neq 0$. $\mathbf{y}(\ell)$, acts like the \emph{trajectory} for a straight line through space, and both $\mathbf{z}$ and $\mathbf{y}$ are coordinate curves on $\mathcal{S}$.

The RS is a developable surface (DS) $\Leftrightarrow$ $\big|\dot{\mathbf{y}} \ {\mathbf{z}} \ \dot{\mathbf{z}} \big| = 0$ (Theorem 58.1 in~\cite[p.182]{Kreyszig_DiffGeom91}).\hspace{1cm}$\square$
\end{definition}

For any DS, $g_{ii}$ can be made constant and equal to 1 $\forall x_i$ by arc length parametrization of $\mathbf{y}$. An example of DS is shown in Fig.~\ref{fig:RCASD_Surface} in Section~\ref{ssec:RCASD_Comp} (a straight line moved along the Archimedes spiral). However, as we show next, any DS will be sub-optimal at high SNR.

When constructing mappings based on surfaces, one simplifying assumption is to construct several parallel and independent systems based on curves (i.e., $1$:$N$ or $M$:$1$ mappings), each one representing a coordinate curve on the resulting surface. This approach was taken in~\cite{Hu_garcia_lamarca_tcom}. This provides a simple way of constructing higher dimensional mappings. However, one cannot obtain optimal performance at high SNR in this way: Take a $m+n$:$2$ mapping when $M>N$ and a $2$:$m+n$ mapping when $M<N$, both realized as two parametric curve-based systems in parallel: a $m$:$1$ and $n$:$1$ mapping when $M>N$ and a $1$:$m$ and $1$:$n$ mapping when $M<N$.


\begin{proposition}\label{prop:Map_Split}\emph{Sub-optimality of decomposable mappings}\\
Any $m+n$:$2$ or $2$:$m+n$ mapping composed of lower dimensional (curve-based) sub-mappings will always have SDR$\sim{\text{SNR}}^{\tilde{r}}$ as SNR$\rightarrow\infty$, with $\tilde{r}$, the dimension change factor of the sub-system with the highest distortion. Therefore, such mappings will diverge from OPTA at high SNR.
\end{proposition}

\emph{Proof:} See Appendix~\ref{sec:app_pf_MapSplit}.\hspace{11.5cm}$\square$

\begin{remark}  A statement for general $M$:$N$ follows from Proposition~\ref{prop:Map_Split} considering several such systems in parallel using power allocation over all sub-systems with water filling~\cite[p.277]{cover06}.
\end{remark}

\begin{remark} As all DS can be seen as a straight line ($1$:$1$ system) moved along a curve $\mathbf{y}$ (a $M$:$1$ or $1$:$N$ system), any DS will diverge from the OPTA bound as SNR grows large, including the suggestion for higher dimensional mappings in~\cite{Hu_garcia_lamarca_tcom}.
\end{remark}


For dimension reducing  mappings, we also have

\begin{corollary}\label{cor:Delta_exponent}
For any uniform dimension reducing $\mathcal{S}$, then $\bar{\varepsilon}_{ch}^2\sim \Delta^{-(M-N)/N}$ to obtain the same slope as OPTA as SNR$\rightarrow\infty$.
\end{corollary}

\emph{Proof:} Follows from the proof of Proposition~\ref{th:asympt_dim_red} in Appendix~\ref{sec:App_SKDist_Assymptotic}, Eqn~(\ref{e:d_tot_init}).\hspace{3cm}$\square$

\begin{ex} Take a uniform $3$:$2$ S-K mapping where $\bar{\varepsilon}_{a}^2\sim \Delta^2$ according to Proposition~\ref{th:apx_dist}. Then we need $\bar{\varepsilon}_{ch}^2\sim {1}/{\Delta}$, in order to obtain $\text{SDR}\sim \text{SNR}^{2/3}$ as SNR$\rightarrow\infty$.
\end{ex}

\begin{remark} A last important condition for S-K mappings~\cite[p.103]{lervik-thesis}: To minimize channel power and reduce the effect of noise it is important that source vectors with highest probability are allocated to channel representations with low power.
\end{remark}

To avoid the problem of non-optimal slope one has to widen the set of mappings beyond DS, keeping a similar type of G as in~(\ref{e:diagonal_indep_MT_pdfopt}). The most direct generalization are surfaces that can be mapped in an angle preserving way, or \emph{conformally}, to the Euclidean plane. The metric between two surfaces $\mathcal{S}$ and $\mathcal{S}^\ast$  are then proportional, i.e. $g^\ast_{\alpha\beta}=\eta(u^1,u^2)g_{\alpha\beta}$~\cite[pp. 193-194]{Kreyszig_DiffGeom91}, with $\eta$ some proportionality factor. A subset of surfaces conformal to the Euclidean plane are shape preserving.

In the rest of this section several $3$:$2$ and $2$:$3$ mappings are evaluated in order to illustrate the results of this paper. There are myriads of known surfaces exemplified in the \emph{Encyclopedia of Analytical Surfaces}~\cite{Surf_Encyclopedia}. The criteria laid down in this paper rules out most of them as potential candidates for S-K mappings.

\subsection{Examples on $3$:$2$ mappings}\label{ssec:Ex_SK_surf_32}
Three $3$:$2$ mappings selected based on intuition obtained through previous sections are evaluated: 1) A DS-based mapping which is simple but decomposable. 2) A mapping which is not decomposable. 3) A hybrid discrete-analog mapping constructed to satisfy all requirements needed to obtain the slope of OPTA at high  SNR.

To evaluate performance of the example mappings we compare them with OPTA and \emph{block pulse amplitude modulation} (BPAM)~\cite{LeePetersen76} which is the optimal linear mapping. Obviously, any choice of nonlinear mapping should rise well above BPAM as the SNR increases. At the end of the section all suggested schemes are compared to existing superior  mappings.

\subsubsection{Right Cylinder with Archimedes Spiral Directrix (RCASD)}\label{ssec:RCASD_Comp}
\begin{figure}[h]
    \begin{center}
        \subfigure[]{
            \includegraphics[width=0.45\columnwidth]{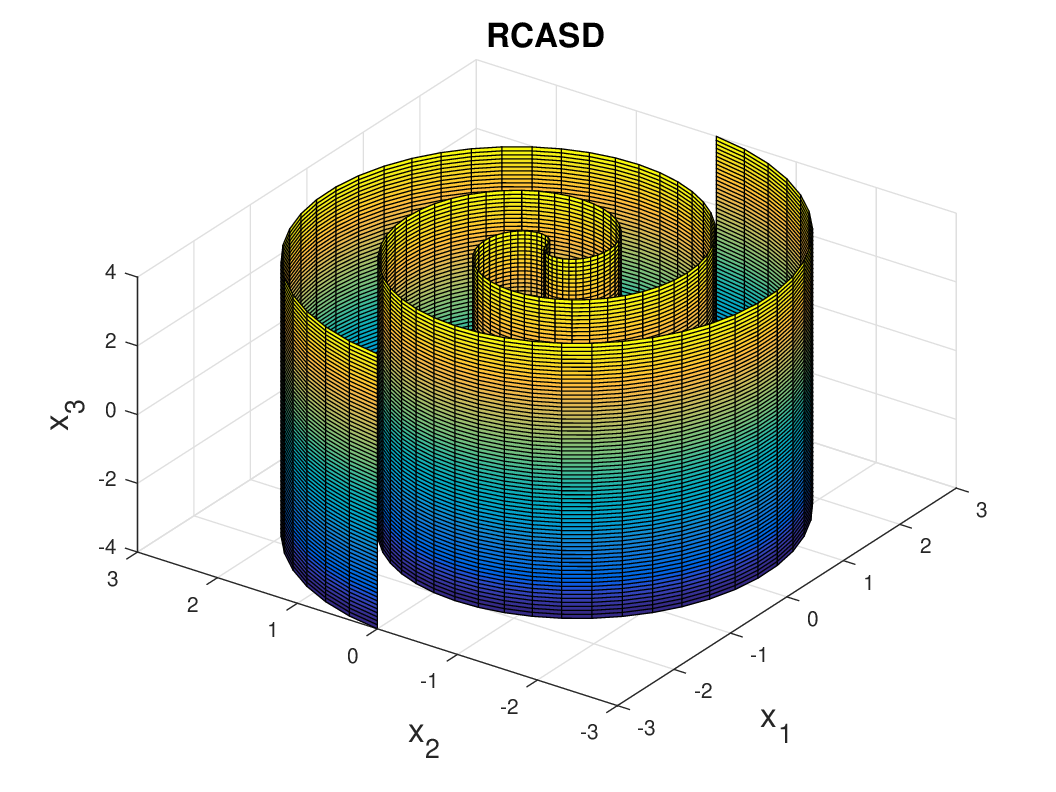}
        \label{fig:RCASD_Surface}}
        \subfigure[]{
            \includegraphics[width=0.45\columnwidth]{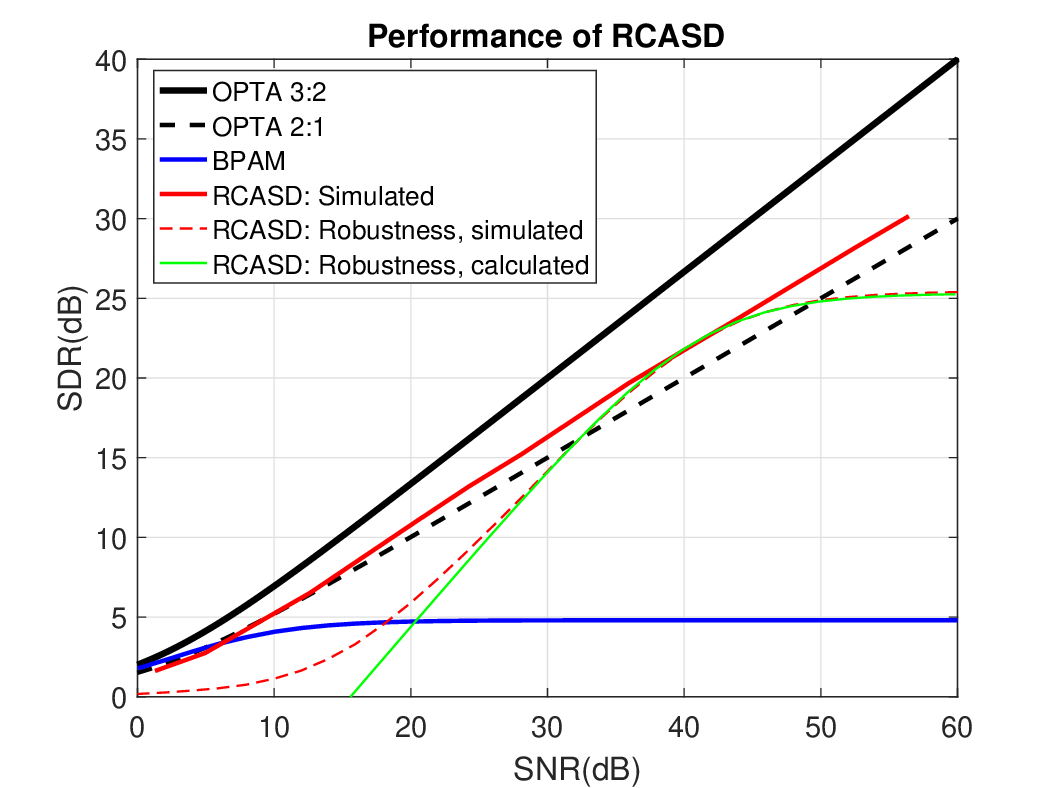}
        \label{fig:RCASD_Performance}}
    \end{center}
    \caption{(a) RCASD in source space with LoC coordinate grid. (b) Performance of RCASD compared to OPTA and BPAM. }\label{fig:RCASD}
\end{figure}

Fig.~\ref{fig:RCASD_Surface} depicts the RCASD in the source space.
The parametric equation for the RCASD is given by~\cite[p. 51]{Surf_Encyclopedia}
\begin{equation}\label{e:S_RCASD_ParEq}
\mathbf{S}(\mathbf{x})=\big[\pm a\varphi(z_1)\cos(\varphi(z_1)), \pm a\varphi(z_1)\sin(\varphi(z_1)),\alpha_2 z_2 \big],
\end{equation}
with $\alpha_2$ some amplification factor and $a=\Delta/\pi$, where $\pm$ refers to positive and negative channel values. $\Delta$ is the smallest distance between the two ``spiral surfaces'' seen in Fig.~\ref{fig:RCASD_Surface}. The RCASD is a DS with Archimedes spiral as directrix (a $2$:$1$ sub-mapping~\cite{hekland_floor_ramstad_T_comm}).

The components of FFF (metric) are
\begin{equation}\label{e:RCASD_FFF_coef_Opt}
g_{11}=\bigg(\frac{\Delta}{\pi}\bigg)^2 \big(\varphi'(z_1)\big)^2 \big({1+\varphi^2(z_1)}\big), \ g_{22}=\alpha_2^2, \ g_{12}=g_{21}=0,
\end{equation}
and the components of the SFF are
\begin{equation}\label{e:RCASD_SFF_coef}
b_{11}=-a(\varphi'(z_1))^2\frac{2+\varphi^2(z_1)}{\sqrt{1+\varphi^2(z_1)}}, \ b_{12}=b_{21}=b_{22}=0.
\end{equation}
The components in~(\ref{e:RCASD_FFF_coef_Opt}) are computed from $g_{\alpha\beta}=\mathbf{S}_\alpha \cdot \mathbf{S}_\beta$,  and the components of the SFF are computed from~(\ref{e:compute_b_ab}) in Appendix~\ref{sec:app_FundForms_Einstein} (see~\cite{floor2021tools} for details). With~(\ref{e:RCASD_FFF_coef_Opt}) and~(\ref{e:RCASD_SFF_coef}) one can from Theorem~\ref{th:LoC_Coordinates} conclude that the coordinate curves are LoC as $g_{12}=b_{12}=0$, and so~(\ref{e:ChannelDist_2ndOrder_Final}) describes the 2nd order behavior of this mapping. 

\textbf{Evaluation of curvature:} With LoC coordinates, the principal curvatures are found from the above fundamental forms as:
\begin{equation}\label{e:RCASD_PrincCurvat}
\kappa_1=\frac{b_{11}}{g_{11}}=-\frac{2+\varphi^2(z_1)}{a(\sqrt{1+\varphi^2(z_1)})^{3/2}}, \ \kappa_2=\frac{b_{22}}{g_{22}}=0.
\end{equation}
By choosing
\begin{equation}\label{e:RCASD_Mapping_func}
\varphi(z_1)=\pm \sqrt{{\alpha_1 z_1}/{(\eta\Delta)}},
\end{equation}
with $\alpha_1$ some amplification factor, one approximates arc length parametrization along the directrix as shown in~\cite{hekland_floor_ramstad_T_comm}.
Evaluation of $\kappa_1$ as function of the free parameters $\Delta$ and $\alpha_1$, is provided in~\cite[p.23]{floor2021tools}, Fig. 18(a). Generally, the curvature is relatively small for this mapping. By inserting optimized parameters for 30dB SNR found by the optimization procedure below ($\Delta^\ast=0.608$, $\alpha_1^\ast=3.33$) one obtains $|\bar{\kappa}_1|<1$ averaged over the relevant range of $z_1$. By considering the distortion terms in~(\ref{e:WeakNoiseDist_2ndOrder_M_N}), with total transmission power 1, then $\sigma_n^2=0.001$, and one can see that the 1st order term is in the  order of about $10^{-3}/(10^{-3})^2 = 1000$ over the 2nd order term. Therefore, RCASD is a mapping following Definition~\ref{def:weak_noise_exp} at high SNR.

\textbf{Optimization of RCASD as $3$:$2$ mapping:}
The RCASD's performance is made scalable with SNR through the factor $a=\Delta/\pi$ where $\Delta$ is adapted to $\sigma_n^2$.

\emph{Distortion:} With $\tilde{z}_i = z_i + n_i$ mapped through~(\ref{e:S_RCASD_ParEq}), the channel distortion is computed from~(\ref{e:mseort_mean_DimRed}). With $\varphi$ as in~(\ref{e:RCASD_Mapping_func}), $g_{11}\approx \alpha_1^2, \ \forall z_1,z_2$. Similarly, since $x_3=\alpha_2 z_2$, then $g_{22}=\alpha_2^2$, and so $G$ is diagonal with constant $g_{ii}$'s, which was one of the criteria sought. Therefore
\begin{equation}\label{e:RCASD_ChannelDist}
\bar{\varepsilon}_{ch}^2 = \frac{\sigma_n^2}{3} \iint\sum_{i=1}^2 g_{ii}(\mathbf{z}) f_\mathbf{z}(\mathbf{z})\mbox{d} \mathbf{z}
=  \frac{\sigma_n^2}{3} \big(\alpha_1^2 + \alpha_2^2 \big) f_\mathbf{z}(\mathbf{z})\mbox{d} \mathbf{z}
= \frac{\sigma_n^2\big(\alpha_1^2 + \alpha_2^2 \big)}{3}.
\end{equation}

From Fig.~\ref{fig:RCASD_Surface} one can see that we have a uniform S-K mapping (Definition~\ref{def:uniform_sk}), implying that Eq.~(\ref{e:a_noise}) with $N=2$ and $M=3$ applies:
\begin{equation}\label{e:RCASD_approx}
\bar{\varepsilon}_{q}^2 \geq {\Delta^2}/{36}.
\end{equation}

\emph{Power:}
The directrix, Archimedes' spiral, was applied as $2$:$1$ mapping in~\cite{hekland_floor_ramstad_T_comm}. With $\varphi$ as in~(\ref{e:RCASD_Mapping_func}) it was shown that a Laplace distribution over $z_1$ is obtained with variance $\sigma_{y_1}^2 = 2(2\eta\sigma_x^2{\pi}/{(\Delta}\alpha_1))^2, \ \eta=0.16$, at high SNR. As $z_2 = x_3/\alpha_3$, $z_2$ has a Gaussian distribution with variance $\sigma_{z_2}^2=\sigma_x^2/\alpha_2^2$.
Therefore, the total channel power becomes
\begin{equation}\label{e:RCASD_PowTot}
P_t = \frac{1}{2}\bigg(2\bigg(\frac{2\eta\sigma_x^2\pi}{\Delta\alpha_1}\bigg)^2+\frac{\sigma_{x}^2}{\alpha_2^2}\bigg).
\end{equation}

\emph{Optimization:}
With constraint $C_t = P_{\text{max}} - P_t(\Delta,\alpha_1,\alpha_2) \geq 0$, the objective function
\begin{equation}\label{e:RCASD_3_2_ObjFunc}
\mathcal{L} (\Delta,\alpha_1,\alpha_2) = \bar{\varepsilon}_{q}^2(\Delta) + \bar{\varepsilon}_{ch}^2 (\alpha_1,\alpha_2)-\lambda C_t(\Delta,\alpha_1,\alpha_2),
\end{equation}
is obtained. The optimal parameters are found by a numerical approach as in~\cite[pp. 87]{FloorThesis}.

The performance of the optimized RCASD is shown in Fig.~\ref{fig:RCASD_Performance} (red curve). The RCASD clearly improves with SNR, rising well above BPAM as SNR increases, and is also robust to varying SNR, having both graceful improvement and reduction for a fixed set of parameters (red dashed curve). The calculated performance is also shown (green curve) in order to demonstrate the accuracy of the theoretical analysis in Section~\ref{sec:mn_dimred}. The distortion contributions in Section~\ref{sec:mn_dimred} can be observed from the robustness graphs: $\bar{\varepsilon}_{q}^2$ dominates above the optimal SNR point, whereas below, $\bar{\varepsilon}_{ch}^2$ dominates. Simulated and calculated performance correspond well, confirming that RCASD follows Definitions~\ref{def:weak_noise_red} for large deviations around the optimal point, even at finite SNR, inline with the curvature evaluation above. However, the slope at high SNR follows that of $2$:$1$ OPTA (black dashed curve) which is expected from Proposition~\ref{prop:Map_Split} as the RCASD is a DS consisting of a $2$:$1$ system and a $1$:$1$ system.  This is explicitly shown in~\cite[p.19]{floor2021tools}. The RCASD is also equivalent to the $3$:$2$ scheme proposed in~\cite{Hu_garcia_lamarca_tcom}.

\subsubsection{Snail Surface}\label{ssec:SnaSu}

\begin{figure}[h]
    \begin{center}
        \subfigure[]{
            \includegraphics[width=0.45\columnwidth]{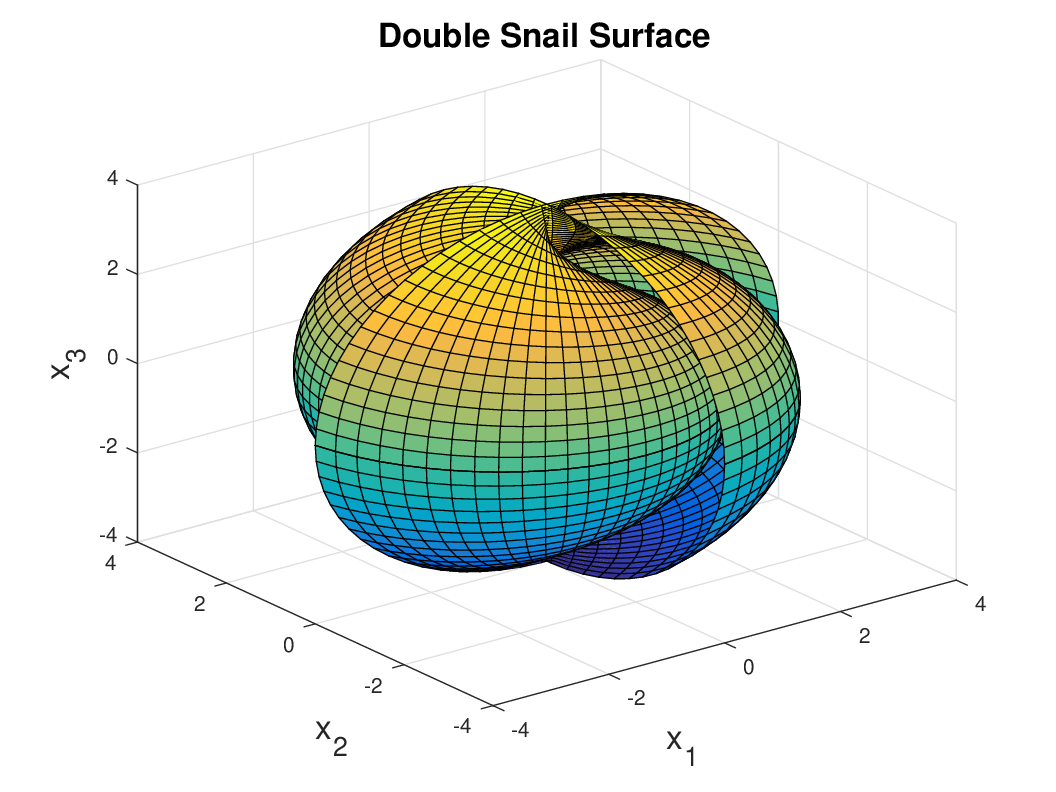}
        \label{fig:Surf_SnaSu}}
        \subfigure[]{
            \includegraphics[width=0.45\columnwidth]{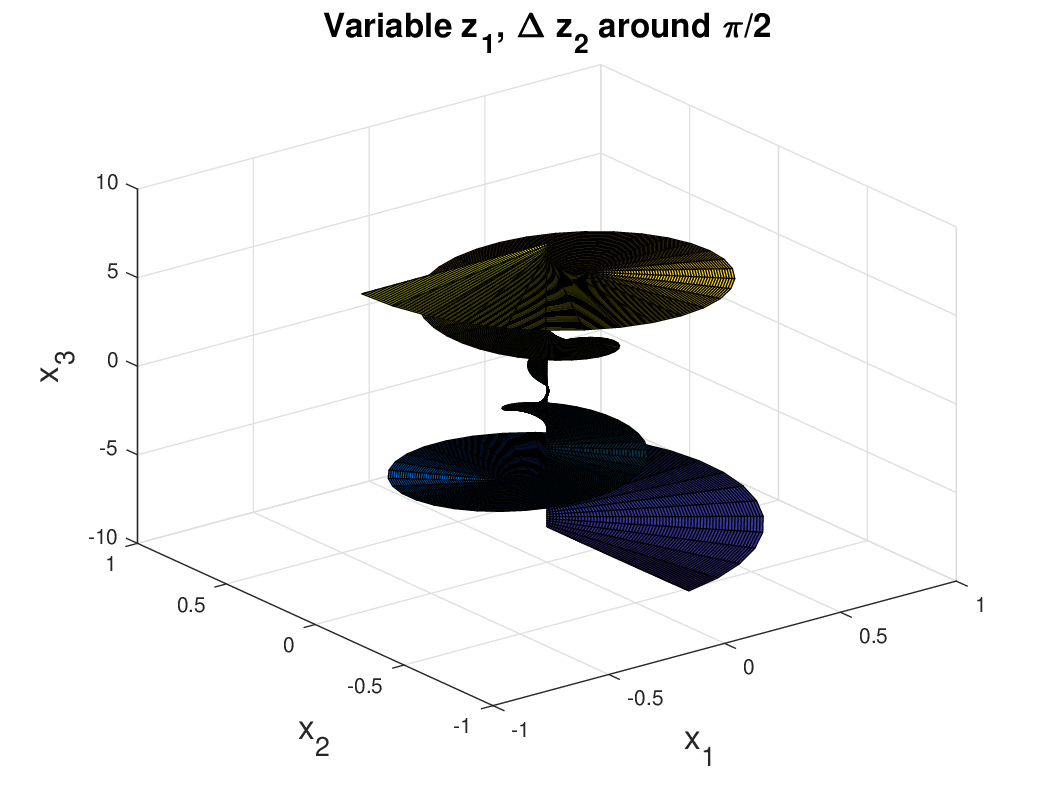}
        \label{fig:SnaSu_Var_z1_dz2}}
        \subfigure[]{
            \includegraphics[width=0.45\columnwidth]{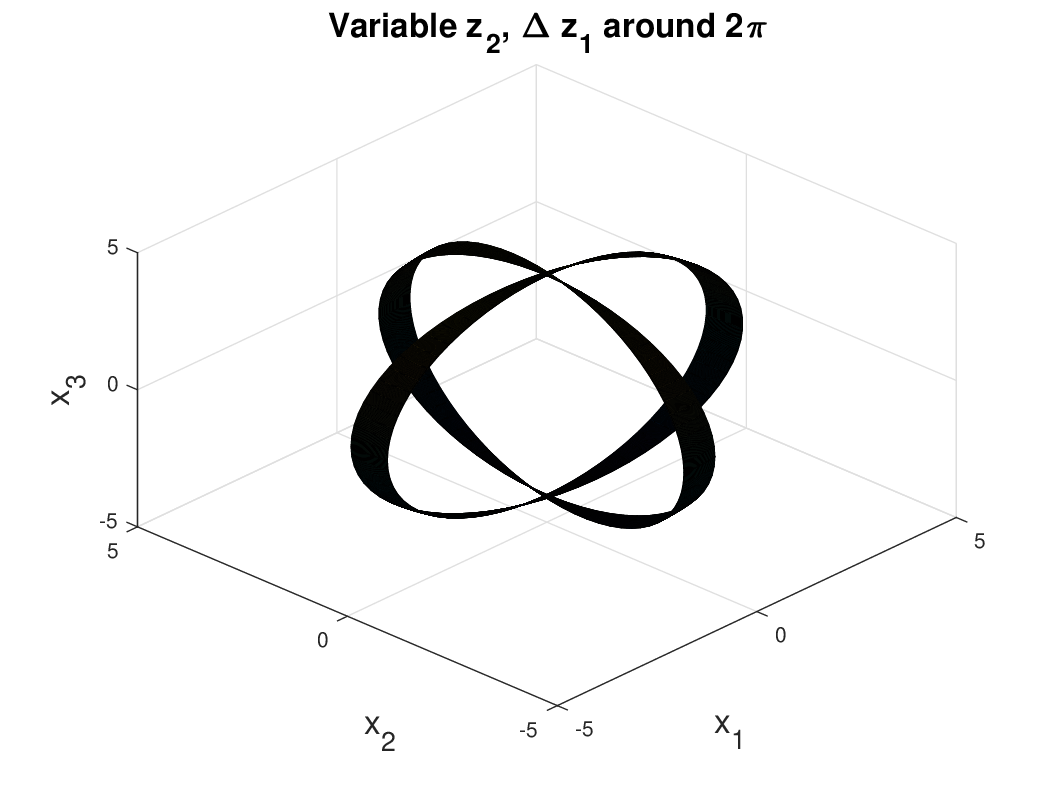}
        \label{fig:SnaSu_Var_z2_dz1}}
        \subfigure[]{
           \includegraphics[width=0.4\columnwidth]{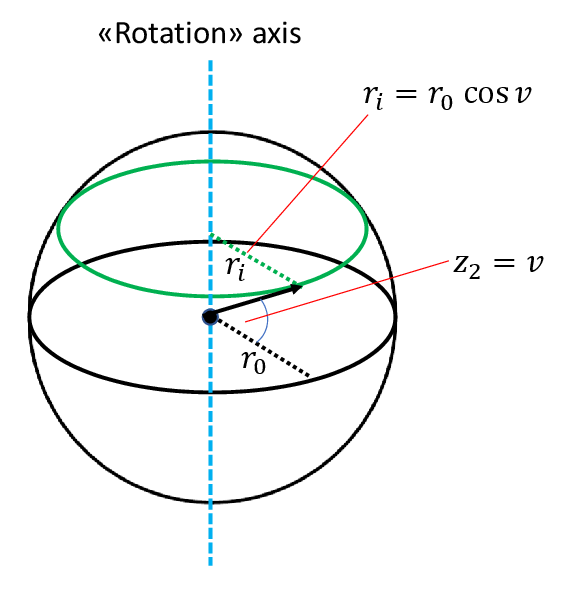}
       \label{fig:SnaSu_z2_sphere}}
     \end{center}
    \caption{(a) The DSS ($a=b=c=2\Delta/\pi$). (b) DSS with $z_1$ variable and $z_2=\pi/2\pm\epsilon_1$. (b) DSS with $z_2$ variable and $z_1=2\pi\pm\epsilon_2$. (d) Virtual spherical shell applied to compute the channel pdf of $z_2$. 
      }\label{fig:SnaSu_summary}
\end{figure}

The \emph{snail surface} cannot be decomposed into sub-mappings and covers a spherical subset of the source space properly, avoiding bends with high curvature (see curvature evaluation below). Its parametrization has components~\cite[p. 280]{Surf_Encyclopedia}
\begin{equation}\label{e:SnaSU_ParEq_pos}
\begin{split}
S_1(z_1,z_2) &=  a\varphi(z_1)\sin(\varphi(z_1))\cos(\alpha_2 z_2+\phi),\\
S_2(z_1,z_2) &=  b\varphi(z_1)\cos(\varphi(z_1))\cos(\alpha_2 z_2+\phi),\\
S_3(z_1,z_2) &= -c \varphi(z_1)\sin(\alpha_2 z_2+\phi),
\end{split}
\end{equation}
which are valid for $0\leq z_1\leq k\pi$, $-\pi\leq z_2\leq\pi$. To include negative values of $z_1$, i.e., $-k \pi\leq z_1\leq 0$, one simply flips the sign of all components in~(\ref{e:SnaSU_ParEq_pos}), and obtain a \emph{double} snail surface (DSS), depicted in Fig.~\ref{fig:Surf_SnaSu}.
By choosing $\psi=\pi/2$ and $a=b=c=2\Delta/\pi$ one obtains a spherical symmetry which leads to a (close to) uniform S-K mapping  (Definition~\ref{def:uniform_sk}), and so~(\ref{e:RCASD_approx}) is a lower bound for $\bar{\varepsilon}_{q}^2$. $\phi$ will be decided later.
%

For a general $\varphi(z_1)$, the metric tensor is found to be~\cite{floor2021tools}
\begin{equation}\label{e:SnaSu_FFF_coef_KL}
g_{11} =(a\varphi'(z_1))^2 \big(1+ \varphi^2(z_1)\cos^2(\alpha_2 z_2 +\phi) \big), g_{22} = a^2\alpha_2^2 \varphi^2(z_1), \ g_{12} = g_{21}=0.
\end{equation}
By inserting $\varphi(z_1)=\alpha_1 z_1$ in~(\ref{e:SnaSu_FFF_coef_KL}) one observes that $g_{ii}\sim z_1^2$, $i=1,2$, implying  that $\bar{\varepsilon}_{ch}^2$ increases with $z_1^2$. One can compensate this for both $g_{ii}$ components simultaneously by choosing $\varphi\sim \sqrt{z_1}$. As the RCASD also has $g_{11}\sim z_1^2$ when $\varphi(z_1)=\alpha_1 z_1$, and that DSS scales with $\Delta$ like RCASD, it makes sense to use~(\ref{e:RCASD_Mapping_func}) for DSS as well, the choice of $\eta$ being arbitrary.

\textbf{Evaluation of curvature:} The components of the SFF, derived in~\cite[p.23]{floor2021tools}, are
\begin{equation}\label{e:SnaSu_SFF}
b_{11}=-\frac{a(\varphi'(z_1))^2 \varphi^2(z_1)\cos^3\theta}{\sqrt{1+\varphi^2(z_1)\cos^2\theta}}, b_{22}=-\frac{a\alpha_2^2 \varphi^2(z_1)\cos\theta}{\sqrt{1+\varphi^2(z_1)\cos^2\theta}}, b_{12}=\frac{a\alpha_1\alpha_2\varphi^2(z_1)\sin\theta}{\sqrt{1+\varphi^2(z_1)\cos^2\theta}},
\end{equation}
As the coordinates are not LoC, the principal curvatures are the roots of~(\ref{e:PrincipalCurvature_Gen}) in Appendix~\ref{sec:app_FundForms_Einstein}
\begin{equation}\label{e:SnaSu_PrincCurvat}
\kappa_{1/2}=\frac{1}{2}\bigg(\frac{b_{11}}{g_{11}}+\frac{b_{22}}{g_{22}} +/- \sqrt{\bigg(\frac{b_{11}}{g_{11}}+\frac{b_{22}}{g_{22}} \bigg)^2  - 4\frac{b_{11}b_{22}-b_{12}^2}{g_{11}g_{22}}}\bigg).
\end{equation}
Evaluation of $\kappa_i$ as function of the free parameters $\Delta$, $\alpha_1$ and $\alpha_2$ is provided in~\cite[p.23]{floor2021tools}, Fig. 18(b). Not surprisingly, the curvature is larger than for RCASD in general, particularly when $\Delta$ is large and $\alpha_1$ is small (corresponding to low SNR case). However, when $\Delta$ is small and $\alpha_1$ is large, corresponding to high SNR case, the curvature is relatively small: By inserting optimized parameters for 30dB SNR found by the optimization procedure below ($\Delta^\ast=0.539$, $\alpha_1^\ast=4.76$, $\alpha_2^\ast=2.57$) one obtains maximal curvature $|\bar{\kappa}_2|<1$ averaged over the relevant range of $z_1$. By considering the distortion terms in~(\ref{e:WeakNoiseDist_2ndOrder_M_N}), with total transmission power 1, then $\sigma_n^2=10^{-3}$, and one can see that the 1st order term is in the order of about $10^{-3}/(10^{-3})^2 = 1000$ over the 2nd order term. Therefore, the DSS is also a mapping following Definition~\ref{def:weak_noise_exp} at high SNR.

\textbf{Optimization of DSS as $3$:$2$ mapping:}

\emph{Channel Power and Density Function:}
To evaluate the channel input from DSS it is convenient to analyze variation for each channel separately, resulting in the geometrical configurations in Figs.~\ref{fig:SnaSu_Var_z1_dz2} and~\ref{fig:SnaSu_Var_z2_dz1} (see~\cite{floor2021tools} for more details).

To derive the pdf of $z_1$, consider Fig.~\ref{fig:SnaSu_Var_z1_dz2}. By perturbing $z_2$ with $\pm \epsilon_1$ around some constant value (here $\pi/2$) with $z_1$ free, we get a \emph{cork screw}-like structure. In the limit $\epsilon_1\rightarrow 0$ we get a spiral with torsion $\tau\neq 0$, rising from the $x_1 x_2$-plane at a \emph{rate} depending on $z_2$: Whenever $z_2=\pm (2m+1)\pi/2, \ m\in\mathbb{N}$, $\tau$ is maximal, whereas when $z_2=\pm m\pi, \ m\in\mathbb{N}$, $\tau=0$, and the spiral is plane. Therefore, the mapping from DSS to $z_1$ can be approximated as the \emph{radius}, $\rho=\sqrt{x_1^2 +  x_2^2 + x_3^2}$, tracing out points inside a sphere as $z_1$ and $z_2$ vary over their domains. Then, with $\Delta$ small, one can approximate the mapping $\mathbf{x} \rightarrow z_1$ as a continuous function $h:\mathbb{R}^3 \rightarrow \mathbb{R}$. This assumption becomes more accurate as SNR grows, i.e., as $\Delta$ decreases. By choosing $\varphi=(\gamma z_1)^n$, $n\in \mathbb{Q}^+$, then $z_1=h(x_1,x_2,x_3)=\pm \gamma a^{-n}(x_1^2+x_2^2+x_3^2)^{n/2}=\pm \gamma a^{-n} \rho^n=\ell(\rho)$. We have:

\begin{lemma}\label{lem:SnaSu_Pdf_z1}
At high SNR, with $\varphi=(\gamma z_1)^n$, the pdf for $z_1$ when $\mathbf{S}$ is a DSS, is given by
\begin{equation}\label{e:SnaSu_Z1_pdf_gen}
f_{z_1}(z_1)=\frac{n a^3 \gamma^3 |z_1|^{3n-1}}{\sqrt{2\pi}\sigma_x^3} e^{-\frac{a^2 \varphi^2(z_1)}{2\sigma_x^2}}.
\end{equation}
\end{lemma}

\emph{Proof:} See Appendix~\ref{sec:app_pf_SnaSu_z1}.\hspace{11.0cm}$\square$

Now assume that $\varphi(z_1)$ is given by~(\ref{e:RCASD_Mapping_func}), then $\gamma =\sqrt{\alpha_1/(\eta \Delta)}$, and thus
\begin{equation}\label{e:SnaSu_Z1_pdf}
f_{z_1}(z_1)=\frac{a^3 \alpha_1^{3/2} \sqrt{|z_1|}}{2\sqrt{2\pi}\sigma_x^3(\eta\Delta)^{3/2}} e^{-\frac{a^2 \alpha_1|z_1|}{2\sigma_x^2\eta\Delta}}.
\end{equation}
According to~\cite[p.87,154]{papoulis02} a Gamma distribution has the form $f_\Gamma (x)=u(x){x^{c-1} e^{-\frac{x}{b}}}/{(\Gamma(c)b^c)}$ with second moment  $\mbox{E}\{x^2\}=c(c+1)b^2$. Therefore~(\ref{e:SnaSu_Z1_pdf}) is a \emph{double Gamma distribution} with $c=3/2$ and $b=(2\eta\Delta \sigma_x^2)/(a^2\alpha_1)$. Since~(\ref{e:SnaSu_Z1_pdf}) has
zero mean, the power of channel 1 becomes
\begin{equation}\label{e:SnaSu_Pow_z1}
P_1 = \text{Var}\{z_1\} = \frac{15(\eta\Delta\sigma_x^2)^2}{a^4\alpha_1^2}=\frac{15(\eta\pi^2\sigma_x^2)^2}{16\alpha_1^2\Delta^2}.
\end{equation}

To derive the pdf of $z_2$, consider Fig.~\ref{fig:SnaSu_Var_z2_dz1}. By perturbing $z_1$ with $\pm \epsilon_2$ around some constant value (here $2\pi$) with $z_2$ free, we get two M\"{o}bius strips. In the limit $\epsilon_2 \rightarrow 0$ we get a circle "rotating" about an axis whose radius increases as $2\Delta z_1/\pi$. Consider $\phi=0$. Then the rotation  axis is at $\pi/2$. The radius of the rotating circle is insignificant as $z_2\in [-\pi,\pi]$, independent of $z_1$. From the perspective of $z_2$, as the joint pdf of $\mathbf{x}$ is spherically symmetric, we have a uniform mass distribution over a virtual spherical shell of arbitrary radius, $r_0$, as depicted in Fig.~\ref{fig:SnaSu_z2_sphere}. To find the probability mass associated with different values of $z_2$, one considers the sum of all points along circles resulting from intersections of this virtual sphere by planes perpendicular to the rotation axis (green circle in Fig.~\ref{fig:SnaSu_z2_sphere}).
%
The radius, $r_i$, of such a circle is  $r_i = r_0 \cos(\upsilon)$, where $\upsilon=z_2$ is the angle from the equatorial plane, $\upsilon = \pi/2 - \varrho$, and $\varrho$ the polar angle. The circumference as a function of $z_2$ is $O(z_2)=2\pi r_0 \cos(z_2)$. Since $r_0$ is arbitrary, one can set $r_0 = 1/(2\pi)$ implying that $f_{z_2}(z_2)\sim |\cos(z_2)|, \ z_2\in[-\pi,\pi]$. To avoid high probability for the largest channel amplitude values, one can set $\phi=\pi/2$ to obtain zero probability there. I.e., a \emph{sine distribution} results. To normalize, as $\int_0^\pi \sin(z_2)\mbox{d}z_2 = 2$, then
\begin{equation}\label{e:pdf_SnaSu_ch2}
f_{z_2}(z_2) = \frac{\alpha_2}{4}|\sin(\alpha_2 z_2)|.
\end{equation}
Since $f_{z_2}(z_2)$ is proportional to \emph{Gilberts sine distribution} $f_x(x)=\sin(2x)$, which according to~\cite{GilbertSine_Edwards} has variance $E\{x^2\}=(\pi^2/4-1)/2$, the power for channel 2 becomes
\begin{equation}\label{e:RCASD_PowZ2}
P_2 = \text{Var}\{z_2\} = \frac{2}{\alpha_2^2}\bigg(\frac{\pi^2}{4}-1\bigg).
\end{equation}

\emph{Distortion:}
Using~(\ref{e:SnaSu_FFF_coef_KL}), we obtain
\begin{equation}\label{e:SnaSU_WeakChannelDist_I2}
I_2 =  \iint g_{22}(\mathbf{z}) f_\mathbf{z}(\mathbf{z})\mbox{d} \mathbf{z}
= 2(a\alpha_2)^2 \frac{\alpha_1}{\eta\Delta}\int_{-\infty}^\infty z_1 f_{z_1}(z_1)\mbox{d}z_1 = 3\alpha_2^2\sigma_x^2.
\end{equation}
The last equality comes from the fact that $z_1$ is gamma distributed, therefore the integral in~(\ref{e:SnaSU_WeakChannelDist_I2}) becomes~\cite[p.154]{papoulis02} $b c/2 = 3\eta \Delta \sigma_x^2 / (a^2 \alpha_1)$. Further, we have using~(\ref{e:SnaSu_FFF_coef_KL}) (see~\cite{floor2021tools} for details),
\begin{equation}\label{e:SnaSU_WeakChannelDist_I1_pre}
I_1 =  \iint g_{11}(\mathbf{z}) f_\mathbf{z}(\mathbf{z})\mbox{d} \mathbf{z}= \frac{\alpha_1\Delta}{\eta\pi^2}E\{z_1^{-1}\}+\frac{2\alpha_1^2}{3\pi^2\eta^2},
\end{equation}
Through power series expansion one can show that $E\{z_1^{-1}\}\approx 4(1+\text{Var}\{z_1\})$ up to 3rd order (see~\cite{floor2021tools}).
Inserting this into~(\ref{e:SnaSU_WeakChannelDist_I1_pre}), the channel distortion results from~(\ref{e:mseort_mean_DimRed}),
\begin{equation}\label{e:SnaSU_WeakChannelDist}
\bar{\varepsilon}_{ch}^2 = \frac{\sigma_n^2\big(I_1 + I_2\big)}{3} \approx \frac{\sigma_n^2}{3}\bigg(\frac{4\alpha_1\Delta}{\eta\pi^2}\bigg(1+\frac{15(\eta\pi^2\sigma_x^2)^2}{16\alpha_1^2\Delta^2}\bigg)+\frac{2\alpha_1^2}{3\pi^2\eta^2} + 3\alpha_2^2\sigma_x^2\bigg).
\end{equation}

\emph{Optimization:}
With constraint $C_t = P_{\text{max}} - P_t(\Delta,\alpha_1,\alpha_2) \geq 0$, we get a similar objective function as in~(\ref{e:RCASD_3_2_ObjFunc}) which is found numerically.

The performance of the optimized DSS is plotted in Fig.~\ref{fig:SnaSu_Performance}.
Comparing with Fig.~\ref{fig:RCASD_Performance} its clear that DSS has better performance than the RCASD at high SNR, which is expected from Proposition~\ref{prop:Map_Split}. The DSS is also noise robust (green dashed curve), and the magenta line confirms that the theoretical model derived for DSS above is quite accurate. However, one can see that the gap to OPTA increases somewhat above 40dB, the reason being that $\bar{\varepsilon}_{ch}^2\sim 1/\Delta^2$ instead of $\bar{\varepsilon}_{ch}^2\sim 1/\Delta$, which is required according to Corollary~\ref{cor:Delta_exponent}. Therefore, the DSS will eventually diverge from OPTA (albeit at a higher SNR than a decomposable mapping). One option that has not yet been investigated that may lead to the right slope is a change of coordinates curves. 


\subsubsection{$3$:$2$ Hybrid Vector Quantizer Linear Coder (HVQLC)}\label{ssec:HVQLC3_2}
We construct a mapping satisfying all necessary criteria for obtaining SDR$\sim \text{SNR}^{2/3}$ as SNR$\rightarrow\infty$. To simplify the problem a HDA approach is taken: Consider approximating $\mathbf{x}$ to planes parallel to the $x_1,x_2$-plane in $\mathbb{R}^3$ with distance $\Delta$ between them. 
One then obtains a uniform $\mathcal{S}$, and so $\bar{\varepsilon}_{q}^2\sim\Delta^2$ according to Proposition~\ref{th:apx_dist}. With parallel planes $G=\alpha I$, with $\alpha$ some scaling factor. To make the mapping non-decomposable, we map the planes onto the channel with their center placed on the Archimedes spiral, thereby obtaining a \emph{mix} of several sources on each channel. A spiral is chosen for two reasons: 1) The condition $\bar{\varepsilon}_{ch}^2\sim1/\Delta$ in Corollary~\ref{cor:Delta_exponent} is obtained, as shown in Proposition~\ref{prop:HVQLC3_2Slope}. 2) To easily scale the mapping with SNR: By choosing $\varphi$ as in~(\ref{e:RCASD_Mapping_func}), an equal distance, $\Delta$, between the spiral arms as well as between the centroids along each arm results (a uniform VQ on a disc, as illustrated in Fig.~\ref{fig:SpiralVQ}, Section~\ref{ssec:HVQLC2_3}).

\begin{figure}[h!]
\begin{center}
\includegraphics[width=0.95\columnwidth]{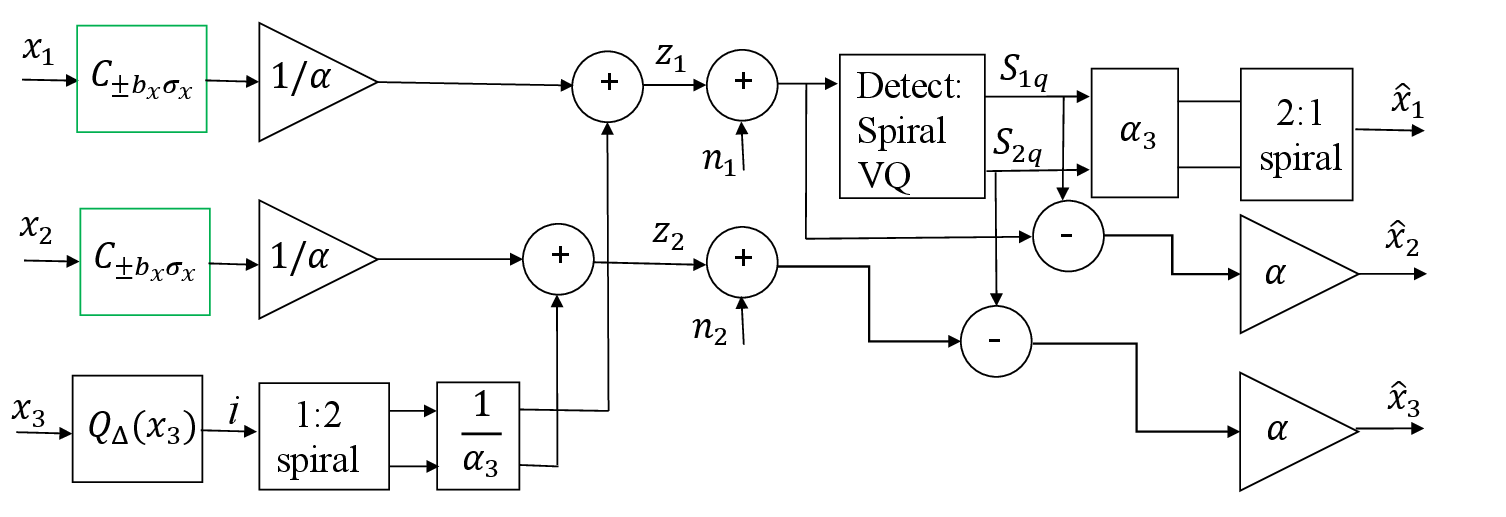}
    \caption{$3$:$2$ HVQLC block diagram. Green blocks are optional.}\label{fig:hvqlc3_2}
\end{center}
\end{figure}

The block diagram for the $3$:$2$ HVQLC is depicted in Fig.~\ref{fig:hvqlc3_2}.

\textbf{Optimization of $3$:$2$ HVQLC:}

\emph{Distortion:}
A drawback of this mapping is that it introduces anomalous errors when centroids are mis-detected. This happens with probability $Pr\{\|\mathbf{y_{12}}+\mathbf{n}\|\geq \Delta/(2\alpha_3)\}$, where $\mathbf{y_{12}}=[x_1\ x_2]/\alpha$. The error is bounded by $2 b_x\sigma_x$, with $b_x$ depending on the \emph{limiting} of $x_1$ and $x_2$ described below. The pdf of $\mathbf{y_{12}}+\mathbf{n}$ is the product the distributions of $x_i+n_i,\ i=1,2$, each with pdf $\mathcal{N}(0,\sigma_y^2+\sigma_n^2)$~\cite[pp.181-182]{papoulis02}. The variable $w=\|\mathbf{y_{12}}+\mathbf{n}\|$ is then Rayleigh distributed~\cite[pp. 202-203]{papoulis02}, $f_w(w)=w/(\sigma_y^2+\sigma_n^2)\exp(-w^2/2(\sigma_y^2+\sigma_n^2))$. Therefore
\begin{equation}\label{e:HVQLC3_2an_err}
\bar{\varepsilon}_{an}^2 = 4 b_x^2\sigma_x^2 Pr\bigg\{w\geq\frac{\Delta}{2\alpha_3}\bigg\}= 4 b_x^2\sigma_x^2\int_{\Delta/(2\alpha_3)}^\infty f_w(w)\mbox{d}w = 4 b_x^2\sigma_x^2 e^{-\frac{\Delta^2}{8\alpha_3^2(\sigma_x^2/\alpha^2+\sigma_n^2)}}.
\end{equation}
In order to obtain a mapping where anomalous errors happen with low probability one will either condition $\Delta/\alpha_3$ to be small, or limit $x_1$ and $x_2$ at some value (green blocks in Fig.~\ref{fig:hvqlc3_2}). By limiting $x_i$ at the value $b_x\sigma_x$, one introduces a distortion~\cite{Floor_Kim_Wernersson11_TCOM}
\begin{equation}\label{e:HVQLC3_2_limit}
\bar{\varepsilon}_{\kappa}^2 =\frac{4}{3} \int_{b_x\sigma_x}^\infty (x_i - b_x\sigma_x)^2f_x(x_i)\mbox{d}x_i,\  i=1,2.
\end{equation}
By limiting each source separately, the HVQLC will result in parallel planes in $\mathbb{R}^3$, whereas by limiting $\sqrt{x_1^2+x_2^2}$, parallel discs are obtained.

To make the probability of anomalous errors small, the following constraint is needed 
\begin{equation}\label{e:HVQLC3_2_constraint}
\frac{\Delta}{\alpha_3} > \frac{2 b_x \sigma_x}{\alpha_1} + 2 b_n\sigma_n.
\end{equation}
The probability is adjusted with the $b_x$ parameter. With $b_x > 4$ then $99.99 \%$ of all source values are still present. 

As mentioned above, $g_{ii}=\alpha^2$ and so the channel distortion  becomes $\bar{\varepsilon}_{ch}^2 = {2\sigma_n^2\alpha^2}/{3}$. We also have a uniform S-K mapping. Therefore, the total distortion becomes 
\begin{equation}\label{e:HVQLC3_2 TotDist}
D_t=  \frac{\Delta^2}{36} + \frac{2 \sigma_n^2 \alpha^2}{3} + 4 b_x^2\sigma_x^2 e^{-\frac{\Delta^2}{8\alpha_3^2(\sigma_x^2/\alpha^2+\sigma_n^2)}}.
\end{equation}

\emph{Power:} Since $x_1$ and $x_2$ are scaled Gaussians, their transmission power becomes $P_1+P_2= 2\sigma_x^2 /\alpha^2$. As $x_3$ is mapped through a discretisized version of the $1$:$2$ mapping in~\cite{hekland_floor_ramstad_T_comm}, the same power expression applies for small $\Delta$ (high SNR)\footnote{A factor appearing in~\cite{hekland_floor_ramstad_T_comm} is removed as we assume $x_3$ to take on values over $\mathbb{R}$. In~\cite{hekland_floor_ramstad_T_comm} the source was limited to $[-1,1]$.}: $P_3 \approx {2\Delta\sigma_x}/({\eta\sqrt{2\pi^5}\alpha_3^2})$.
The fact that $P_3\sim\Delta/\alpha_3^2$ gives $\bar{\varepsilon}_{ch}^2\sim 1/\Delta$ as required by Corollary~\ref{cor:Delta_exponent}. The total power is then
\begin{equation}\label{e:HVQLC3_2 TotPow}
P_t = \frac{2\sigma_x^2}{\alpha^2} + \frac{2\Delta\sigma_x}{\eta\sqrt{2\pi^5}\alpha_3^2}.
\end{equation}

\emph{Optimization:}To determine optimal performance we consider the Lagrangian
\begin{equation}\label{e:HVQLC_3_2_ObjFunc}
\mathcal{L} (\Delta,\alpha,\alpha_3) = D_t(\Delta,\alpha) - \lambda_1 C_1(\Delta,\alpha,\alpha_3)-\lambda_2 C_2(\Delta,\alpha,\alpha_2),
\end{equation}
where $C_1=P_{\text{max}}-(P1+P2+P3)$ and $C_2(\Delta,\alpha,\alpha_2) = \Delta - {2 b_x \sigma_x\alpha_3}/{\alpha_1} - 2 b_n\sigma_n\alpha_3$.
The slight difference from the constraint in~(\ref{e:HVQLC3_2_constraint}) is for better numerical stability when solving~(\ref{e:HVQLC_3_2_ObjFunc}).

The optimized performance of HVQLC, ignoring limitation, is shown in Fig.~\ref{fig:All3_2Performance}, magenta curve. 
The HVQLC follow the OPTA slope at high SNR (as shown below), and it is noise robust (magenta dashed curve) despite of anomalous errors. However, anomalies are likely the reason why HVQLC backs off fom OPTA compared to DSS for SNR $< 40$ dB.
%

\emph{High SNR analysis:} We prove that $3$:$2$ HVQLC has the same slope as OPTA as SNR$\rightarrow \infty$.

\begin{proposition}\label{prop:HVQLC3_2Slope}\emph{$3$:$2$ HVQLC at high SNR:}
At high SNR the SDR of $3$:$2$ HVQLC follows
\begin{equation}\label{e:hvqlc3_2Dist_highSNR_Optimal}
\text{SDR}=\bigg(\frac{9\eta \sqrt{\pi^5}}{8b_x^2}\bigg)^\frac{2}{3} \text{SNR}^\frac{2}{3}.
\end{equation}
\end{proposition}

\emph{Proof:} See Appendix~\ref{sec:app_pf_HVQLC3_2Slope}.\hspace{11.1cm}$\square$

This shows that $3$:$2$ HVQLC follows the OPTA slope. From~(\ref{e:hvqlc3_2Dist_highSNR_Optimal}), with $\sigma_x=1$, $b_x=4$ and $\eta = 0.16$, the loss from OPTA becomes $\text{SDR}_{\text{loss}} = - 10\lg (9 \cdot 0.16\sqrt{\pi^5}/(8 \cdot 16))^{2/3}\approx 4.7$dB,  corresponding to the gap seen in Fig.~\ref{fig:All3_2Performance}.

\subsubsection{Comparison of different $3$:$2$ schemes}\label{ssec:compare3_2syst}
\begin{figure}[h]
    \begin{center}
        \subfigure[]{
            \includegraphics[width=0.45\columnwidth]{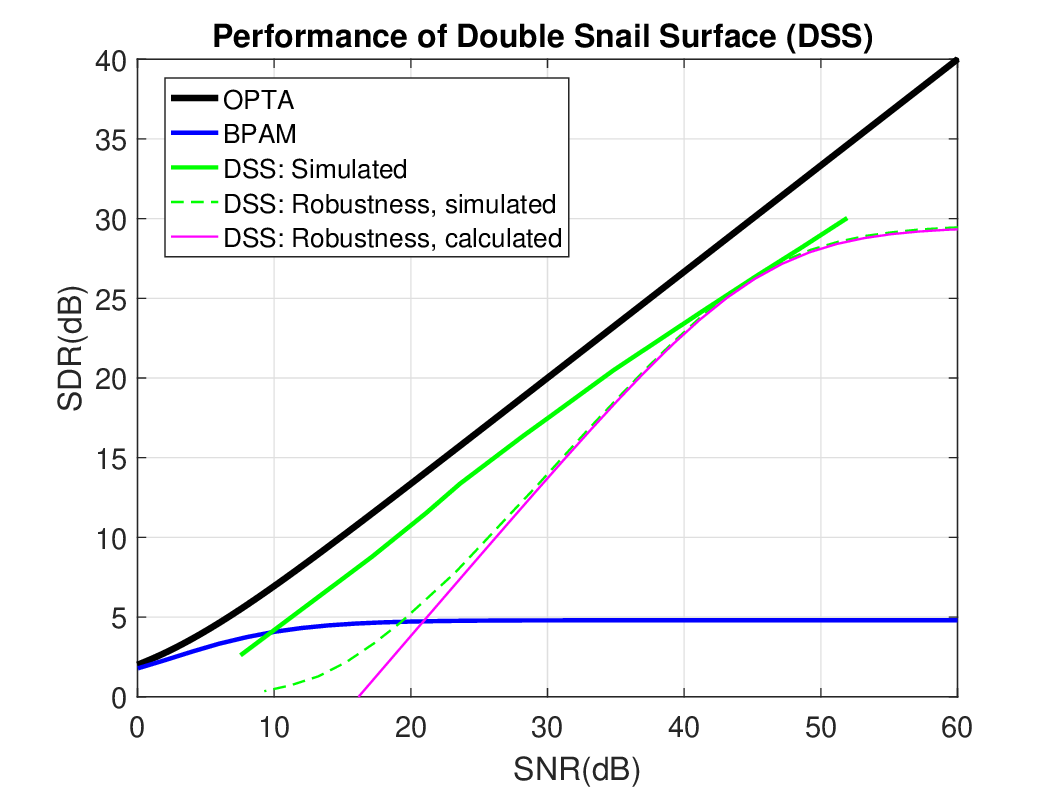}
        \label{fig:SnaSu_Performance}}
        \subfigure[]{
            \includegraphics[width=0.45\columnwidth]{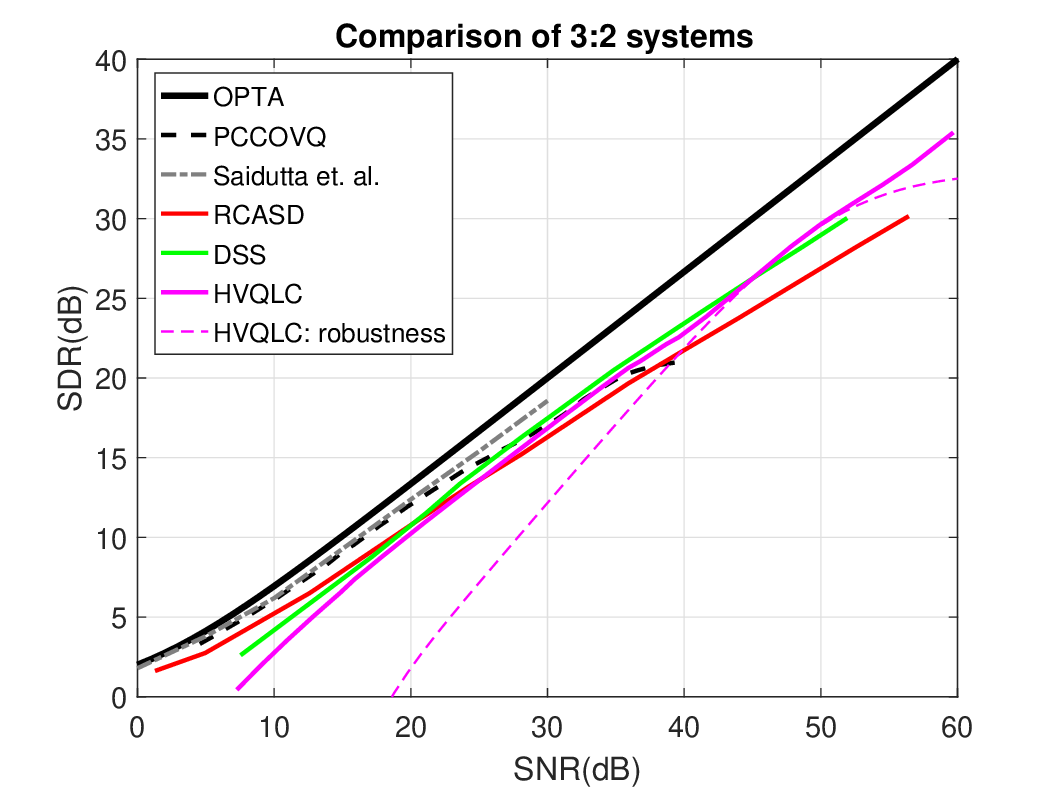}
        \label{fig:All3_2Performance}}
    \end{center}
    \caption{(a) Performance of Double Snail Surface (DSS) compared to OPTA and BPAM. (b) Performance of all suggested $3$:$2$ schemes compared to key mappings in literature. }\label{fig:3_2systems}
\end{figure}

The performance of all mappings proposed in this section are compared in Fig.~\ref{fig:All3_2Performance}. We also include the performance of Saidutta et. al.'s $3$:$2$ mapping~\cite{Saidutta_JSAC2021} found using deep learning, a method named \emph{variational auto encoders} (VAE), as well as \emph{power constrained channel optimized vector quantizer} (PCCOVQ)~\cite{fulds97a,fuldsethThesis}. PCCOVQ is a numerically optimized discrete mapping for any $r=N/M \in \mathbb{Q}$, replicating continuous or piece-wise continuous mappings when the number of source- and channel symbols in the mapping is large. Approaching (or beating) the VAE or PCCOVQ system is a good indication of a well performing mapping as these are properly optimized mappings. 

Not surprisingly, Saidutta's VAE mapping (gray dash-dot curve) and PCCOVQ (black dashed curve) have superior performance in the SNR range they have been optimized for\footnote{The reason why the PCCOVQ system declines above 22dB is because $4096$ symbols are used during optimization, which is too small a number at higher SNR.}. However, the proposed mappings of this paper is only about 1dB inferior to the reference systems. The RCASD is best at low SNR, the DSS has the best performance between 20 and 45 dB, while the HVQLC is best from 45dB and above, and is the only system that  does not diverge from OPTA at high SNR. 

Its is interesting to see that different configurations provide well performing mappings. However, any such configuration will need to comply with the conditions presented in this paper. Although DS-based mappings, like RCASD, diverge from OPTA at high SNR, decomposable mappings have their \emph{virtue} as as simple alternative that perform well at low to medium SNR, and which is easy to generalize to higher dimensions. 

Although the mappings proposed are inferior to the two optimized schemes, the loss is small, and they have the advantage of being a parametric representation, providing one codebook that only needs to be scaled in order to adapt to varying SNR. Thus lowering complexity. 

\subsection{Examples on $2$:$3$ mappings}\label{ssec:Ex_SK_surf_23}
We analyze two mappings: i) A hybrid discrete analog scheme, \emph{hybrid vector quantizer linear coders} (HVQLC) suggested in~\cite[pp. 89-93]{FloorThesis}. ii) The RCASD treated in Section~\ref{ssec:RCASD_Comp}.

\subsubsection{HVQLC}\label{ssec:HVQLC2_3}
\begin{figure}[h!]
\begin{center}
\includegraphics[width=0.95\columnwidth]{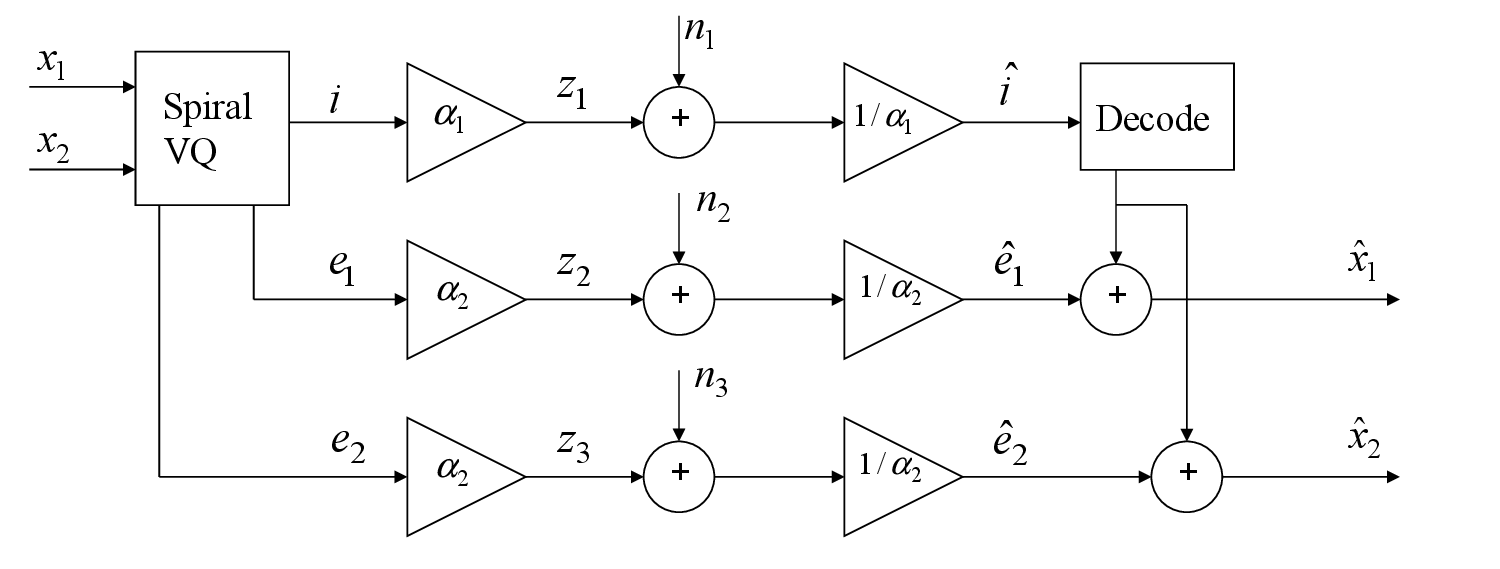}
    \caption{$2$:$3$ HVQLC block diagram.}\label{fig:hvqlc_sys}
\end{center}
\end{figure}

This is a generalization of the hybrid scalar qunatizer linear coder
(HSQLC) proposed in~\cite{Coward00b}. The block diagram is depicted in Fig.~\ref{fig:hvqlc_sys}.

\begin{figure}[h!]
    \begin{center}
        \subfigure[]{
            \includegraphics[width=0.47\columnwidth]{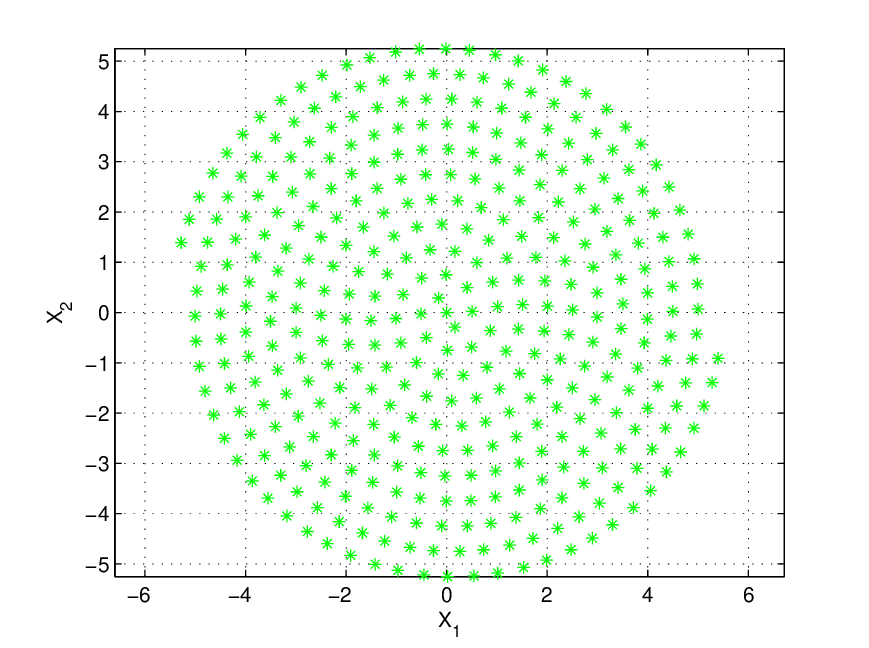}
           \label{fig:SpiralVQ}
           }
        \subfigure[]{
            \includegraphics[width=0.47\columnwidth]{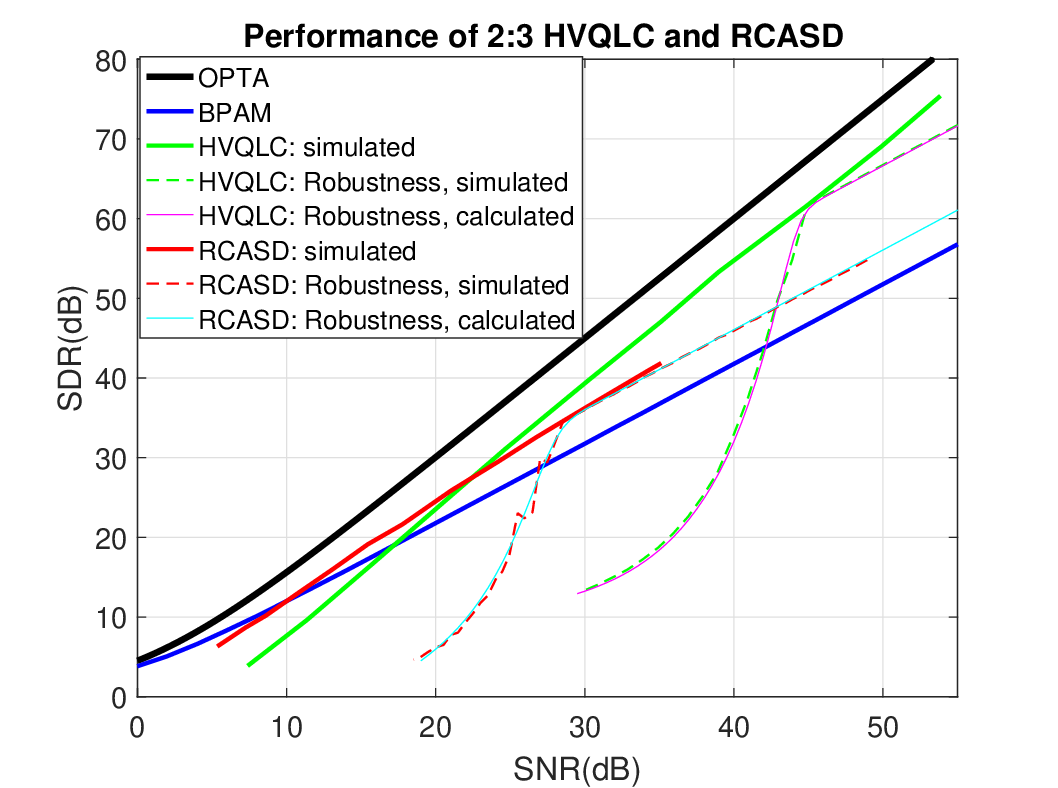}
          \label{fig:HVQLC_RCASD}}
    \end{center}
     \caption{(a) The Spiral VQ applied in HVQLC mapping. (b) Performance of $2$:$3$ HVQLC and RCASD compared to OPTA and BPAM.}\label{fig:2_3Mappings}
\end{figure}

Here VQ centroid indices are denoted by $i$, $e_1$ and $e_2$ denote the two error components from the VQ, and $\alpha_1$, $\alpha_2$ are scaling factors to adjust channel power.
To make the VQ adaptable to varying SNR, its centroids are placed on Archimedes' spiral as shown in Fig.~\ref{fig:SpiralVQ}. Arc length parametrization is chosen along the spiral for the same reason as for the $3$:$2$ HVQLC. 
The scaled VQ indices are transmitted as PAM symbols on channel 1, while the scaled error components are transmitted on channels 2 and 3, leading to a ``mix'' of both sources on all three channels. Geometrically, the $2$:$3$ HVQLC consists of  planes parallel to the $z_1,z_2$-plane in channel space (as illustrated in~\cite[p.90]{FloorThesis}) making it similar to $3$:$2$ HVQLC (parallel planes in source space).

%

\emph{Distortion}: $\bar{\varepsilon}_{wn}^2$ can be
found from (\ref{e:wn_uniform}), as the HVQLC is shape preserving. Only the error components $e_1,e_2$ contribute, and so $\bar{\varepsilon}_{wn}^2 = {\sigma_n^2}/{\alpha_2^2}$.

As the VQ indices are scaled by $\alpha_1$, the distance between each plane in channel space is
$\alpha_1$. Therefore, the anomalous error probability is $p_{th}=Pr\{n_1\geq \alpha_1/2\}$. Since $n_1$ is Gaussian, $p_{th}=(1-\mbox{erf}({\alpha_1}/{2\sqrt{2}\sigma_n})$ (see~\cite[p.90]{FloorThesis}). The error made when anomalous errors occur is $\Delta$, as this is the distance to the nearest neighbor for any given centroid. Therefore,
\begin{equation}
\bar{\varepsilon}_{an}^2 =
\frac{\Delta^2}{2}\bigg(1-\mbox{erf}\bigg(\frac{\alpha_1}{2\sqrt{2}
\sigma_n}\bigg)\bigg).
\end{equation}

\emph{Power}: As centroids are placed on the Archimedes' spiral in an equidistant manner, the pdf
of $z_1$ will be a discretized version of the pdf for  RCASD directrix, a discretized Laplace pdf. For small $\Delta$, its variance can be approximated by the variance of a Laplace pdf. Therefore, the power on channel 1 can be approximated as
\begin{equation}\label{e:hvqlc_PowCh1_2}
P_1=\text{Var}\{z_1\} \approx 2\alpha_1^2\left(2 \eta
\frac{\pi^2}{\Delta^2}\sigma_{x}^2\right)^2.
\end{equation}
Generally, the right hand side will be somewhat smaller than the real power, but the smaller $\Delta$ is (higher SNR) the better they coincide. Note particularly that $\sigma_{z_1}^2 \sim 1/\Delta^4$, different from the $3$:$2$ RCASD directrix where $\sigma_{z_1}^2 \sim 1/\Delta^2$. The reason is that indices are sent on the channel and so the length measured along the spiral is independent of $\Delta$. This difference in exponent is crucial to make HVQLC obtain the same slope as $2$:$3$ OPTA.

For channels 2 and 3, assuming that $\Delta$ is small, $e_1$ and $e_2$ are uniformly
distributed over $(\Delta/2)\times(\Delta/2)$. Therefore, the power on channel 2 and 3 can be approximated by $P_2=P_3 \approx {\alpha_2^2\Delta^2}/{12}$.
The total channel power is then
\begin{equation}\label{e:ch_pov_hvqlc}
P_t=\frac{2}{3}\bigg(\bigg(2\alpha_1\eta\frac{\pi^2}{\Delta^2}\sigma_x^2\bigg)^2+\frac{\alpha_2^2\Delta^2}{12}\bigg).
\end{equation}

\emph{Optimization:} The Lagrangian is $\mathcal{L}(\Delta,\alpha_1,\alpha_2,\lambda)=\bar{\varepsilon}_{wn}^2(\alpha_2)+\bar{\varepsilon}_{th}^2(\Delta,\alpha_1)-\lambda
c_t(\Delta,\alpha_1,\alpha_2)$, where $c_t(\Delta,\alpha_1,\alpha_2)=P_{\text{max}}-P_t(\Delta,\alpha_1,\alpha_2)\geq 0$, with $P_t(\Delta,\alpha_1,\alpha_2)$ as in~(\ref{e:ch_pov_hvqlc}), and $P_{\text{max}}$ the maximum power per channel. Numerical optimization is applied (see~\cite[p. 92]{FloorThesis}).

The performance of the optimized $2$:$3$ HVQLC system is
shown in Fig.~\ref{fig:HVQLC_RCASD} (green curve). The HVQLC has  decent performance, about 5dB from OPTA above 20dB SNR. It also follows the slope of OPTA at high SNR as will be shown in Proposition~\ref{prop:HVQLC2_3Slope}. Both simulated (green dashed curve) and calculated (magenta curve) robustness performance are shown, where the two distortion contributions can be seen: $\bar{\varepsilon}_{wn}^2$  dominates above the optimal SNR point and behaves like a linear scheme (having the same slope as BPAM) which is to be expected from Definition~\ref{def:weak_noise_exp} and Proposition~\ref{th:weak_noise}. Below the optimal SNR, $\bar{\varepsilon}_{an}^2$ dominates, and is observed to diverge faster from OPTA than $\bar{\varepsilon}_{wn}^2$. The theoretical model coincides well with simulations at high SNR. Since the HVQLC consists of planes, the weak noise regime of Definitions~\ref{def:weak_noise_exp} will be satisfied exactly.

\emph{High SNR analysis:} We prove that $2$:$3$ HVQLC has the same slope as OPTA as SNR$\rightarrow \infty$. To simplify one can eliminate anomalous errors by choosing $\alpha_1$ sufficiently large. By letting $\alpha_1 \geq 2 b_n \sigma_n$, with $b_n > 4$, $99.99 \% $ of all possible events are included, and the total distortion can be approximated as $D_t \approx \sigma_n^2/\alpha_2^2$.
%

\begin{proposition}\label{prop:HVQLC2_3Slope}\emph{$2$:$3$ HVQLC at high SNR:} At high SNR, the SDR of $2$:$3$ HVQLC follows
\begin{equation}\label{e:hvqlc2_3Dist_highSNR_Optimal}
\text{SDR}= \frac{7\sqrt{3}\sigma_x^2}{6\eta\pi^2\sigma_x^2 b_n}  \text{SNR}^\frac{3}{2}.
\end{equation}
\end{proposition}

\emph{Proof:} See Appendix~\ref{sec:app_pf_HVQLC2_3Slope}.\hspace{11.1cm}$\square$

With $\sigma_x=1$, $b_n=4$, $\eta = 0.16$, the loss from OPTA is $\text{SDR}_{\text{loss}} = - 10\lg (7\sqrt{3}/(6\eta \pi^2 b_n))\approx 4.95$dB, corresponding to the performance gap in Fig.~\ref{fig:HVQLC_RCASD}.

\subsubsection{$2$:$3$ RCASD}\label{ssec:RCASD_exp}
The equation for this mapping is the same as in~(\ref{e:RCASD_Mapping_func}), but now as a function of the source vectors $\mathbf{x}.$

The distortion and power for this mapping is easily derived using results from preceding sections and existing papers:

To compute $\bar{\varepsilon}_{wn}^2$ we assume arc length parametrization and obtain the same $G$ as for the $3$:$2$ RCASD in Section~\ref{ssec:RCASD_Comp}. Then~(\ref{e:mseort_mean_exp}) is reduced to $\bar{\varepsilon}_{wn}^2 = 0.5 {\sigma_n^2}(\alpha_1^{-2} + \alpha_2^{-2})$.  Furthermore, $\bar{\varepsilon}_{an}^2$ is the same as for the $1$:$2$ mapping in~\cite{hekland_floor_ramstad_T_comm}, Eqn.(25), scaled by $0.5$.

The power on channels 1 and 2 is also the same as for the $1$:$2$ mapping in~\cite{hekland_floor_ramstad_T_comm}, and is given by $ P_1+P_2 = 2\Delta\sigma_x \alpha_2/ (\eta\sqrt{2\pi})$. As $x_3$ is scaled with $\alpha_3$ and sent on channel 3, the power is $P_3 = \sigma_x^2 \alpha_3^2$. Then $P_t=(P_1+P_2+P_3)/3$.

The performance of the optimized RCASD is shown in Fig.~\ref{fig:HVQLC_RCASD}. As expected from Proposition~\ref{prop:Map_Split}, the RCASD diverges from OPTA, following the slope of a $1$:$1$ system at high SNR. However, between 10-22 dB SNR, the RCASD outperforms the HVQLC. As for $3$:$2$ case, a DS based approach brings advantages at low to medium SNR. The correspondence between the calculated and simulated robustness curves indicate that the theoretical model fits well with reality at high SNR.

\subsection{Remarks for both $3$:$2$ and $2$:$3$ mappings:}
From the analysis and simulations in Sections~\ref{ssec:Ex_SK_surf_32} and~\ref{ssec:Ex_SK_surf_23} it may appear like fully continuous mappings based on surfaces may be sub-optimality in the sense that they cannot follow the slope of OPTA at high SNR. This is in contrast to S-K mappings realized by curves where such divergence is  not observed~\cite{hekland_floor_ramstad_T_comm,FloorThesis,HeklandThesis}. However, we still have not investigated the optimal choice of coordinate system on non-decomposable mappings like DSS. Just as curve-based mappings will diverge from OPTA if $\varphi$ is chosen non-wisely~\cite{FloorThesis}, it may be that the divergence observed for surfaces is due to wrong choice of coordinate system.  This is indicated by~\cite{Saidutta_JSAC2021} where fully continuous mappings resulting from a deep learning approach, a structure quite similar to the DSS proposed in this paper, seem to follow OPTA at high SNR. However, this is not conclusive as~\cite{Saidutta_JSAC2021} only show performance up to 30dB, where also the DSS follows the slope of OPTA.  An intuitive choice of coordinate curves for dimension reducing mappings are geodesics (see~\cite[p.11]{floor2021tools} or~\cite[pp.162-168]{Kreyszig_DiffGeom91}) as they minimize the length between any two points on $\mathcal{S}$, and thereby the $g_{ii}$'s. However, determining the optimal coordinate system  in general is difficult even for 2D surfaces, and should be followed up in future effort(s).

\section{Summary, discussion and extensions}\label{sec:dis_con}
In this paper a theoretical framework for analyzing and constructing analog mappings used for joint source-channel coding has been proposed. A general set of continuous or piecewise continuous mappings named Shannon-Kotel'nikov (S-K) mappings have been considered for the case of memoryless sources and channels.

Generally, S-K mappings are nonlinear direct mappings between source- and channel space. In this paper we focused on spaces of different dimensions. The distortion framework  introduced describes S-K mapping behaviour in general, that is, without reference to a specific mapping realization. Also, the framework provides guidelines for construction of well performing mappings for both low and arbitrary complexity and delay.

Two propositions (Proposition~\ref{th:asympt_dim_exp} and~\ref{th:asympt_dim_red})  indicate under which conditions S-K mappings may achieve the information theoretical bounds (OPTA) for Gaussian sources.
Not surprisingly, the dimensionality of a mapping must be infinite to achieve optimality when the source and channel dimension do not match.
This is because the optimal \emph{space utilization} with such mappings is obtained only in the limit of infinite dimensionality.

When it comes to construction of mappings it is shown that any mapping which can be decomposed into combinations of lower dimensional sub-mappings cannot obtain the same slope as the information theoretical bounds at high SNR.  We also apply the provided theory to construct mappings for $2$:$3$ and $3$:$2$ cases. These mappings have decent performance. Albeit some of them are inferior to mappings found by machine learning and other numerical optimization methods, the loss is small (about 1dB), and the mappings found can easily be adapted to varying channel conditions simply by scaling one given structure, thereby reducing complexity.

The condition stated can provide constraints on numerical approaches~\cite{Akyol2014_TIT,Mehmetoglu_Akyol_Rose} that may provide mappings closer to the global optimum without having to input a pre-determined, close to optimal mapping. The conditions presented may also provide a deeper understanding of why certain configurations are favored by machine learning approaches~\cite{Saidutta_JSAC2021}.

\emph{Future Extensions:}

\textbf{1) Global (Manifold) structure:} Although the main results of this paper provides indications on the global structure for S-K mappings, they do not necessarily provide the exact optimal solution. Several approaches for finding the global structure exists, like the PCCOVQ algorithm~\cite{fuldsethThesis,floor_iswcs07}, approach using variational calculus~\cite{Akyol_rose_ramstad,Mehmetoglu_Akyol_Rose} and machine learning~\cite{Saidutta_JSAC2021}. All these works rely on numerical methods, and there is no guarantee that the optimal mappings have been found. Constraining solutions based on conditions determined throughout this paper may be one step towards obtaining globally optimal mappings.

\textbf{2) Low SNR:} Further analysis is necessary in order to deal properly with the low SNR case. We considered ML decoding here, but MMSE decoding is needed at low SNR to obtain optimal performance. However, deriving analytical expressions under MMSE decoding is not necessarily feasible for nonlinear mappings in general.

\textbf{3) Correlated sources:} The results of this paper can be extended to correlated sources. 
For example, the special case of two correlated sources transmitted on two channels  was treated in~\cite{Floor_Kim_2013_Entropy} where it was shown that a \emph{ruled surface} can utilize correlation to obtain significant gains. The approach in~\cite{fulds97a} also indicate how to extend these mappings to correlated sources.

\textbf{4) Multiple access networks:} Attempts have been made to extend some of the results of this paper to correlated
Gaussian sources communicated on a Gaussian multiple access channel with both orthogonal and
simultaneous transmission~\cite{Floor_Kim_2013_Entropy,Floor_Kim_Wernersson11_TCOM,Floor_et_al_TCOM_MGMAC_2015}. These are heuristic approaches, however.

\appendices
\section{Concepts from differential geometry}\label{sec:app_DiffGeom_Details}

\subsection{Arc length parametrization, differential geometry of curves and formula of Frenet.}\label{sec:app_usp}
Let $\mathbf{S}: u\in [a,b]\subseteq \mathbb{R}\rightarrow \mathbf{S}(u)\in\mathbb{R}^N$ with $\mathbf{S}(u)\in C^1$ be a parametrization for the curve $\mathcal{C}$ w.r.t. $\ell(u)$. Let $\ell(u)$ denote the arc length of $\mathbf{S}$ as defined in~(\ref{e:CurveLength}) and $\varphi$ its inverse.

\begin{theorem}\label{th:clpar}
Let $\mathbf{y}(\ell)$ be a parametrization of $\mathcal{C}$. Then
$\mathbf{y}(\ell)$ and $\mathbf{s}(\varphi(u))$ will have the same image, and
$\|\mathbf{y}'(\ell)\|=\| \mathbf{s}'(\varphi(\ell))\|\equiv1,\hspace{0.5cm}
\forall\ell $.\hspace{9cm}$\square$
\end{theorem}
\begin{pf}
See\cite[pp. 115-116]{callahan00}.\hspace{10.5cm}$\square$
\end{pf}

There are three unit vectors connected to any curve $\mathcal{C}:\mathbf{S}\in\mathbb{R}^3$: The \emph{unit tangent vector} $\mathbf{t}=\dot{\mathbf{S}}=\mathbf{S}'/\|\mathbf{S}'\|$, the \emph{unit principal normal vector} $\mathbf{p}=\dot{\mathbf{t}}/\|\dot{\mathbf{t}}\|= \ddot{\mathbf{S}}/\|\ddot{\mathbf{S}}(x_0)\|$ , and the \emph{unit binormal vector} $\mathbf{b} =\mathbf{t} \times \mathbf{p}$.
The vectors $\mathbf{t}$, $\mathbf{p}$ and $\mathbf{b}$ make out a vector space of mutually orthogonal vectors named \emph{moving trihedron} which is so defined at each point along $C$. This is illustrated in~\cite[pp. 36-37]{Kreyszig_DiffGeom91}. These vectors further define three mutually orthogonal planes: i) \emph{Osculating Plane} spanned by $\mathbf{t}$ and $\mathbf{p}$, ii) \emph{normal plane} spanned by $\mathbf{p}$ and $\mathbf{b}$, and iii) \emph{rectifying plane} spanned by $\mathbf{t}$ and $\mathbf{b}$.

For a parametric curve, $\mathbf{S}(u)$,  curvature w.r.t. arch length is defined as  $\kappa_0=\|\ddot{\mathbf{S}}(u_0)\|$~\cite[p. 34]{Kreyszig_DiffGeom91}. Then we also have $\mathbf{p} = (1/\kappa)\ddot{\mathbf{S}} =\rho \ddot{\mathbf{S}}$. The \emph{torsion}~\cite[p. 37-40]{Kreyszig_DiffGeom91} is defined as $\tau(x) = -\mathbf{p}\cdot\mathbf{b} = {\big|\dot{\mathbf{S}} \ \ddot{\mathbf{S}} \ \dddot{\mathbf{S}}\big|}/{\|\ddot{\mathbf{S}}\|^2}.$
For a general parametrization we have~\cite[pp.35,39]{Kreyszig_DiffGeom91}
\begin{equation}\label{e:CurvatureTorsion_curve_gen_coord}
\kappa = \frac{\sqrt{\|\mathbf{S}'\|^2\|\mathbf{S}''\|^2-(\mathbf{S}' \cdot \mathbf{S}'')^2}}{\|\mathbf{S}'\|^\frac{3}{2}}, \ \ \tau = \frac{\big|\mathbf{S}' \ \mathbf{S}'' \ \mathbf{S}'''\big|}{\|\mathbf{S}'\|^2\|\mathbf{S}''\|^2-(\mathbf{S}' \cdot \mathbf{S}'')^2}.
\end{equation}
For scaled arc length parametrization, $\|{\mathbf{S}}'(u_0)\|=\alpha\|\dot{\mathbf{S}}(u_0)\|=\alpha$, $\forall u_0$, and so $\kappa$ in~(\ref{e:CurvatureTorsion_curve_gen_coord}) reduces to $\kappa(u_0)=\|\mathbf{S}_0''(u_0)\|/\|\mathbf{S}_0'\|^2$ as $\mathbf{S}_0' \perp \mathbf{S}_0''$.

The curvature can locally be interpreted as a circle of radius $\rho = 1/\kappa$, named \emph{radius of curvature}, lying in the osculating plane of $\mathbf{s}$. The corresponding circle is named \emph{osculating circle} and its center named \emph{centre of curvature}. I.e., the curvature in a neighborhood of $u_0$ is equivalent to that of a circle with radius $\rho$ (illustration is provided in~\cite{floor2021tools}, Fig. 3). This concept is also valid for curves in $\mathbb{R}^M, M\geq3$.

\begin{definition}\label{def:sk_mapp}\emph{Formula of Frenet} (FoF)~\cite[p. 41]{Kreyszig_DiffGeom91} relates the derivatives $\dot{\mathbf{t}}$, $\dot{\mathbf{p}}$ and $\dot{\mathbf{b}}$ to linear combinations of $\mathbf{t}$, $\mathbf{p}$, and $\mathbf{b}$ of  curve $\mathcal{C}$ defined in Section~\ref{ssec:intr_sk_mapp} as follows:
\begin{equation}\label{e:FoF}
\dot{\mathbf{t}} = \kappa\mathbf{p}, \ \ \dot{\mathbf{p}} = -\kappa \mathbf{t} + \tau \mathbf{b}, \ \ \dot{\mathbf{b}} =  -\tau \mathbf{p},
\end{equation}
where $\kappa$ is the curvature and $\tau$ the torsion. \hspace{9cm}$\square$
\end{definition}

\subsection{Einstein summation convention, surfaces, fundamental forms and curvature}\label{sec:app_FundForms_Einstein}

\subsubsection{Summation convention}\label{ssec:app_EinsteinSum}
To efficiently express multiple sum-operations resulting when analyzing surfaces, \emph{Einstein summation convention} is convenient~\cite[p.84]{Kreyszig_DiffGeom91}:\\
\textbf{If in a product a letter figures twice, once a superscript and once a subscript, summation should be carried out from $1$ to $N$ w.r.t. this letter.}

For example, for simple and double sums we have
\begin{equation}\label{e:summ_conv_vector}
\sum_{\alpha = 1}^N a^\alpha b_\alpha = a^\alpha b_\alpha, \ \ \ \sum_{\alpha = 1}^N \sum_{\beta = 1}^N a_{\alpha\beta} u^\alpha u^\beta = a_{\alpha\beta} u^\alpha u^\beta.
\end{equation}

\subsubsection{Fundamental forms}\label{ssec:app_FFF_SFF}

\textbf{First fundamental form (FFF)}: Consider a hyper surface $\mathcal{S}$ realized by~(\ref{e:par_surf_eq}) or~(\ref{e:par_surf_eq2}). In order to measure lengths, angles and areas on $\mathcal{S}$, a metric is needed. A length differential of a curve $C \in\mathcal{S}$ is given by~\cite[p.82]{Kreyszig_DiffGeom91}\footnote{We look at a 2D surface here for better readability. The general case is straight forward to determine.}:
\begin{equation}\label{e:arc_length_element_surf}
\mbox{d}\ell^2 = (\mathbf{S}_1 \mbox{d}u^1 + \mathbf{S}_2 \mbox{d}u^2)\cdot(\mathbf{S}_1 \mbox{d}u^1 + \mathbf{S}_2 \mbox{d}u^2)= \mathbf{S}_1\cdot\mathbf{S}_1 (\mbox{d}u^1)^2 + 2\mathbf{S}_1\cdot\mathbf{S}_2 \mbox{d}u^1 \mbox{d}u^2 + \mathbf{S}_2\cdot\mathbf{S}_2 (\mbox{d}u^2)^2.
\end{equation}
The quantities $g_{\alpha\beta}=\mathbf{S}_\alpha\cdot\mathbf{S}_\beta$ are components of a \emph{2nd order covariant tensor} (see~\cite[pp.88-105]{Kreyszig_DiffGeom91} or~\cite{floor2021tools} for definition of covariant and contravariant tensor) named \emph{metric tensor}.
By the summation convention, $\mbox{d}\ell^2  = g_{\alpha\beta} \mbox{d}u^\alpha \mbox{d}u^\beta$, named \emph{first fundamental form} (FFF). For a smooth embedding $\mathbf{S}$ in $\mathbb{R}^N$ ($M\leq N$) the metric tensor is a symmetric, positive definite $M\times M$ matrix $G=J^T J$~\cite[pp.301-343]{Spivak99}, with $J$ the \emph{Jacobian}~\cite[p.47]{munkr91} of $\mathcal{S}$, a $N\times M$ matrix with entries $J_{ij}=\partial \mathbf{S}_i / \partial u_j$, $i\in 1,\cdots,N, \ j\in 1,\cdots,M$. I.e., $g_{ii}$ is the squared norm of the tangent vector along the $i$'th coordinate curve of $\mathbf{S}$.  All \emph{cross terms} $g_{ij}$, are inner products of tangent vectors along the $i$'th and $j$'th coordinate curve of $\mathbf{S}$.

The \emph{contravariant metric tensor} is a tensor with components $g^{\alpha\beta}$ satisfying $g_{\alpha\beta}g^{\alpha\beta}=\delta_\alpha^\beta$. Let $g=\mbox{det}(G)$. For a 2-dimensional $\mathcal{S}$, the covariant and contravariant metrics are related as
\begin{equation}\label{e:contravar_covar_rel}
g^{11}=\frac{g_{22}}{g}, \ g^{12}=g^{21}=- \frac{g_{12}}{g}, \ g^{22}=\frac{g_{11}}{g}.
\end{equation}

\textbf{Second fundamental form (SFF)}:
 For any point, $P$, of a curve $\mathcal{C}\in \mathcal{S}\in\mathbb{R}^3$, the corresponding unit normal to $\mathcal{S}$, $\mathbf{n}={\mathbf{S}_1\times \mathbf{S}_2}/{\|\mathbf{S}_1\times \mathbf{S}_2\|}$,
lies in the normal plane of $\mathcal{C}$ which also contains its principal normal $\mathbf{p}$. The angle between $\mathbf{n}$ and $\mathbf{p}$ denoted $\gamma$, depends on the geometry of both $\mathcal{C}$ and $\mathcal{S}$ in a neighborhood of $P$. We have two extremes:  1) $\gamma=\pi/2$, $\forall u_0 \in \mathcal{C}$, implying that  $\mathbf{p}\perp \mathbf{n}$, and $\mathcal{C}$ is a plane curve. I.e., $\mathcal{S}$ is a plane. 2) $\gamma = 0$, $\forall u_0 \in \mathcal{C}$, implying that  $\mathbf{p}|| \mathbf{n}$. Then $\mathcal{C}$ is a \emph{geodesic} on $\mathcal{S}$, i.e., the arc with the shortest possible length between two points on $\mathcal{S}$~\cite[pp.160-162]{Kreyszig_DiffGeom91}\footnote{Geodesics are solutions to the \emph{Euler-Lagrange Equations} in Variational Calculus~\cite[pp.13-17]{troutman96}.}. Examples are straight lines in the plane and great circles on a sphere.

Assume that curve $\mathcal{C}\in \mathcal{S}$ is represented by arc length parametrization $u^1(\ell), u^2(\ell)$, with $\gamma$ the angle between $\mathbf{n}$ and $\mathbf{p}$. As these are unit vectors, $\cos \gamma = \mathbf{p}\cdot\mathbf{n}$, which will generally vary along $\mathcal{C}$. From FoF~(\ref{e:FoF}) we have $\mathbf{p}=\ddot{\mathbf{S}}/\kappa$ and so $\kappa \cos \gamma = \ddot{\mathbf{S}}\cdot \mathbf{n}$.
From the product rule
\begin{equation}\label{e:CurveSurf_arcderivative}
\dot{\mathbf{S}}= \frac{\partial \mathbf{S}}{\partial u^1} \frac{\mbox{d}u^1}{\mbox{d}\ell}+\frac{\partial \mathbf{S}}{\partial u^2} \frac{\mbox{d}u^2}{\mbox{d}\ell}=\mathbf{S}_\alpha \dot{u}^\alpha.
\end{equation}
Differentiating w.r.t. $\ell$ again, then $\ddot{\mathbf{S}}=\mathbf{S}_{\alpha\beta} \dot{u}^\alpha\dot{u}^\beta + \mathbf{S}_\alpha\ddot{u}^\alpha.$
Since $\mathbf{S}_\alpha \cdot \mathbf{n} = 0$, then $\kappa \cos \gamma = (\mathbf{S}_{\alpha\beta} \cdot \mathbf{n})\dot{u}^\alpha\dot{u}^\beta$.
The term in the parentheses,
\begin{equation}\label{e:b_ab_coefs1}
b_{\alpha\beta}=\mathbf{S}_{\alpha\beta} \cdot  \mathbf{n}, \  \alpha,\beta=1,\cdots,N,
\end{equation}
depend on $\mathcal{S}$ only (independent of $\mathcal{C}$) and are symmetric due to the symmetry of $\mathbf{S}_{\alpha\beta}$. $b_{\alpha\beta}$ are components of a 2nd order covariant tensor, where the quadratic form $b_{\alpha\beta} \mbox{d}u^\alpha \mbox{d}u^\beta$ is the \emph{second fundamental form} (SFF).
To compute $b_{\alpha\beta}$ for surfaces in $\mathbb{R}^3$, the following relation is convenient,
\begin{equation}\label{e:compute_b_ab}
b_{\alpha\beta}= \mathbf{S}_{\alpha\beta} \cdot  \mathbf{n} = \mathbf{S}_{\alpha\beta}\cdot \bigg(\frac{\mathbf{S}_1\times \mathbf{S}_2}{\sqrt{g}}\bigg)=\frac{1}{\sqrt{g}}|\mathbf{S}_1\ \mathbf{S}_2\ \mathbf{S}_{\alpha\beta}|.
\end{equation}

\subsubsection{Normal curvature, principal curvature, lines of curvature}\label{ssec:app_curvature}
Let $t$ be any allowable parameter for curve $\mathcal{C}$. Then $\dot{u}^\alpha = ({\mbox{d} u^\alpha}/{\mbox{d} t})({\mbox{d} t}/{\mbox{d} \ell}) = {{u^\alpha}'}/{\ell'}$.
and therefore
\begin{equation}\label{e:normals_CurveSurf_angle3}
\kappa \cos \gamma = b_{\alpha\beta} \dot{u}^\alpha\dot{u}^\beta  = \frac{b_{\alpha\beta} {u^\alpha}' {u^\beta}'}{\ell'}=\frac{b_{\alpha\beta} {u^\alpha}' {u^\beta}'}{g_{\alpha\beta} {u^\alpha}' {u^\beta}'}{=\frac{b_{\alpha\beta} \mbox{d}{u^\alpha} \mbox{d}{u^\beta}}{g_{\alpha\beta} \mbox{d}{u^\alpha} \mbox{d}{u^\beta}}}.
\end{equation}
It is shown in~\cite[pp. 121-124]{Kreyszig_DiffGeom91} that $\kappa_n=\kappa  \cos \gamma$ is the curvature at point $P\in\mathcal{S}$ of the \emph{normal section} $\mathcal{C}\in\mathcal{S}$, named \emph{normal curvature} at $P$. Further, the \emph{theorem of Meusnier}~\cite[p.122]{Kreyszig_DiffGeom91} states that one can restrict the consideration of curvature at any point $P\in\mathcal{S}$ to that of normal sections without loss of generality.

The directions where $\kappa_n$ has extremal values can be determined except when $b_{\alpha\beta}\sim g_{\alpha\beta}$ (named \emph{umbilic} point): Eq.~(\ref{e:normals_CurveSurf_angle3}) can be rewritten as $(b_{\alpha\beta}- \kappa_n g_{\alpha\beta}) \mbox{d}{u^\alpha} \mbox{d}{u^\beta}=0$. By differentiating w.r.t. $\mbox{d}{u^\gamma}$, treating $\kappa_n$ as a constant, one obtains~\cite[pp.128-129]{Kreyszig_DiffGeom91},
\begin{equation}\label{e:Principal_curvature_DiffEq1}
 (b_{\alpha\gamma}- \kappa_n g_{\alpha\gamma}) \mbox{d}{u^\alpha}=0, \ \gamma = 1,2.
\end{equation}
The roots of~(\ref{e:Principal_curvature_DiffEq1}) are directions for which $\kappa_n$ is extreme, named \emph{principal directions of normal curvature}  at $P$. The corresponding curvatures, $\kappa_1$ and $\kappa_2$, are the \emph{principal normal curvatures} of $\mathcal{S}$, corresponding the maximal and minimal curvature of $\mathcal{S}$ at $P$, respectively.

It is proven in~\cite[p.129]{Kreyszig_DiffGeom91} that the roots of~(\ref{e:Principal_curvature_DiffEq1}) are real, and at every point (not an umbilic), the principal directions are orthogonal. Further, a curve on $\mathcal{S}$ whose direction at every point is a principal direction is a \emph{line of curvature} (LoC) of $\mathcal{S}$. It is proven in~\cite[p.130]{Kreyszig_DiffGeom91} that LoC on any surface $\mathcal{S}\in C^r$, $r\geq 3$, are real curves, and if $\mathcal{S}$ has no umbilics, the LoC form an orthogonal net everywhere on $\mathcal{S}$. One may always choose coordinates $u^1,u^2$ on $\mathcal{S}$ so that the LoC are allowable coordinates (see~\cite[p. 2]{floor2021tools}) at any point of $\mathcal{S}$ (not umbilic). Then~\cite[p.130]{Kreyszig_DiffGeom91}:

\begin{theorem}\label{th:LoC_Coordinates}
The coordinate curves of any allowable coordinate system on $\mathcal{S}$  coincide with the LoC $\Leftrightarrow$ $g_{12}=0 \ \ \text{and} \ \ b_{12}=0,$
at any point where those coordinates are allowable.
\end{theorem}

\emph{Proof:} See~\cite[p.130]{Kreyszig_DiffGeom91}.\hspace{12.0cm}$\square$

When the coordinate curves are LoC,~(\ref{e:Principal_curvature_DiffEq1}) holds with $\kappa_n=\kappa_1$, $\mbox{d}u^2=0$ and again with $\kappa_n=\kappa_2$, $\mbox{d}u^1=0$. Therefore $\kappa_1=b_1^1, \ b_1^2=0, \ \kappa_2=b_2^2, \ b_2^1=0$, and one obtains $\kappa_i=b_{ii}/g_{ii}$. Generally, with $B$ the matrix of $b_{\alpha\beta}$, $\kappa_i$ are the roots of~\cite[p. 130]{Kreyszig_DiffGeom91} 
\begin{equation}\label{e:PrincipalCurvature_Gen}
\kappa_i^2 - b_{\alpha\beta}g^{\alpha\beta}\kappa_i + \frac{\det(B)}{\det(G)}=0.
\end{equation}

\section{Proofs for Section~\ref{sec:Dist_SK}}\label{sec:App_SKDist_FiniteDim}
\subsection{Proofs for Section~\ref{sec:mn_dimexp}}

\subsubsection{Proof, Proposition~\ref{th:weak_noise}}\label{sec:app_pf_th1}
Assume that $S_i\in C^1(\mathbb{R}^M),\hspace{2mm}  i=1,..,N$.
The tangent space at $\mathbf{x}_0$ is given by~(\ref{e:lin_apr_exp}). Applying ML detection, then $\mathbf{S}(\mathbf{x}_{ML})=\mathbf{S}(\mathbf{x}_0)+P_{proj}\mathbf{n}$ (see Fig.~\ref{fig:smaln_gen}), where $P_{proj}$ is a projection matrix given by~\cite[p.158]{Strang86}
\begin{equation}\label{e:projm}
P_{proj}=J(\mathbf{x}_0)(J(\mathbf{x}_0)^TJ(\mathbf{x}_0))^{-1}J(\mathbf{x}_0)^T=J(\mathbf{x}_0)G(\mathbf{x}_0)^{-1}J(\mathbf{x}_0)^T,
\end{equation}
Setting $\mathbf{S}_{\text{lin}}(\mathbf{x})=\mathbf{S}(\mathbf{x}_{ML})$, with $\mathbf{S}_{\text{lin}}$ as in~(\ref{e:lin_apr_exp}) and from~(\ref{e:projm}), we get
\begin{equation}\label{e:proj}
J(\mathbf{x}_0)(\mathbf{x}_{\scriptsize{\mbox{ML}}}-\mathbf{x}_0)=J(\mathbf{x}_0)G(\mathbf{x}_0)^{-1}J(\mathbf{x}_0)^T\mathbf{n}.
\end{equation}
Multiplying both sides from the left with $J^T$, using the fact that $G$ is invertible, then $(\mathbf{x}_{\scriptsize{\mbox{ML}}}-\mathbf{x}_0)=G(\mathbf{x}_0)^{-1}J(\mathbf{x}_0)^T\mathbf{n}$. The MSE given that $\mathbf{x}_0$ was transmitted is then
\begin{equation}\label{e:mse}
\varepsilon_{wn}^2=\frac{1}{M}E\{(\mathbf{x}_{\scriptsize{\mbox{ML}}}-\mathbf{x}_0)^T(\mathbf{x}_{\scriptsize{\mbox{ML}}}-\mathbf{x}_0)\}.
\end{equation}

\begin{lemma}\label{lm:weak_n_mse}
With ML detection, the minimum  MSE in~(\ref{e:mse}) is achieved with a diagonal $G$
\begin{equation}\label{e:mse3ort}
\varepsilon_{wn}^2=\frac{\sigma_n^2}{M}\sum_{i=1}^M\frac{1}{g_{ii}},
\end{equation}
where $g_{ii}$ are the diagonal components of $G$ at $\mathbf{x}_0$. \hspace{7.5cm}$\square$\\
\end{lemma}

Lemma~\ref{lm:weak_n_mse} implies that the smallest possible weak noise MSE is obtained with orthogonal coordinate curves. Expectation over $\mathcal{D}$ gives the wanted result.\hspace{5.5cm}$\square$

\emph{Proof, Lemma~\ref{lm:weak_n_mse}:}
Consider the MSE in~(\ref{e:mse}). To avoid matrix multiplication, the $N$-dimensional
noise vector $\mathbf{n}$ is, without loss of generality, replaced
by its $M$ dimensional projection $\mathbf{n}_P$, which is also Gaussian since $P_{proj}$ is a linear
transformation~\cite[p.117]{Strang86}. Let $J=J(\mathbf{x}_0)$. Assume that a hypothetical
inverse $\mathbf{B}=J^{-1}$ exists. Let
$\mathbf{S}_t$ denote the (M-dimensional) tangent space of $\mathbf{S}$ at $\mathbf{x}_0$. Under Definition~\ref{def:weak_noise_exp} the linear approximation to $\mathbf{S}^{-1}$ can be applied, and so $\hat{\mathbf{x}} = \mathbf{S}^{-1}(\mathbf{S}_t(\mathbf{x}_0)+\mathbf{n}_{P})\approx \mathbf{S}^{-1} (\mathbf{S}_t(\mathbf{x}_0)) + \mathbf{B}
\mathbf{n}_P = \mathbf{x}_0 + \mathbf{B}\mathbf{n}_P$. Then, using Einstein summation convention, Eqn.~(\ref{e:mse}) becomes  $\varepsilon_{wn}^2 = E \{\mathbf{n}_P^T \mathbf{B}^{T}\mathbf{B} \mathbf{n}_P \}/M =\mathbf{b}_i^T \mathbf{b}_j E\{n^i n^j\}/{M}$, with $\mathbf{b}_i$ column vector no. $i$ of $\mathbf{B}$.
With i.i.d. noise, $E\{n_i n_j\}=\sigma_n^2\delta_{ij}$, then
\begin{equation}\label{e:end_res}
\varepsilon_{wn}^2 =\frac{1}{M} E \{\mathbf{n}_P^T \mathbf{B}^{T}
\mathbf{B} \mathbf{n}_P \}= \frac{\sigma_n^2}{M}\sum_{i=1}^M
\mathbf{b}_i^T \mathbf{b}_i=\frac{\sigma_n^2}{M}\sum_{i=1}^M
\|\mathbf{b}_i\|^2.
\end{equation}
Since $\mathbf{B}=J^{-1}$, and since any orthogonal matrix has an inverse, the above result implies that the basis of the
tangent space of $\mathcal{S}$ can
be chosen orthogonal without any loss. An orthogonal $J$ results in a diagonal $G$ (see Appendix~\ref{ssec:app_FFF_SFF}). Therefore
$G^{-1}$, as well as $G^{-n})$, are also diagonal with elements $1/g_{ii}$. With $G^{-2}$ diagonal, $E\{n_i n_j\}=\sigma_n^2
\delta_{ij}$, and with~(\ref{e:end_res}) in mind,~(\ref{e:mse}) leads to
\begin{equation}
\bar{\varepsilon}_{wn}^2 =\frac{1}{M}
E\{(G^{-1}J^T\mathbf{n})^T(G^{-1}J^T\mathbf{n})\}=\frac{1}{M}
E\{(J^T\mathbf{n})^T G^{-2}(J^T\mathbf{n})\}=
\frac{\sigma_n^2}{M}\sum_{i=1}^M \frac{1}{g_{ii}^2} \|J_i\|^2,
\end{equation}
where $J_i$ is column vector no. $i$ of $J$ and $\|J_i\|^2\equiv g_{ii}$.\hspace{7cm}$\square$

\subsubsection{Proof, Proposition~\ref{th:DimExp_2ndOrder}}\label{sec:app_pf_prop_WN_2nd_order}
\begin{figure}[h]
    \begin{center}
           \includegraphics[width=0.35\columnwidth]{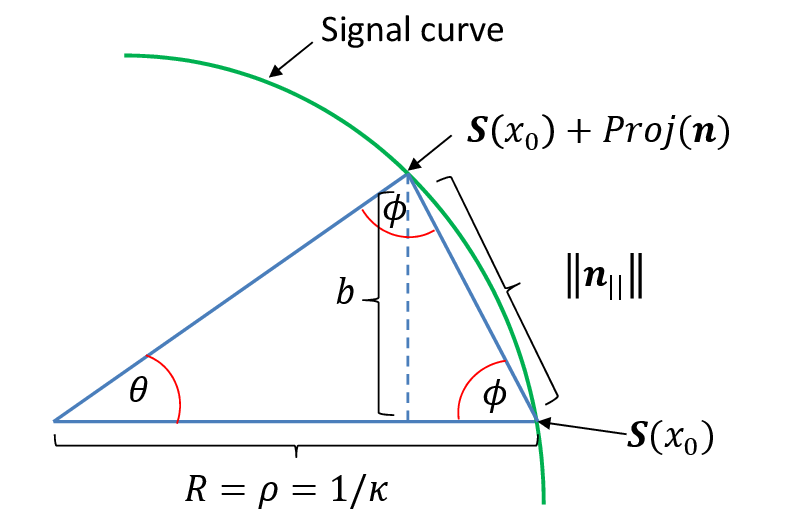}
    \end{center}
    \caption{Circle approximation for calculation of error up to 2nd order.}\label{fig:PPM_curvature}
\end{figure}

The ML-estimate of this problem using 2'nd order Taylor approximation has no simple solution~\cite{floor2021tools}. However, we can apply the analysis for \emph{pulse position modulation} (PPM) in~\cite[pp. 703-704]{wozandj65}. Geometrically, PPM is a curve on a hyper sphere where the arc between any two coordinate axes is like the circle segment depicted in Fig.~\ref{fig:PPM_curvature}.
As the curvature of any curve can be described locally by the osculating circle, the analysis done for PPM is also valid locally for any $1$:$N$ mapping under arc length parametrization. In the following, the curve segment in Fig.~\ref{fig:PPM_curvature} is named \emph{circle approximation}.

We divide the noise into the components $\text{Proj}(\mathbf{n})=\mathbf{n}_{||}$, i.e., the  projection onto the closest point on the circle in Fig.~\ref{fig:PPM_curvature}, and its normal, $\mathbf{n}_\perp$. For the circle approximation $R(x)= \rho(x) = {1}/{\kappa(x)}={1}/{\|\ddot{\mathbf{S}}_0\|}$
iff $\mathbf{t}=\|\dot{\mathbf{S}}_0\|=1$ (under arc length parametrization). In polar coordinates, $\mathbf{S}(x)\approx [R(x),\theta(x)]=[\rho,\theta]$. Then $R(x)=R=\rho, \ \forall x$ and  we have $\mathbf{S}=[R \cos(\theta(x)), R\sin(\theta(x))]$. Then
\begin{equation}\label{e:derivative_COC}
\frac{\mbox{d}\mathbf{S}}{\mbox{d} x} =\bigg[R\sin(\theta(x))\frac{\mbox{d}\theta(x)}{\mbox{d} x}, -R\cos(\theta(x))\frac{\mbox{d}\theta(x)}{\mbox{d} x} \bigg].
\end{equation}
By taking the norm we find that $\|{\mbox{d}\mathbf{S}}/{\mbox{d} x}\|^2=R^2 ({\mbox{d}\theta(x)}/{\mbox{d} x})^2$.
From this we get
\begin{equation}\label{e:PPM_theta_derivative}
\frac{\mbox{d}\theta(x)}{\mbox{d} x} = \frac{1}{R}\bigg\|\frac{\mbox{d}\mathbf{S}}{\mbox{d} x}\bigg\| = \frac{\|\mathbf{S}_0'\|}{\rho}=\alpha \kappa= \alpha \|\ddot{\mathbf{S}}_0\|.
\end{equation}
That is, ${\mbox{d}\theta(x)}/{\mbox{d} x} \sim \kappa$. The two last equalities in~(\ref{e:PPM_theta_derivative}) are valid under scaled arc length parametrization. To determine the ML estimate for the circle approximation we rewrite~(\ref{e:PPM_theta_derivative}) as $\mbox{d}x = \mbox{d}\theta(x)/(\|\mathbf{s}_0'\|\|\mathbf{s}_0''\|)=\mbox{d}\theta(x)\rho(x_0)/\|\mathbf{s}_0'\|$. Then
\begin{equation}\label{e:ML_est_1_N_CircEst}
x_0 - \hat{x}_{ML} = \frac{\rho(x_0)}{\|\mathbf{s}_0'\|}\theta = \frac{\theta }{\kappa(x_0)\|\mathbf{s}_0'\|}.
\end{equation}
$\theta$ must be expressed in terms of $\mathbf{n}_{||}$ and $\rho$ (or $\kappa$). In Fig.~\ref{fig:PPM_curvature} we have a right-legged triangle where $\phi = (\pi-\theta)/2$, and with $b=\rho \sin(\theta)$ its right normal. Therefore $\sin(\phi) = {b}/{\|\mathbf{n}_{||}\|}$ $\Rightarrow$ $\|\mathbf{n}_{||}\| = {b}/{\sin(\phi)}={\rho \sin(\theta)}/{\sin(\phi)}$. Furthermore,  $\sin(\phi) = \sin({\pi}/{2}-{\theta}/{2})=\cos({\theta}/{2})$.
Since $\sin 2y = 2\sin y \cos y$, with $y=\theta/2$, then $\sin(\theta) = 2\sin({\theta}/{2})\cos({\theta}/{2})$, implying that $\|\mathbf{n}_{||}\| =2\rho\sin({\theta}/{2}).$ Therefore $\theta = 2\sin^{-1}({\|\mathbf{n}_{||}\|}/{(2\rho(x_0))})$. By the 2nd order expansion~\cite[p. 117]{Rottmann91}   $\sin^{-1}(x)= x + {x^3}/{(2\cdot 3)} +{ x^5 1\cdot 3}/{2\cdot 4 \cdot 5} +\cdots$, we get
\begin{equation}\label{e:Theta_afo_noise_approx}
\theta \approx \frac{\|\mathbf{n}_{||}\|}{\rho(x_0)}\bigg(1+\frac{\|\mathbf{n}_{||}\|^2}{24\rho^2(x_0)}\bigg).
\end{equation}
One can then compute the error up to second order from~(\ref{e:ML_est_1_N_CircEst}) and~(\ref{e:Theta_afo_noise_approx})
\begin{equation}\label{e:WeakNoiseDist_2ndOrder_pre}
\begin{split}
{\varepsilon}_{wn}^2 &= E\big\{(x-\hat{x}_{ML})^2 | x=x_0\big\} =E\bigg\{\bigg(\frac{\rho(x_0)}{\|\mathbf{s}_0'\|}\theta \bigg)\bigg\}= \frac{\rho^2(x_0)}{\|\mathbf{s}_0'\|^2} E\bigg\{\bigg(\frac{\|\mathbf{n}_{||}\|}{\rho(x_0)}\bigg(1+\frac{\|\mathbf{n}_{||}\|^2}{24\rho^2(x_0)}\bigg)\bigg)^2\bigg\}\\
 &=\frac{1}{\|\mathbf{s}_0'\|^2} E\bigg\{\|\mathbf{n}_{||}\|^2  + \frac{2\|\mathbf{n}_{||}\|^4}{24\rho^2(x_0)}+\frac{\|\mathbf{n}_{||}\|^6}{24^2\rho^4(x_0)}\bigg\}.
\end{split}
\end{equation}
Since $E\{n^a\}= 1\cdot3\cdots (a-1)\sigma_n^a, \ a$ even, and zero otherwise~\cite[p.148]{papoulis02}, we get
\begin{equation}\label{e:WeakNoiseDist_2ndOrder_1}
{\varepsilon}_{wn}^2 = \frac{\sigma_n^2}{\|\mathbf{s}_0'\|^2}\bigg(1+\frac{\sigma_n^2}{4\rho^2(x_0)}+\frac{5\sigma_n^4}{48\rho^4(x_0)}\bigg)=\frac{\sigma_n^2}{\|\mathbf{s}_0'\|^2}\bigg(1+\frac{1}{4}\sigma_n^2 \kappa^2(x_0)+\frac{5}{48}\sigma_n^4 \kappa^4(x_0)\bigg).
\end{equation}
In general, $\kappa$ is given by~(\ref{e:CurvatureTorsion_curve_gen_coord}). Under scaled arc length parametrization, $\|{\mathbf{S}}'(x_0)\|=\alpha\|\dot{\mathbf{S}}(x_0)\|=\alpha$, $\forall x_0$, then~(\ref{e:CurvatureTorsion_curve_gen_coord}) reduces to $\kappa(x_0)=\|\mathbf{S}_0''(x_0)\|/\|\mathbf{S}_0'\|^2$. Then the error can be expressed in terms of the signal curve's derivatives as
\begin{equation}\label{e:WeakNoiseDist_2ndOrder_2}
{\varepsilon}_{wn}^2 = \frac{\sigma_n^2}{\|\mathbf{S}_0'\|^2}\bigg(1+\frac{1}{4}\sigma_n^2 \frac{\|\mathbf{S}_0''\|^2}{\|\mathbf{S}_0'\|^4}+\frac{5}{48}\sigma_n^4 \frac{\|\mathbf{S}_0''\|^4}{\|\mathbf{S}_0'\|^8}\bigg).
\end{equation}

As will be shown in Lemma~\ref{lem:canal_surf_cond_surfaces}, one should let $\sigma_n^2 << \rho^2(x_0)$ to avoid larger errors (at least at high SNR).
At high SNR ($\sigma_n^2 << 1$) one can therefore make the approximation
\begin{equation}\label{e:WeakNoiseDist_2ndOrder_Final}
{\varepsilon}_{wn}^2 \approx \frac{\sigma_n^2}{\|\mathbf{S}_0'\|^2}\bigg(1+\frac{1}{4}\sigma_n^2 \kappa^2(x_0)\bigg)= \frac{\sigma_n^2}{\|\mathbf{S}_0'\|^2}\bigg(1+\frac{1}{4}\sigma_n^2 \frac{\|\mathbf{S}_0''\|^2}{\|\mathbf{S}_0'\|^4}\bigg)
\end{equation}
At high SNR the $\mathbf{S}$ can be significantly stretched, i.e.,  $\|\mathbf{S}_0'\|=\alpha >> 1$. The the 1st order term in~(\ref{e:WeakNoiseDist_2ndOrder_Final}) will dominate more and more over the 2nd order term. As shown in Section~\ref{sec:app_pf_prop_WN_2nd_order_red}, higher order terms will contribute even less, and therefore the circle approximation above will do. 

To see that the circle approximation above is valid locally for any $1$:$N$ mapping:
For a general curve in polar coordinates, the product rule gives
\begin{equation}\label{e:diff_polar_coord_gen_curve}
\frac{\mbox{d}\mathbf{S}}{\mbox{d}x}=\bigg[R'(x)\cos(\theta(x))+R(x)\sin(\theta(x)),R'(x)\sin(\theta(x))-R(x)\cos(\theta(x))\bigg].
\end{equation}
Locally, $R'(x)$ is small if the curvature of $\mathbf{S}$ changes slowly with $x$. Therefore, the smaller the derivative $\kappa'(x)$, the more accurate the circle approximation is. As will become clear in Lemma~\ref{lem:canal_surf_cond_surfaces}, it is not convenient to use a curve (or surface in general) with a rapidly changing curvature.

For generalization to surfaces, consider first $2$:$N$ mappings. Assume that $x_1$ and $x_2$ are parameters in a LoC coordinate representation. Then, according to Theorem~\ref{th:LoC_Coordinates} in Section~\ref{ssec:app_curvature}, $g_{12}=b_{12}=0$, therefore the curvature is $\kappa_i(\mathbf{x}_0)={b_{ii}(\mathbf{x}_0)}/{g_{ii}(\mathbf{x}_0)}$, $i=1,2$.
Since $x_1$, $x_2$, $n_1$ and $n_2$ are independent and i.i.d., and we have two orthogonal coordinate curves,
\begin{equation}\label{e:WeakNoiseDist_2ndOrder_2_N}
{\varepsilon}_{wn}^2(\mathbf{x}_0) \approx \frac{\sigma_n^2}{2} \sum_{i=1}^2 \bigg\{\frac{1}{g_{ii}(\mathbf{x}_0)}\bigg(1+\frac{\sigma_n^2}{4} \kappa_i^2(\mathbf{x}_0)\bigg)\bigg\} = \frac{\sigma_n^2}{2}\sum_{i=1}^2 \bigg\{\frac{1}{g_{ii}(\mathbf{x}_0)}\bigg(1+\frac{\sigma_n^2}{4} \frac{b_{ii}^2(\mathbf{x}_0)}{g_{ii}^2(\mathbf{x}_0)}\bigg)\bigg\},
\end{equation}
which is a straight forward generalization of the result for $1$:$N$ mappings. For $M$:$N$ mappings, the generalization follows directly as we have $M$ orthogonal curves forming a coordinate grid, and the result is obtained by letting the sum in~(\ref{e:WeakNoiseDist_2ndOrder_2_N}) run from 1 to $M$.  \hspace{5cm}$\square$

\subsubsection{Proof, Lemma~\ref{lem:canal_surf_cond_surfaces}}\label{sec:app_pf_th_Casu}
Condition i) is obvious, and can be seen directly from Fig.~\ref{fig:dim_exp_anom}.

Condition ii): We begin with $1$:$N$ mappings (see~\cite[pp.266-267]{Kreyszig_DiffGeom91}): The spheres $\mathbb{S}^{N-2}\in F$ can be represented as $S_c(z_i,x)=(\mathbf{z}-\mathbf{S}(x))\cdot(\mathbf{z}-\mathbf{S}(x))-r^2 = 0$, $i=1,\cdots,{N}$. Further, ${\partial S_c}/{\partial x} = - 2 \dot{\mathbf{S}}\cdot(\mathbf{z}-\mathbf{S})=0$, $\dot{\mathbf{S}}={\partial  \mathbf{S}}/{\partial x}$, implying that $(\mathbf{z}-\mathbf{S})\perp \dot{\mathbf{S}}$, and
\begin{equation}\label{e:Diff2_congruent_sphere}
\frac{\partial^2 S_c}{\partial x^2} = - 2 \ddot{\mathbf{S}}\cdot(\mathbf{z}-\mathbf{S})+2 = -2\kappa_s \mathbf{p}\cdot(\mathbf{z}-\mathbf{S})+2 = 0.
\end{equation}
The last equality is due to the FoF~(\ref{e:FoF}). With $\rho_s=1/\kappa_s$ we get $\mathbf{p}\cdot(\mathbf{z}-\mathbf{S})-\rho_s = 0.$

$\Rightarrow:$ This condition follows directly from~(\ref{e:Diff2_congruent_sphere}). $\Leftarrow:$ Since $\|\mathbf{p}\cdot (\mathbf{z}-\mathbf{S})\| = \|\mathbf{z}-\mathbf{S}\|=r$, then if $\rho_s > r\ \forall x$, the last equation in~(\ref{e:Diff2_congruent_sphere}) will not have a real solution (or characteristic  points).

With the definition of principal curvature in Appendix~\ref{ssec:app_curvature} it is straight forward to extend the proof to $M$:$N$ mappings (canal hyper surfaces): The result follows directly from $1$:$N$ case by letting a curve $\mathcal{C}$ be LoC with maximal principal curvature for all points of $\mathcal{S}$. That is, $\mathcal{C}$ is always in the direction of the maximal curvature on $\mathcal{S}$. \hspace{8.5cm}$\square$

\subsection{Proofs Section~\ref{sec:mn_dimred}}

{\subsubsection{Proposition~\ref{th:channel_dist}}\label{sec:app_pf_ChDist_lin}
Under Definition~\ref{def:weak_noise_red}, the received signal
$\hat{\mathbf{x}}=\mathbf{S}(\hat{\mathbf{z}})$ can be approximated
by~(\ref{e:lin_appr_red}), where $J(\mathbf{z}_0)\mathbf{n}$ contributes to the distortion. The MSE per source component
given that $\mathbf{z}_0$ was transmitted is then
\begin{equation}\label{e:ev_mn}
\begin{split}
\varepsilon^2_{ch}
&=\frac{1}{M}E\big{\{}(J(\mathbf{z}_0)\mathbf{n})^T(J(\mathbf{z}_0)\mathbf{n})\big{\}}\\
&=\frac{1}{M}E\bigg{\{}\bigg(\frac{\partial{S_1}}{\partial{z_1}}n_1+\cdots+\frac{\partial{S_1}}{\partial{z_N}}n_N\bigg)^2
+\cdots+\bigg(\frac{\partial{S_M}}{\partial{z_1}}n_1+\cdots+\frac{\partial{S_M}}{\partial{z_N}}n_N\bigg)^2\bigg{\}}.
\end{split}
\end{equation}
Since noise on each sub-channel is independent,
$E\{n_i n_j\}=\sigma_n^2\delta_{ij}$. After some rearrangement,
\begin{equation}\label{e:mse_g}
\begin{split}
\varepsilon^2_{ch}
&=\frac{\sigma_n^2}{M}\bigg(\bigg[\bigg(\frac{\partial{S_1}}{\partial{z_1}}\bigg)^2+\cdots
+\bigg(\frac{\partial{S_M}}{\partial{z_1}}\bigg)^2\bigg]+\cdots+\bigg[\bigg(\frac{\partial{S_1}}{\partial{z_N}}\bigg)^2+\cdots+\bigg(\frac{\partial{S_M}}{\partial{z_N}}\bigg)^2\bigg]\bigg)\\&=
\frac{\sigma_n^2}{M}\big(g_{11}+g_{22}+\cdots
+g_{NN}\big)=\frac{\sigma_n^2}{M}\sum_{i=1}^N g_{ii}.
\end{split}
\end{equation}
Expectation w.r.t. $\mathbf{z}$ gives the wanted result.\hspace{9cm}$\square$

\subsubsection{Proof, Proposition~\ref{prop:DimRed_2ndOrder}}\label{sec:app_pf_prop_WN_2nd_order_red}
With $z_0$ transmitted, and noise $n$, we have:
\begin{equation}\label{e:Taylor_M_1}
\mathbf{S}(z_0 + n) \approx \mathbf{S}(z_0) + n \mathbf{S}'(z_0) + \frac{n^2}{2} \mathbf{S}''(z_0)+\frac{n^3}{3!} \mathbf{S}'''(z_0),
\end{equation}
from the 3rd order Taylor expansion. From this we can derive the \emph{channel distortion} as
\begin{equation}\label{e:Derive_ChDist_M1_3rdOrder}
{\varepsilon}_{ch}^2(z_0) = E\bigg\{\bigg\| n \mathbf{S}'(z_0) + \frac{n^2}{2} \mathbf{S}''(z_0)+\frac{n^3}{3!} \mathbf{S}'''(z_0) \bigg\|^2\bigg\}.
\end{equation}
To expand this expression and take the expectation, it is advantages to use arc length parametrization (see Appendix~\ref{sec:app_usp}). Then, $\dot{\mathbf{S}} \cdot \ddot{\mathbf{S}} = 0$, $\dot{\mathbf{S}} \cdot \dddot{\mathbf{S}} = 0$ and $\ddot{\mathbf{S}} \cdot \dddot{\mathbf{S}} = 0$~\cite[pp. 36-37]{Kreyszig_DiffGeom91}. With this in mind, using the fact that $E\{n^a\}= 1\cdot3\cdots (a-1)\sigma_n^a, \ a$ even, and zero otherwise~\cite[p.148]{papoulis02},
 the expectation in~(\ref{e:Derive_ChDist_M1_3rdOrder}) can be found from straight forward calculations
\begin{equation}\label{e:ChDist_M1_3rdOrder_pf}
{\varepsilon}_{ch}^2(z_0) = \sigma_n^2\|\dot{\mathbf{S}}(z_0)\|^2 + \frac{3\sigma_n^4}{4} \|\ddot{\mathbf{S}}(z_0)\|^2 + \frac{5\sigma_n^6}{12} \|\dddot{\mathbf{S}}(z_0)\|^2,
\end{equation}
where $\|\dot{\mathbf{S}}(x_0)\|=1$ according to Theorem~\ref{th:clpar} in Appendix~\ref{sec:app_usp}. The first term in~(\ref{e:ChDist_M1_3rdOrder_pf}) dominates when $\sigma_n$ is small if the $\kappa$ and $\tau$  are sufficiently small (see Eq.~(\ref{e:ChDist_M1_3rdOrder})).

Consider scaled arc length parametrization, i.e., $\|{\mathbf{S}}'(z_0)\|=\alpha\|\dot{\mathbf{S}}(z_0)\|=\alpha$, $\forall z_0$. Then~(\ref{e:CurvatureTorsion_curve_gen_coord}) reduces to $\kappa(z_0)=\|\mathbf{S}_0''(z_0)\|/\|\mathbf{S}_0'\|^2$ (since $\mathbf{S}_0' \perp \mathbf{S}_0''$, still) and the channel error can be expressed up to 2nd order in terms of the signal curves derivatives as in~(\ref{e:ChannelDist_2ndOrder_Final}).

For general hyper surfaces the Taylor expansion for vector valued functions leads to a complicated expression, making it hard to draw conclusions. It is more convenient to consider LoC: By choosing LoC as coordinates on a $M$:$2$ mapping $\mathcal{S}$, then, as for the $M<N$ case, the most direct generalization of the $M$:$1$ case results. Assume that $z_1$ and $z_2$ are along the LoC. Then, according to Theorem~\ref{th:LoC_Coordinates}, $g_{12}=b_{12}=0$.  Therefore $\kappa_i(\mathbf{z}_0)={b_{ii}(\mathbf{z}_0)}/{g_{ii}(\mathbf{z}_0)}$ with $g_{ii}$ and $b_{ii}$ as defined before, but now evaluated w.r.t. the channel variables $z_i$. Then~(\ref{e:ChannelDist_2ndOrder_M_N}) follows.\hspace{2cm}$\square$

\subsubsection{Proof, Proposition~\ref{th:apx_dist}}\label{sec:app_pf_Approx_UniSphere}
Consider an $m$-dimensional vector quantizer (VQ) with equal the distance among each neighboring centroids, $\Delta$, named \emph{uniform} VQ. The distortion of a uniform VQ is lower bounded by assuming $m-1$-spheres as Voronoi regions~\cite{gersho79,Conway_Sloane_99}, a sphere bound. Let the radius of these be $\rho_m$. Then

\begin{lemma}\label{le:vq_dist}
The distortion for a m-dimensional uniform VQ is lower bounded by\footnote{Note that~(\ref{e:approx}) differs from the bound derived in~\cite{gersho79}, since that bound is invariant with respect to size of the quantizer cells. Here we need the distortion to scale with the cell size so it can depend on the SNR.} 
\begin{equation}\label{e:approx}
\bar{\varepsilon}_a^2 \geq E\{\|\mathbf{x}-\mathbf{q}(\mathbf{x})\|^2\}=\frac{m}{4(m+2)}\Delta^2,
\end{equation}
where $\mathbf{q}(\mathbf{x})$ denote its centroids\hspace{11cm}$\square$
\end{lemma}

Since the decision borders of a uniform VQ become spherical as $m\rightarrow \infty$~\cite{gersho79,Conway_Sloane_99,Zamir_Feder_96}, equality is obtained in~(\ref{e:approx}) as $m\rightarrow \infty$ when $\Delta$ is sufficiently small.

Eqn.~(\ref{e:approx}) must be modified to entail S-K mappings. Assuming uniform S-K mapping (Definition~\ref{def:uniform_sk}), then for each point $\mathbf{S}_0 \in \mathbf{S}$ the decision borders for approximating values in $\mathbb{R}^M$ to this point is an $M-N-1$-sphere, $\mathbb{S}^{M-N-1}$ (the family of such sphere $\forall \mathbf{S}_0\in \mathbf{S}$ is a canal surface). The $N-M$ dimensional space where this sphere lies, is orthogonal to $\mathbf{S}$ at $\mathbf{S}_0$, implying that the approximation to an N-dimensional uniform S-K mapping results in the same distortion as that of an $M-N$ dimensional VQ. By substituting $m=M-N$ in~(\ref{e:approx}) and dividing by $M$ the wanted result is obtained.\hspace{13cm}$\square$

\emph{Proof, Lemma~\ref{le:vq_dist}:}
Consider first the special case of high dimensional VQ with spherical Voronoi regions of unit radius $\rho_m=1$ with one centroid at the origin. If the source variance $\sigma_x >> \rho_m$, then the pdf of $\|\mathbf{x}-\mathbf{q}(\mathbf{x})\|$, $f_\mathbf{q}$, will be approximately uniform~\cite{gersho79}. Then one may consider any of the centroids to quantify $\bar{\varepsilon}_a^2$. Let $\mathbf{q}(\mathbf{x})=0$ for simplicity, and let $B_m$ denote the volume contained within the relevant Voronoi region (see~(\ref{e:unit_volum_gamma}) for analytical expression). Then $f_\mathbf{q}=1/B_m, q\in[0,\rho_m]$, and $0$ otherwise. Then
\begin{equation}
\bar{\varepsilon}_a^2 = \int\cdots\int_{B_m}\|\mathbf{x}\|^2 f_\mathbf{q} \mbox{d}\mathbf{x}= \frac{1}{B_m}\int\cdots\int_{B_m}\|\mathbf{x}\|^2  \mbox{d}\mathbf{x}.
\end{equation}
From~\cite[p. 375]{gersho79}, we have  that the moment of inertia
\begin{equation}
\int\cdots\int_{B_m}\|\mathbf{x}\|^2  \mbox{d}\mathbf{x} = \frac{m}{m+1} B_m.
\end{equation}
Therefore $\bar{\varepsilon}_a^2 = m/(m+1)$. As the quantization error scales with the radius of the Voronoi regions, $\rho_m$, the distortion $\bar{\varepsilon}_a^2$, will scale with $\rho_m=\Delta^2/4$.  \hspace{6cm}$\square$

\section{Proofs for Section~\ref{sec:AsymptAnalysis_SK}}\label{sec:App_SKDist_Assymptotic}

\subsection{Proof, Proposition~\ref{th:asympt_dim_exp}:}\label{sec:app_pf_th2}
Assume that the channel signal and the noise are normalized
with $N$. For a power constrained
Gaussian channel, the received vector will lie within an
$N-1$ sphere of radius
\begin{equation}\label{e:rho_n}
\rho_N=\sqrt{P_N+b_N^2\sigma_n^2},
\end{equation}
with high probability. $P_N$ and $\sigma_n^2$  channel signal power and
noise variance per dimension, respectively. With $b_N$ one takes into consideration that $\rho_N$ exceeds
$\sqrt{P_N+\sigma_n^2}$ for finite $N$, so $b_N\rightarrow 1$ as $N\rightarrow \infty$.  Let $B_n$ denote the volume inside an $(n-1)$- sphere of unit
radius~\cite{N_sphere_wiki} 
\begin{equation}\label{e:unit_volum_gamma}
B_n=\frac{\pi^\frac{n}{2}}{\Gamma\big(\frac{n}{2}+1\big)}.
\end{equation}
The volume of the canal surface, $\mathbf{S}^M\times \mathbb{S}^{N-M-1}$, must be smaller than or equal to the channel space volume in order to to satisfy the channel power constraint. If  $\mathbf{S}^M\times \mathbb{S}^{N-M-1}$ has no characteristic points, then locally we have $\mathbb{B}^M\times \mathbb{S}^{N-M-1}$, for all points on $\mathbf{S}$. Therefore
\begin{equation}\label{e:radM}
B_{M} \rho_{M}^M B_{N-M} \rho_{MN}^{N-M}\leq B_{N} \rho_N^N,
\end{equation}
where $\rho_{MN}$ is the canal hyper surface radius  and  $\rho_{M}$ is the radius of the source space.

Further, assume the same decomposition of $\mathbf{n}$ as in Section~\ref{ssec:sph} (see Fig.~\ref{fig:dim_exp_norm}), the $M$-dimensional tangent to $\mathcal{S}$, $\mathbf{n}_{wn}$ and the $N-M$ dimensional normal $\mathbf{n}_{an}$.
To avoid anomalous errors, Proposition~\ref{prop:asympt_an_dist} states that $\rho_{MN}\geq \|\mathbf{n}_{an}\|=\sqrt{((N-M)/N) b_{NM}^2\sigma_n^2}$, where $b_{NM}\rightarrow 1$ as $M,N \rightarrow \infty$. When $M,N$ is large enough,~(\ref{e:radM}) can be written ($b_{N}=b_{NM}=1$ as $M,N\rightarrow \infty$)
\begin{equation}\label{e:ineq_s2}
B_{M} \rho_M^M B_{N-M} \bigg(\frac{N-M}{N}\sigma_n^2
\bigg)^\frac{N-M}{2}\leq B_N (P_N+\sigma_n^2)^\frac{N}{2}.
\end{equation}

With a shape preserving mapping, $\bar{\varepsilon}_{wn}^2$ is determined from $\rho_M$.
Solving~(\ref{e:ineq_s2}) w.r.t. $\rho_M$,
\begin{equation}\label{e:rho}
\rho_{M}\leq \sqrt[M]{\tilde{B}} \sigma_n
 \bigg(\frac{1}{1-M/N}\bigg)^\frac{N-M}{2 M}
\bigg(1+\frac{P_N}{\sigma_n^2}\bigg)^\frac{N}{2 M},
\end{equation}
where from~(\ref{e:unit_volum_gamma})
\begin{equation}\label{e:B}
\tilde{B} = \frac{B_N}{B_M B_{N-M}} =
{\Gamma{\bigg(\frac{N-M}{2} +1\bigg)} \Gamma{\bigg(\frac{M}{2}
+1\bigg)} }\bigg/{\Gamma{\bigg(\frac{N}{2} +1\bigg)}}.
\end{equation}
Eqn.~(\ref{e:B}) can be
expressed through the \emph{Beta function} using~\cite[p. 9]{Bateman53}
\begin{equation}\label{e:beta_func}
\mathcal{B}(\varrho,\varsigma) = \int_0^1 t^{\varrho-1}
(1-t)^{\varsigma-1}\mbox{d}t = \frac{\Gamma(\varrho)
\Gamma(\varsigma)}{\Gamma(\varrho+\varsigma)},
\end{equation}
and the \emph{Functional relation} $\Gamma(a+1)= a\Gamma(a)$~\cite[p.
3]{Bateman53}.
Letting $\varrho= (N-M)/2 +1$ and $\varsigma=M/2+1$, using the
above relations, we obtain
\begin{equation}
\tilde{B} = \bigg(\frac{N}{2} +1 \bigg)
\mathcal{B}\bigg(\frac{N-M}{2}+1,\frac{M}{2}+1\bigg) =
\bigg(\frac{N}{2} +1 \bigg)\mathcal{B}_{(N,M)}.
\end{equation}
As $M$ of $N$ noise components ($\mathbf{n}_{wn}$) contribute to weak noise distortion, we get
\begin{equation}\label{e:s_mse}
\bar{\varepsilon}_{wn}^2 = \frac{E\{\|\mathbf{n}_{wn}\|^2\}}{\rho_M^2} =\frac{M \sigma_n^2}{N \rho_M^2},
\end{equation}
from~(\ref{e:wn_uniform}). With $\rho_{MN}>\|\mathbf{n}_{an}\|$, $\bar{\varepsilon}_{an}^2=0$ from Proposition~\ref{prop:asympt_an_dist}. Then $\bar{\varepsilon}_{wn}^2$ is the total distortion $D_t$.
Assume a fixed $r=N/M$. Substituting
$M=N/r$ and inserting~(\ref{e:rho}) into~(\ref{e:s_mse}) then
\begin{equation}\label{e:weak_noise_asympt}
 D_t = \frac{1}{r}\bigg(1-\frac{1}{r}\bigg)^{r-1} \bigg(\frac{N}{2} +1
\bigg)^{-\frac{2 r}{N}}\mathcal{B}_{(N,r)}^{-\frac{2 r}{N}}
\bigg(1+\frac{P_N}{\sigma_n^2}\bigg)^{-r},
\end{equation}
where
\begin{equation}\label{e:beta_int}
\mathcal{B}_{(N,r)} = \int_0^1 t^{\frac{N}{2r}(r-1)}
(1-t)^{\frac{N}{2r}}\mbox{d}t.
\end{equation}
What is left to show is then
\begin{equation}\label{e:limit1}
\lim_{N\rightarrow \infty } \bigg(\frac{N}{2} +1 \bigg)^{-\frac{2
r}{N}} \mathcal{B}_{(N,r)}^{-\frac{2 r}{N}} =
r\bigg(1-\frac{1}{r}\bigg)^{1-r}.
\end{equation}
Using the product rule for limits~\cite[p.68]{calculus98}, the first term on the left in~(\ref{e:limit1})  is eliminated
since its limit equals 1.  Further, using
H\"{o}lders inequality~\cite[p. 135-136]{Gasquet99}, we get
\begin{equation}\label{e:ineq_exp}
 \mathcal{B}_{(N,r)}\leq\big\|t^{\frac{N}{2r}(r-1)}
(1-t)^{\frac{N}{2r}}\big\|_{\infty},
\end{equation}
with equality as $N\rightarrow \infty$. By differentiation one find that $t_{max} = 1-1/r$ maximizes the norm in~(\ref{e:ineq_exp}), and so
\begin{equation}\label{e:ineq_exp2}
 \mathcal{B}_{(N,r)}=\bigg(1-\frac{1}{r}\bigg)^{\frac{N}{2r}(r-1)}\bigg(\frac{1}{r}\bigg)^\frac{N}{2r},
\end{equation}
as $N\rightarrow \infty$.
Raising both sides of~(\ref{e:ineq_exp2}) to the power $-2r/N$ gives the wanted result\footnote{The above result does not contain $\sigma_x$, but can easily be included by setting $\rho_M = \alpha\sigma_x$, solving~(\ref{e:rho}) with respect to $\alpha$, and substituting
$\alpha$ for $\rho_M$ in~(\ref{e:s_mse}).}.\hspace{1cm}$\square$

\textbf{Comments on finite dimensionality:}
For finite $M,N$ anomalous errors will always have some probability of occurrence as $\|\tilde{\mathbf{n}}\|$ has
nonzero variance. For finite $M,N$, $b_N$ and $b_{NM}$ must be included to account for a nonzero variance around the mean length of both source- and channel vectors. Given a certain probability for anomalous errors, $b_{MN}$ is found from~(\ref{e:mspdf}) by substituting $N-M$ for $N$. Some further elaboration on finite dimensionality was given in~\cite{floor_itw07}.


\subsection{Proof, Proposition~\ref{th:asympt_dim_red}:}\label{sec:app_pf_th_red}
To make $\bar{\varepsilon}_q^2$ small,
$\mathbf{S}\times \mathbb{S}^{M-N-1}$ should \emph{cover} the
source space. With no characteristic points, we are under Definition~\ref{def:local_hycyl_red}, and the following inequality should be satisfied
\begin{equation}\label{e:ineq_comp}
B_N \rho_N^N B_{M-N}\rho_{MN}^{M-N} \geq B_M \rho_{M}^M.
\end{equation}
$\rho_M = \|\mathbf{x}\|=\sqrt{M b_M^2\sigma_x^2}$ is the
radius of the source-space, $\rho_N = \alpha\sqrt{N(P_N +
b_N^2\sigma_n^2)}$ is the radius of the channel space (these are not normalized here), where
$\alpha$ is an amplification factor, $\rho_{MN}=\Delta/2$ is the canal surface radius, and $b_M,b_N \rightarrow
1$ as $M,N \rightarrow \infty$. As in Appendix~\ref{sec:app_pf_th2} these are set to one in what follows. Inserting the above in~(\ref{e:ineq_comp}) and solving w.r.t. $\alpha$, we obtain
\begin{equation}\label{e:K_ineq}
\alpha \geq \sqrt{\frac{M^\frac{M}{N}}{N}}\tilde{B}^\frac{1}{N}
\bigg(\frac{\Delta}{2}\bigg)^{-\frac{M-N}{N}} \sigma_x^\frac{M}{N}
\sigma_n^{-1} \bigg(1+\frac{P_N}{\sigma_n^2}\bigg)^{-\frac{1}{2}},
\end{equation}
where
\begin{equation}\label{e:beta_rel}
\tilde{B} = \bigg(\frac{M}{2} +1 \bigg)
\mathcal{B}\bigg(\frac{M-N}{2}+1,\frac{N}{2}+1\bigg) =
\bigg(\frac{M}{2} +1 \bigg)\mathcal{B}_{(M,N)},
\end{equation}
derived in a similar way as in Appendix~\ref{sec:app_pf_th2}. Assuming a shape preserving mapping and inserting~(\ref{e:K_ineq})
into~(\ref{e:e_ch_uniform}), an expression for $\bar{\varepsilon}_{ch}^2$ is found.
Furthermore, $\bar{\varepsilon}_q^2 $ and $\bar{\varepsilon}_{ch}^2$ can be considered independent under Definition~\ref{def:local_hycyl_red} as they are perpendicular, thus
\begin{equation}\label{e:d_tot_init}
D_{t} =\bar{\varepsilon}_q^2 + \bar{\varepsilon}_{ch}^2 = \frac{M-N}{4M(M-N+2)}\Delta^2 + M^{\frac{M}{N}-1}\tilde{B}^\frac{2}{N}\bigg(\frac{\Delta}{2}\bigg)^{-2\frac{M-N}{N}}
\sigma_x^{2\frac{M}{N}}
\bigg(1+\frac{P_N}{\sigma_n^2}\bigg)^{-1}.
\end{equation}
Differentiating~(\ref{e:d_tot_init}) with respect to $\Delta$,
equating to zero and solving for $\Delta$, we obtain
\begin{equation}\label{e:delt_opt}
\Delta_{opt} =
M^{\frac{M-N}{2M}}\bigg(\frac{4M(M-N+2)}{M-N}\bigg)^\frac{N}{2
M}\bigg(\frac{M-N}{N}\bigg)^\frac{N}{2M}
2^{1-\frac{N}{M}}\tilde{B}^\frac{1}{M} \sigma_x
\bigg(1+\frac{P_N}{\sigma_n^2}\bigg)^{-\frac{N}{2 M}}.
\end{equation}
Inserting~(\ref{e:delt_opt}) and~(\ref{e:beta_rel})
into~(\ref{e:d_tot_init}) and using the relation $N=M r$, with $r\in\mathbb{Q}[0,1]$, we get
\begin{equation}
D_{t}=\bigg(1+\frac{r}{1-r}\bigg)\bigg(\frac{1-r}{1-r+2/M}\bigg)^{1-r}
\bigg(\frac{1-r}{r}\bigg)^r\bigg(\frac{M}{2}
+1\bigg)^\frac{2}{M}\mathcal{B}_{(M,r)}^\frac{2}{M} \sigma_x^2
\bigg(1+\frac{P_N}{\sigma_n^2}\bigg)^{-r}.
\end{equation}
where
\begin{equation}
\mathcal{B}_{(M,r)} = \int_0^1 t^{\frac{M}{2}(1-r)}
(1-t)^{\frac{Mr}{2}}\mbox{d}t.
\end{equation}
We get rid of two terms as
\begin{equation}
\lim_{M\rightarrow \infty }\bigg[\bigg(\frac{M}{2} +1\bigg)^\frac{2}{M},\bigg(\frac{1-r}{1-r+2/M}\bigg)^{1-r}\bigg]=[1,1],
\end{equation}
further using the product rule for limits~\cite[p.68]{calculus98}. From H\"{o}lders inequality~\cite[p. 135-136]{Gasquet99}
\begin{equation}\label{e:ineq}
 \mathcal{B}_{(M,r)}\leq (1-r)^{\frac{M}{2} (1-r)} r^\frac{Mr}{2},
\end{equation}
with equality when $M \rightarrow \infty$, and so
\begin{equation}
\lim_{M\rightarrow \infty }\bigg(1+\frac{r}{1-r}\bigg)
\bigg(\frac{1-r}{r}\bigg)^r\mathcal{B}_{(M,r)}^\frac{2}{M}=\bigg(1+\frac{r}{1-r}\bigg) \bigg(\frac{1-r}{r}\bigg)^r(1-r)^{
(1-r)} r^r=1
\end{equation}
\hspace{16cm}$\square$

\section{Proofs for Section~\ref{sec:Mapping_constr}}\label{sec:App_SK_Examples}

\subsubsection{Proof, Proposition~\ref{prop:Map_Split}}\label{sec:app_pf_MapSplit}

Take the dimension reduction case: The minimal distortion for a $n$:$1$ system is found by solving~(\ref{eq:OPTA_BWrelation}) w.r.t. $D_t$, setting $N=1,M=n$,
\begin{equation}\label{e:OptDist_n_1}
D_{n:1} = \bigg(\frac{\sigma_x^2}{1+\frac{P_{n:1}}{\sigma_n^2}}\bigg)^\frac{1}{n}.
\end{equation}
For an $m$:$1$ system, simply substitute $n$ with $m$ in~(\ref{e:OptDist_n_1}).
With $P_t$, the total power of the $(m+n)$:$2$ system, one can allocate power to the two sub-systems through a factor $\kappa\in[0,1]$ so that $P_{n:1}=\kappa P_t$. Let SNR$=P_t/\sigma_n^2$ and use the fact that $1+x\approx x$ as $x$ becomes large. Then,
\begin{equation}\label{e:OptDist_SplitCompMN_2_tot}
\begin{split}
\lim_{\text{SNR}\rightarrow \infty} D_{t(m+n:2)} &= \lim_{\text{SNR}\rightarrow \infty}\frac{\sigma_x^2}{m+n}\bigg[\bigg(1+\kappa \text{SNR}\bigg)^\frac{1}{n} + \bigg(1+(1-\kappa)\text{SNR}\bigg)^\frac{1}{m}\bigg]\\
  &=\frac{\sigma_x^2}{m+n}\lim_{\text{SNR}\rightarrow \infty} \bigg[\frac{1}{\kappa^{1/n}\text{SNR}^{1/n}} + \frac{1}{(1-\kappa)^{1/m}\text{SNR}^{1/m}}\bigg].
\end{split}
\end{equation}
According to the laws of limits $\lim_{x\rightarrow \infty} \kappa x = \kappa \lim_{x\rightarrow \infty} x$, and so the power allocation factor(s) can be moved outside the limit. Then, for $m>n$ since ${\text{SNR}}^{1/m}$ grows more slowly than ${\text{SNR}}^{1/n}$, $D_{t(m+n:2)}$ will be dominated by the second term in~(\ref{e:OptDist_SplitCompMN_2_tot}) as SNR$\rightarrow \infty$.

In the expansion case, a similar derivation leads to
\begin{equation}\label{e:OptDist_SplitExpM_N_tot}
\begin{split}
\lim_{\text{SNR}\rightarrow \infty} D_{t(2:m+n)} =\frac{\sigma_x^2}{m+n}\lim_{\text{SNR}\rightarrow \infty} \bigg[\frac{1}{\kappa\text{SNR}^{n}} + \frac{1}{(1-\kappa)\text{SNR}^{m}}\bigg].
\end{split}
\end{equation}
If now $m>n$, the first term in~(\ref{e:OptDist_SplitExpM_N_tot}) will dominate as SNR$\rightarrow \infty$.
\hspace{5cm}$\square$

\subsubsection{Proof, Lemma~\ref{lem:SnaSu_Pdf_z1}}\label{sec:app_pf_SnaSu_z1}
The cumulative distribution is given by a straight forward generalization of the $h:\mathbb{R}^2 \rightarrow \mathbb{R}$ case in~\cite[pp. 180-181]{papoulis02}:
\begin{equation}\label{e:SnaSu_CDF_z1}
\begin{split}
&F_{z_1}(z_1)=p_r\{Z_1\leq z_1\}=p_r\{(x_1,x_2,x_3)\in \mathcal{D}_{Z_1}^+\cup \mathcal{D}_{Z_1}^-\}\\
&=\iiint_{\mathcal{D}_{Z_1}^+\cup \mathcal{D}_{Z_1}^-} f_{X_1 X_2 X_3}(x_1,x_2,x_3 ) \mbox{d}x_1 \mbox{d}x_2 \mbox{d}x_3,
\end{split}
\end{equation}
where $f_{X_1 X_2 X_3}(x_1,x_2,x_3)$  is the joint Gaussian distribution and
\begin{equation}\label{e:SnaSu_S_domain}
\begin{split}
&\mathcal{D}_{Z_1}^+ = \bigg\{(x_1,x_2,x_3)\big|(x_1^2+x_2^2+x_3^2)^\frac{n}{2}\leq \rho^n, z_1\geq 0 \bigg\},\\
& \mathcal{D}_{Z_1}^- = \bigg\{(x_1,x_2,x_3)\big|(x_1^2+x_2^2+x_3^2)^\frac{n}{2}\geq -\rho^n, z_1< 0 \bigg\}.
\end{split}
\end{equation}
Then $f_{z_1}(z_1)=\mbox{d}F_{z_1}/\mbox{d}{z_1}$. As $f_{z_1}(z_1)$ is symmetric about the origin for the DSS, one can consider $\mathcal{D}_{Z_1}^+$ only. Since $\mathcal{D}_{Z_1}^+$ is spherical, it is convenient to integrate in spherical coordinates~\cite{Richter}
\begin{equation}\label{e:SnaSu_pdf_int}
f_{z_1}(z_1)=\frac{1}{2}
\frac{\mbox{d}}{\mbox{d}z_1}\int_0^{2\pi}\int_0^\pi\int_0^{a\varphi(z_1)}
f_{\rho}(\rho)\rho^{2}\sin(\theta)\mbox{d}\rho\mbox{d}\theta\mbox{d}\phi,
\end{equation}
where  $f_\rho (\rho)=\exp(-\rho^2/(2\sigma_x^2))/((2\pi)^{3/2} \sigma_x^2)$. The integrals over $\theta, \phi$ become $I_{(\theta, \phi)}=\pi$, and 
\begin{equation}\label{e:rho_int}
\frac{\mbox{d}}{\mbox{d}z_1}\int_0^{a \varphi(z_1)}f_{\rho}(\rho)\rho^{2}\mbox{d}\rho = \frac{1}{(2\pi)^{3/2}\sigma_x^3} \frac{\mbox{d}}{\mbox{d}z_1}\int_0^{a \varphi(z_1)} e^{-\frac{\rho^2}{2\sigma_x^2}}\rho^{2}\mbox{d}\rho=\frac{n a^3 \gamma^3 z_1^{3n-1}}{(2\pi)^{3/2} \sigma_x^3}e^{-\frac{a^2\varphi^2(z_1)}{2\sigma_x^2}}.
\end{equation}
Multiplying with $\pi$, further using absolute value to include negative values, the wanted result is obtained. \hspace{14.5cm}$\square$

\subsubsection{Proof, Proposition~\ref{prop:HVQLC3_2Slope}}\label{sec:app_pf_HVQLC3_2Slope}

With $b_x$ sufficiently large, the probability for anomalies becomes small due to the constraint in~(\ref{e:HVQLC3_2_constraint}), and the last term in~(\ref{e:HVQLC3_2 TotDist}) becomes negligible. A constant gap to OPTA at high SNR then implies that   
\begin{equation}\label{e:HVQLC3_2_Dt_high}
D_t=\frac{\Delta^2}{36} + \frac{2 \sigma_n^2 \alpha^2}{3}= C\cdot \text{SNR}^{-\frac{2}{3}},
\end{equation}
with $C$ some constant. We show that such a constant exists, complying with the KKT problem in~(\ref{e:HVQLC_3_2_ObjFunc}) with $D_t$ as in~(\ref{e:HVQLC3_2_Dt_high}). Let $\kappa=\eta\sqrt{2\pi^5}$. As $\alpha_3$ does not occur explicitly in~(\ref{e:HVQLC3_2_Dt_high}), we eliminate it by equating the constraints in~(\ref{e:HVQLC_3_2_ObjFunc}) to zero and solving w.r.t. $\alpha_3$. Then
\begin{equation}
\alpha_3^2=\frac{\Delta\sigma_x\alpha^2}{\kappa(P_t\alpha^2)}=\frac{\Delta^2\alpha^2}{(2b_x\sigma_x+2b_n\sigma_n\alpha)^2}.
\end{equation}
From this an equation for $\alpha$ results, $(4b_n^2 \sigma_n^2-P_t\Delta \kappa)\alpha^2+ 8b_x b_n \sigma_n \alpha +(4b_x^2+\Delta\kappa) = 0$, assuming $\sigma_x=1$.
With SNR$=P_t/\sigma_n^2$, the solution is
\begin{equation}
\alpha = \frac{-4b_x\pm\sqrt{\Delta\kappa(4\text{SNR}b_x b_n^{-2}+\Delta\kappa_n^{-2}\text{SNR}-4)}}{b_n\sigma_n(4-\Delta\kappa b_n^{-2}\text{SNR} )}\underset{\text{SNR}\rightarrow\infty}{\approx} \frac{\pm \sqrt{\Delta\kappa \text{SNR}(4b_x^2 + \Delta\kappa)}}{\Delta\kappa \sigma_n \text{SNR}},
\end{equation}
where we have used $x+\text{constant}\rightarrow x$ for large $x$ in the last approximation.
Then, since only the positive solution is viable
\begin{equation}\label{e:HVQLC_Ch_dist_High}
\bar{\varepsilon}_{ch}^2 = \frac{2\sigma_n^2}{3}\alpha^2 =\frac{2}{3}\frac{4b_x^2 + \Delta\kappa}{\Delta\kappa \text{SNR}}.
\end{equation}
The distortion contributions should \emph{balance} at high SNR~\cite{hekland05}, i.e., $\bar{\varepsilon}_{ch}^2=\bar{\varepsilon}_{q}^2$, and so $\Delta^2/36 = (C/2)\text{SNR}^{-\frac{2}{3}}$. Therefore $\Delta=3\sqrt{2}\text{SNR}^{-\frac{1}{3}}$, and from~(\ref{e:HVQLC3_2_Dt_high}) and~(\ref{e:HVQLC_Ch_dist_High})
\begin{equation}
D_t=2\bar{\varepsilon}_{ch}^2 = \frac{4}{3}\cdot\frac{4b_x^2 + 3\kappa\sqrt{2C}\text{SNR}^{-\frac{1}{3}}}{3\kappa\sqrt{2C}\text{SNR}^{-\frac{1}{3}} \text{SNR}}=C\cdot \text{SNR}^{-\frac{2}{3}}.
\end{equation}
Therefore
\begin{equation}\label{e:3_2HVQLC_Cons}
\frac{4b_x^2}{3\sqrt{2C}\kappa\text{SNR}^{-\frac{1}{3}}}=\frac{3C}{4} \text{SNR}^{\frac{1}{3}}-\kappa\underset{\text{SNR}\rightarrow\infty}{\approx} \frac{3C}{4} \text{SNR}^{\frac{1}{3}}.
\end{equation}
Solving~(\ref{e:3_2HVQLC_Cons}) w.r.t. $C$, then $C=(16b_x^2/(9\sqrt{2}\kappa))^{2/3}$. \hspace{7cm}$\square$

\subsubsection{Proof, Proposition~\ref{prop:HVQLC2_3Slope}}\label{sec:app_pf_HVQLC2_3Slope}
The Lagrangian for the problem is now
\begin{equation}\label{e:hvqlc_lagrangian_high_SNR}
\mathcal{L}(\Delta,\alpha_2,\lambda)=\frac{\sigma_n^2}{\alpha_2^2} +\lambda_1 \bigg(\frac{\kappa^2\alpha_1^2}{\Delta^4}+\frac{\alpha_2^2 \Delta^2}{12}-P_t\bigg) + \lambda_2 (2 b_n\sigma_n -\alpha_1),
\end{equation}
where $\kappa = 2\eta \pi^2 \sigma_x^2$. Equality constraints are assumed, i.e., $P_t=0.5(P_1+P_2)$ and $\alpha_1 = 2 b_n \sigma_n$, as all the available power should be used, and the HVQLC should fill the channel space as properly as possible under the given constraints. By solving~(\ref{e:ch_pov_hvqlc}) w.r.t. $\alpha_2$ we get $\alpha_2 = {12}({3 P_t}/{2} - {\kappa^2 \alpha_1^2}/{\Delta^4})/{\Delta^2}$.
Then,
\begin{equation}\label{e:hvqlc_Dist_highSNR}
\text{SDR}= \frac{\sigma_x^2}{D_t} = \frac{\sigma_x^2\alpha_2^2}{\sigma_n^2} = \frac{12\sigma_x^2}{\sigma_n^2 \Delta^2}\bigg(\frac{3 P_t}{2} - \frac{\kappa^2\alpha_1^2}{\Delta^4}\bigg)
\end{equation}
The constrained problem over $\Delta, \alpha_1, \alpha_2$ is now converted to an unconstrained problem over $\Delta$. By solving $\partial\text{SDR} / \partial \Delta = 0$ we get $\Delta^* = \sqrt[4]{{12 \kappa^2 b_n^2 }/{\text{SNR}}}$, with $\text{SNR} = P_t/\sigma_n^2$. By inserting this into~(\ref{e:hvqlc_Dist_highSNR}), we get the wanted result. \hspace{9.5cm}$\square$


\ifCLASSOPTIONcaptionsoff
  \newpage
\fi



%


\bibliographystyle{IEEEtran}
\bibliography{IEEEabrv,./references}


%








\end{document}